\documentclass[fleqn,usenatbib]{mnras}

\usepackage{newtxtext,newtxmath}

\usepackage[T1]{fontenc}

\usepackage{graphicx}	
\usepackage{amsmath}	

\graphicspath{{./}{figures/}}

\newcommand{\beq}{\begin{equation}}
\newcommand{\eeq}{\end{equation}}
\newcommand{\MSun}{\,\mathrm{M_{\odot}}}
\newcommand{\Rsun}{\,\mathrm{R_{\odot}}}

\newcommand{\acc}{\MSun\mathrm{yr}^{-1}}
\newcommand{\mbh}{\dot{m}_{\mathrm{BH}}}
\newcommand{\medd}{\dot{m}_{\mathrm{Edd}}}

\title[Formation of SMBH seeds in nuclear star clusters]{Formation of supermassive black hole seeds in nuclear star clusters via gas accretion and runaway collisions}
\author[A. Das et al.]{
Arpan Das,$^{1}$\thanks{E-mail: adas45@uwo.ca}
Dominik R. G. Schleicher,$^{2}$
Nathan W. C. Leigh$^{2,3}$
and Tjarda C. N. Boekholt$^{4}$
\\
$^{1}$Department of Physics and Astronomy, University of Western Ontario, London, Ontario N6A 3K7, Canada\\
$^{2}$Departamento de Astronom\'ia, Facultad Ciencias F\'isicas y Matem\'aticas, Universidad de Concepci\'on\\
$^{3}$Department of Astrophysics, American Museum of Natural History, Central Park West at 79th Street, New York, NY 10024, USA\\
$^{4}$Rudolf Peierls Centre for Theoretical Physics, Clarendon Laboratory, Parks Road, Oxford, OX1 3PU, UK
}

\date{Accepted XXX. Received YYY; in original form ZZZ}

\pubyear{2020}

\begin{document}
\label{firstpage}
\pagerange{\pageref{firstpage}--\pageref{lastpage}}
\maketitle

\begin{abstract}
More than two hundred supermassive black holes (SMBHs) of masses $\gtrsim 10^9\MSun$ have been discovered at $z \gtrsim 6$. One promising pathway for the formation of SMBHs is through the collapse of supermassive stars (SMSs) with masses $\sim 10^{3-5}\MSun$ into seed black holes which could grow upto few times $10^9\MSun$ SMBHs observed at $z\sim 7$. In this paper, we explore how SMSs with masses $\sim 10^{3-5}\MSun$ could be formed via gas accretion and runaway stellar collisions in high-redshift, metal-poor nuclear star clusters (NSCs) using idealised N-body simulations. We explore physically motivated accretion scenarios, e.g. Bondi–Hoyle–Lyttleton accretion and Eddington accretion, as well as simplified scenarios such as constant accretions. While gas is present, the accretion timescale remains considerably shorter than the timescale for collisions with the most massive object (MMO). However, overall the timescale for collisions between any two stars in the cluster can become comparable or shorter than the accretion timescale, hence collisions still play a crucial role in determining the final mass of the SMSs. We find that the problem is highly sensitive to the initial conditions and our assumed recipe for the accretion, due to the highly chaotic nature of the problem. The key variables that determine the mass growth mechanism are the mass of the MMO and the gas reservoir that is available for the accretion. Depending on different conditions, SMSs of masses $\sim10^{3-5} \MSun$ can form for all three accretion scenarios considered in this work. 
\end{abstract}




\section{Introduction} \label{sec:intro}
The observation of more than two hundred SMBHs with masses ${\rm \gtrsim 10^9\MSun}$ at redshift $z\, {\rm \gtrsim 6}$ \citep{Fan01,Will10,Mor11,Wu15,Ban18,Mat18,Wan19,She19,Mat19,ono19,Vit19} has challenged our general understanding of black hole growth and formation. How these massive objects formed and grew over cosmic time is currently one of the biggest puzzles in astrophysics \citep{Vol10,Vol12,Lat17,Gal17,Smi19,Ina19,Lat19}. Our current understanding is that the initial populations of black holes seeds were formed at $z \sim 20-30$~\citep{Ren01}, and then they rapidly grow to their final masses by gas accretion and mergers~\citep{Day19,Pac20,Day21}. In depth reviews about the formation and growth of SMBHs in the early universe can be found in~\citet{Ina19,Lat19,Hae20}.

The discovery of billion solar mass black holes at a time when the Universe was less than 1 Gyr old has created the so called ``seeding'' problem. In the standard Eddington limited accretion scenario a black hole of mass $M_\bullet$ grows exponentially over time,
\beq
\label{growth}
M_\bullet(t)=M_{\bullet,0}\exp\left(\frac{1-\epsilon}{\epsilon}\frac{t}{t_{\mathrm{S}}} \right),
\eeq
where $M_{\bullet,0}$ is the initial mass of the black hole or the `seed' mass, $\epsilon$ is the radiation efficiency with standard value 0.1 \citep{Shapiro05} and $t_{\mathrm{S}}$ is the Salpeter time-scale, $t_{\mathrm{S}}=450$ Myr. In the standard light seeding models the SMBH starts growing from a Pop III stellar remnant of mass $M_\bullet\lesssim 100\MSun$ \citep{Mad01,Abe02,Vol06,Yos08,Hir14,Sta16}. To reach  a mass of $\gtrsim 10^9\MSun$ at $z\gtrsim6$ from a $100\MSun$ Pop III star, the growth process requires continuous Eddington accretion (or nearly continuous Bondi accretion \citep[e.g.][]{leigh13}). However this is extremely difficult to achieve as the radiative and kinetic feedback from the stellar winds slow down the accretion of the gas \citep{Tan09,Reg19}. One possible solution to this problem could be galaxy mergers \citep{Cap15,Vol16,Pac20}, which could lead to the mergers of massive BHs. However, during mergers the large recoil induced by gravitational wave emission can unbind the merger remnants from the shallow potential wells of their host galaxies \citep{Hai04}.

An alternate solution relies on super-Eddington accretion \citep{Vol05,Ale14,Mad14,Vol15,Pac15,Pac17,Beg17,Toy19,Tak20}. Unlike the previous scenario, a temporary non-availability of gas is less of a problem in this scenario, as it can be compensated via highly efficient accretion at other times. However, ensuring a sufficiently high streaming efficiency of the gas from the large-scale reservoir down to horizon scales is difficult to achieve because accretion flows onto seed BHs might be reduced by their own radiation~\citep{Mil09,Sug18,Reg19} and due to the inefficient gas angular momentum transfer~\citep{Ina18,Sug18}. 

Another alternate and more optimistic scenario is the direct collapse of black holes (DCBH) \citep{Oh02,Bro03,Beg06,Aga12,Lat13,Dji14,Fer14,Bas19,Cho20,Luo20}. The seed mass produced by this mechanism is $\sim 10^{4-5}\,\MSun$, sufficiently high for the seed BH to grow to a $10^9\,\MSun$ black hole until $z\sim 7$. The basic idea is that a gas cloud in a massive halo with virial temperature $T \gtrsim 8000$ K and no ${\rm H_2}$ molecules or metals present inside the gas cloud can collapse without fragmentation directly into a SMS which can collapse into seed black holes of similar mass after their lifetime~\citep{Ume16}. To prevent the formation of ${\rm H_2}$ molecules, a very high background UV radiation flux is required \citep{Lat15,Wol17}. Another key requirement for this scenario are large inflow rates of $\sim 0.1\MSun\text{yr}^{-1}$ which can be obtained easily in metal free halos~\citep{Aga12,Lat13,Shl16,Reg18,Bec18,Cho18,Aga19,Lat20}. As a result of the high accretion rate, the SMS becomes inflated and the effective temperature of the SMS drops to several 1000 K~\citep{Hos12,Hos13,Sch13,Woo17,Hae18}. Therefore, the radiative feedback becomes inefficient and the rapid accretion flow continues, allowing the SMS to reach the mass of $\sim 10^{4-5}\MSun$. Recent numerical studies by~\citet{Lat20} and~\citet{Reg20a} have confirmed such large inflow rates lasting for millions of years required to form the DCBHs. However, the presence of even tiny amounts of metals can trigger strong fragmentation through metal line cooling \citep{Omu08, Dop11, Lat16, Mor18, Cho20} and reduce the infall rate. Even in metal enriched halos ($Z < 10^{-3}\,Z_\odot$) where fragmentation takes place, the central massive stars could be fed by the accreting gas and grow supermassive~\citep{Cho20}. Above this threshold value, the gas fragments
into lower mass stars and a single object could not be formed. Alternatively, the requirement of a high infall rate could be met by dynamical heating during rapid mass growth of low-mass halos in over-dense regions at high redshifts~\citep{Wis19}, or by massive nuclear inflows in gas-rich galaxy mergers~\citep{May15}.~\citet{Reg20b} have shown that atomic cooling haloes with higher metal enrichment ($Z > 10^{-3}\,Z_\odot$)  can also be possible candidates for the formation of SMSs in the early universe if the metal distribution is inhomogeneous. As another caveat, we note that even tiny amounts of dust can initiate fragmentation through  enhanced cooling. While the presence of such a strong background radiation is rare, it may be possible to find such conditions \citep{Vis14,Dji14,Reg17}.~\citet{Wis19} and~\citet{Reg20} have recently studied atomic cooling haloes which satisfy the above requirements. \citet{Wis19} have shown that the requirement of the high UV flux can be significantly reduced in halos with rapid merger histories \citep{Ina218} and in halos located in regions of high baryonic streaming velocities \citep{Nao13,Hir17}.  In realistic scenarios, there will always be at least some fragmentation happening even if $\mathrm{H}_2$ cooling is suppressed. Using high resolution numerical simulations ~\citet{Lat13} have found that fragmentation occurs even in the atomic-cooling regime in 7 out of 9 simulations, unless a subgrid model is added which can provide additional viscosity in the simulation. The final mass of the collapsed object may further be limited by  X-ray feedback \citep{Ayk14}. Thus, it motivates to explore other mechanisms of formation of SMBH seed at high redshift.

A third possibility is formation of massive black hole seeds in dense stellar clusters either via runaway collisions of stars, leading to the formation of a supermassive star (SMS)  \citep{Zwa02,Zwa04,Fre07,Fre08,Gle09,Moe11,Lup14,Kat15,Sak17,Boe18,Rei18,Sch19,Ali20} or via the mergers of stellar-mass black holes with other black holes or stars~\citep{giersz15,Has16,Riz21}. In dense high redshift dense stellar clusters, black hole seeds of masses $\sim 10^{2-4}\MSun$ could be formed~\citep{Gre20} which could grow either via tidal capture and disruption events~\citep{Sto17,Ale17,Sak19} or via enhanced accretion and collisions~\citep{Ves10,leigh13,Boe18}. A recent study by \citet{Tag20} has shown that seeds of masses $\sim 10^{5-6}\MSun$ can also be produced via frequent stellar mergers.

Even though the  SMSs are one of the most promising progenitors of the observed high redshift quasars, this hypothesis still requires observational verification. Upcoming next generation telescopes e.g. James Webb Space Telescope (\textit{JWST}), \textit{Euclid} and \textit{WFIRST} will be able to detect SMSs at $z\sim 6-20$ \citep{Sur18,Sur19,Woo20,Mar20}. The number of SMSs per unit redshift per unit solid angle, which are expected to be detected at a redshift $z$, is given by:
\beq
\frac{dn}{dzd\Omega}=\dot{n}_{\mathrm{SMS}}t_{\mathrm{SMS}}r^2\frac{dr}{dz},
\eeq
where $\dot{n}_{\mathrm{SMS}}$ is the SMS formation rate per unit comoving volume, $t_{\mathrm{SMS}}$ is the average lifetime of a SMS as given by eq. \ref{lifetimeSMS} , and $r(z)$ is the comoving distance to redshift $z$ given by:
\beq
r(z)=\frac{c}{H_0}\int_0^z \frac{dz^\prime}{\sqrt{\Omega_m(1+z^\prime)^3+\Omega_\Lambda}}. 
\eeq
However, the actual number of SMSs that will be detected will depend on the sensitivity of the instruments (see discussion below). At present the $\dot{n}_{\mathrm{SMS}}$ is very poorly constrained.

SMSs could be observed both to be cool and red or hot and blue \citep{Sur18,Sur19}. Whether an observed SMS will be red or blue depends on how quickly and persistently the SMS is growing, whether it is rapidly accreting gas from an atomic-cooling halo or growing from runaway collisions in a stellar cluster. For rapidly-accreting SMSs, if the star is growing with a rate of $10^{-3}\MSun\mathrm{yr}^{-1}$ \citep{Hae18} or less, or if accretion is halted for longer than the thermal timescale of the envelope \citep{Saku15}, then the SMS will contract to the main sequence and becomes a hyper-luminous Pop III star and will be observed as blue, implying a spectral temperature of $10^4$~K or higher. For a higher accretion rate $\gtrsim 10^{-3}\MSun\mathrm{yr}^{-3}$ the SMS will be red and bloated~\citep{Omu03,Hos12,Hos13,Hae18}, and the atmospheric temperature will be less than $10^4$~K, implying a negligible amount of UV radiation.~\citet{Hae18} have argued that the critical value decreases below $10^{-2}\MSun\mathrm{yr}^{-1}$ if the mass of the SMS is $\gtrsim 600\MSun$. For SMSs growing from runaway collisions in a cluster, it is assumed that a SMS persistently growing via collisions will similarly remain red and bloated as long as it is steadily bombarded \citep{Rei18}. The deposition of the kinetic energy causes the star to bloat post-collision, which then radiates away on a Kelvin-Hemholtz timescale (Eq.~\ref{lifetimeSMS}). However, if after coalescence the star is allowed to thermally relax, then it will contract to become blue. If the star goes through complete thermal relaxation, it will become a very hot source of ionization \citep{Woo20}. The relevant timescales both for collisions and accretion depend on the environment, but will be of order of one million years or so~\citep[e.g.][]{Boe18}; and also the Kelvin-Helmholtz timescale will be of similar magnitude. However the timescale is expected to depend on environmental conditions e.g. the number density of stars~\citep{Lei17} and may also be enhanced in the presence of accretion. If detected via JWST or other missions, the star may be either in the hot or blue state depending on its environment. In general, however, the SMS phase is expected to be short lived, so the most likely observable would be a SMBH in a star cluster, potentially still with gas in its environment.

Using existing data such as the Hubble Ultra Deep Field, it is currently not possible to detect either of these SMSs, as the AB magnitude limit is 29 at 1.38 $\mu m$, well below that expected for either type of SMS even at $z \sim 6$. However, with JWST which has NIRCam AB magnitude limits of 31.5, it will be possible to detect cooler, redder SMSs at $z\sim 18-20$ and hotter, bluer SMSs at $z\lesssim 13-10$ due to quenching by their accretion envelopes \citep{Sur19}. Even though current Euclid and WFIRST detection limits (26 and 28, respectively) are well below the H band magnitudes of both stars at $z \sim 6-20$, both of them would be able to detect blue SMSs as even a slight amount of gravitational lensing will boost the fluxes of these blue SMSs above the detection limits of these two missions. \citet{Mar20} found some important spectral features in the observational properties of SMSs with mass range $\sim 10^3-5\times 10^4\,\MSun$. They have computed the spectra of SMSs for non-local thermal equilibrium spherical stellar atmosphere models. According to their model, cool SMSs with effective temperatures of $\sim\,10^4$ K will exhibit a Balmer break in emission, which is not expected for normal stars. Hotter SMSs with effective temperatures of $\gtrsim4\times10^4$ will exhibit a Lyman break in emission. However, the resonant scattering of Ly$\alpha$ photons by the neutral IGM at $z > 6$ will be a great challenge for observations. It is important to note that the detections of SMSs discussed in ~\citet{Sur18,Sur19} are for high accretion rate $\sim 1\MSun\mathrm{yr^{-1}}$. Also, it will be difficult to observe the hotter blue supergiants at high redshifts due to the quenching effect mentioned above. Moreover, we do not expect the observations to measure anything related to SMSs directly.  It would purely be a color measurement, which would be possible even if the entire cluster is unresolved and observed as a point-source. As our measurement capabilities increase with the upcoming telescopes to probe higher redshifts, a simple color measurement could be quite useful and informative, purely because these processes we quantify would yield a color inconsistent with a single burst of star formation. So potentially one can just compare to stellar population synthesis models. What future telescopes can do is still an open question.

As another potential caveat, we note the importance to distinguish SMS signatures from other types of stars in massive star clusters, such as extended horizontal branch stars or A-type, which could even be brighter than blue stragglers in the same clusters \citep{Leigh16}. Another challenge related to the observation of blue SMSs will be to distinguish them from hot blue dark stars powered by dark matter annihilation \citep[e.g.][]{Free10,Free16,Sur19}. One way to distinguish them is using the prominent continuum absorption features on the spectra redward of Ly$\alpha$ in the rest frame of the SMS due to the high accretion rate, which are absent from the spectra of blue dark stars. A very prominent Ly$\alpha$ line is found in blue SMS spectra due to pumping of the accretion envelope by high-energy UV photons from the star. These spectral features will be really important to distinguish blue SMSs from hot dark stars of similar mass. A dark star will not have these spectral features due to the absence of a dense accretion shroud. 

Studies have found that many galaxies harbor massive NSCs \citep{Car97,Bok02,leigh12,leigh15,Geo16} with masses of $\sim 10^{4-8}\MSun$. Many of these galaxies host a central SMBH \citep{Cor13}. Interestingly, studies have found correlations between both the SMBH mass and the NSCs mass with the galaxy mass \citep{Fer06,Ros06,leigh12,sco13,Set20}. In many galaxies NSCs and SMBHs co-exist \citep{Set08,Gra09,Geo16,Ngu19}. Galaxies like our own \citep{Sco14}, M31 \citep{Ben05} and M32 \citep{Ngu18} host a SMBH and a NSC in the center. It therefore makes sense to consider a link between NSCs and SMBHs. Studies have shown that SMBH seeds could be formed inside NSCs via runaway tidal
encounters~\citep{Sto17} or from core collapse and stellar collisions~\citep{Dev09,Dev10,Dav11} or by gas inflows~\citep{Lup14}. Recent studies by~\citet{Kro20} and~\citet{Nat21} investigated black hole accretion inside NSCs. In depth reviews about NSCs can be found in~\citet{Neu20}.

In this paper, we explore high-redshift, metal-poor NSCs as the possible birthplaces of SMSs via gas accretion and runaway stellar collisions. Low metallicity may favor the formation of very dense clusters, as fragmentation will occur at higher density, while the mass loss from winds becomes negligible \citep{Vin01}, thereby making this formation channel more effective. In addition, in low metallicity environoments the gas will be warmer, thereby contributing to higher accretion rates through a higher speed of sound.\footnote{In the regime where self-gravity regulates fragmentation and at least the initial masses of the clumps, the accretion rate can be estimated as the Jeans mass over the free-fall time, leading to a dependency on sound speed cubed. } Here we explore how runaway collisions between stars and gas accretion onto the stars can lead to the formation of a SMS in such clusters. 

We use N-body simulations to model runaway collisions and gas accretion in dense NSCs. We describe our simulation setup in Sec. \ref{simulation}. Our results are presented in Sec.~\ref{Results} and the final discussion is given in Sec.~\ref{Discussion} along with a summary of our main conclusions.

Similar work had been done by \citet{Boe18,Rei18} for Pop III stellar clusters. The most important distinctions between our work and the previous works are that we study clusters with higher initial stellar mass ($10^5\MSun$), higher number of stars (5000), a Salpeter type initial mass function (IMF), and more sophisticated and physically motivated accretion scenarios. We further provide a more detailed analysis here on the comparison of the timescales for collisions and accretion.



\section{Simulation Setup}
\label{simulation}
The complicated physical processes operating together in stellar clusters include gravitational N-body dynamics, gravitational coupling between the stars and the gas, accretion physics, stellar collisions and mass growth due to accretion and collisions.  Properly accounting for all of these physical processes provides a significant challenge with respect to the modeling. Moreover, it is computationally costly to model clusters consisting mostly of massive main sequence (MS) stars, due to the large number of stars. We use the Astrophysical MUlti-purpose Software Environment (AMUSE) \citep{Por09,Por13,Pel13,Por18} to model the nuclear clusters and include all the physics mentioned above. AMUSE is very efficient in including new physics such as the mass-radius relation, accretion physics, collisional dynamics, and coupling all of these to existing N-body codes. In this section, we present and discuss all the physical processes that we include in our simulation and how we couple them into the numerical model. 

\subsection{Initial Conditions}
We model the NSCs with MS stars embedded in a stationary gas cloud. For simplicity, we assume that the stars and the gas are equally distributed, i.e.,  both the gas mass ($M_{\mathrm{g}}$) and gas radius ($R_{\mathrm{g}}$) are equal to the mass ($M_{\mathrm{cl}}$) and radius ($R_{\mathrm{cl}}$) of the stellar cluster. Both the cluster and gas follow a Plummer distribution with the same characteristic Plummer radius \citep{Plu11}. The Plummer density distribution is given as
\beq
\rho(r)=\frac{3M_{\mathrm{cl}}}{4\pi b^3}\left( 1 + \frac{r^2}{b^2}\right)^{-\frac{5}{2}},
\eeq
where $M_{\mathrm{cl}}$ is the mass of the cluster and $b$ is the Plummer length scale or the Plummer radius that sets the size of the cluster core. The Plummer length scale of the cluster (or the gas) is equal to one fifth of $R_{\mathrm{cl}}$ (or $R_{\mathrm{g}}$). We introduce a cut-off radius, equal to five times the Plummer radius, after which the density is set to zero in order to create a stable cluster with finite total mass and to make sure each star initially is within the gas cloud. The initial parameters for the numerical setup are $M_{\mathrm{cl}}$, $R_{\mathrm{cl}}$, $M_{\mathrm{g}}$, $R_{\mathrm{g}}$, and the number of stars $N$. 
For the initial mass of the stars we consider a Salpeter initial mass function (IMF) \citep{Sal55} given by :
\beq
\xi(m)\Delta m = \xi_0 \left( \frac{m}{\MSun}\right)^{-\alpha}\left( \frac{\Delta m}{\MSun}\right),
\eeq
with three different mass ranges, $10\MSun-100\MSun,\,10\MSun-120\MSun,\,10\MSun-150\MSun$, assuming a power-law slope of $\alpha = 2.35$. Since our main goal is to explore the interplay between accretion physics and the dynamics of the runaway collisions producing a massive object, we do not take into account complicating factors such as the effect of cluster rotation or an initial binary fraction, which would also be highly uncertain in the context considered here, while  significantly increasing the computational expense of our simulations.

The gravitational interactions between the stars are modelled using the N-body code ph4 based on a fourth-order Hermite algorithm. The gravitational effect of the gas cloud is included via an analytical background potential coupled to the N-body code using the BRIDGE method \citep{Fuj07}. This allows us to incorporate the gravitational forces felt by the stars both due to the gas and the stellar component.  In other words, the motions of the stars are controlled by the total combined potential of the gas and stars.  As explained in more detail in the subsequent sections, we consider direct accretion onto stars but do not factor in gas dynamical friction.  This is because in previous works it has been shown to be inefficient, although over sufficiently long timescales it can contribute to cluster contraction \citep{leigh13b,leigh14}.

\subsection{Mass-radius relation}
\begin{figure}
	\includegraphics[width=\columnwidth]{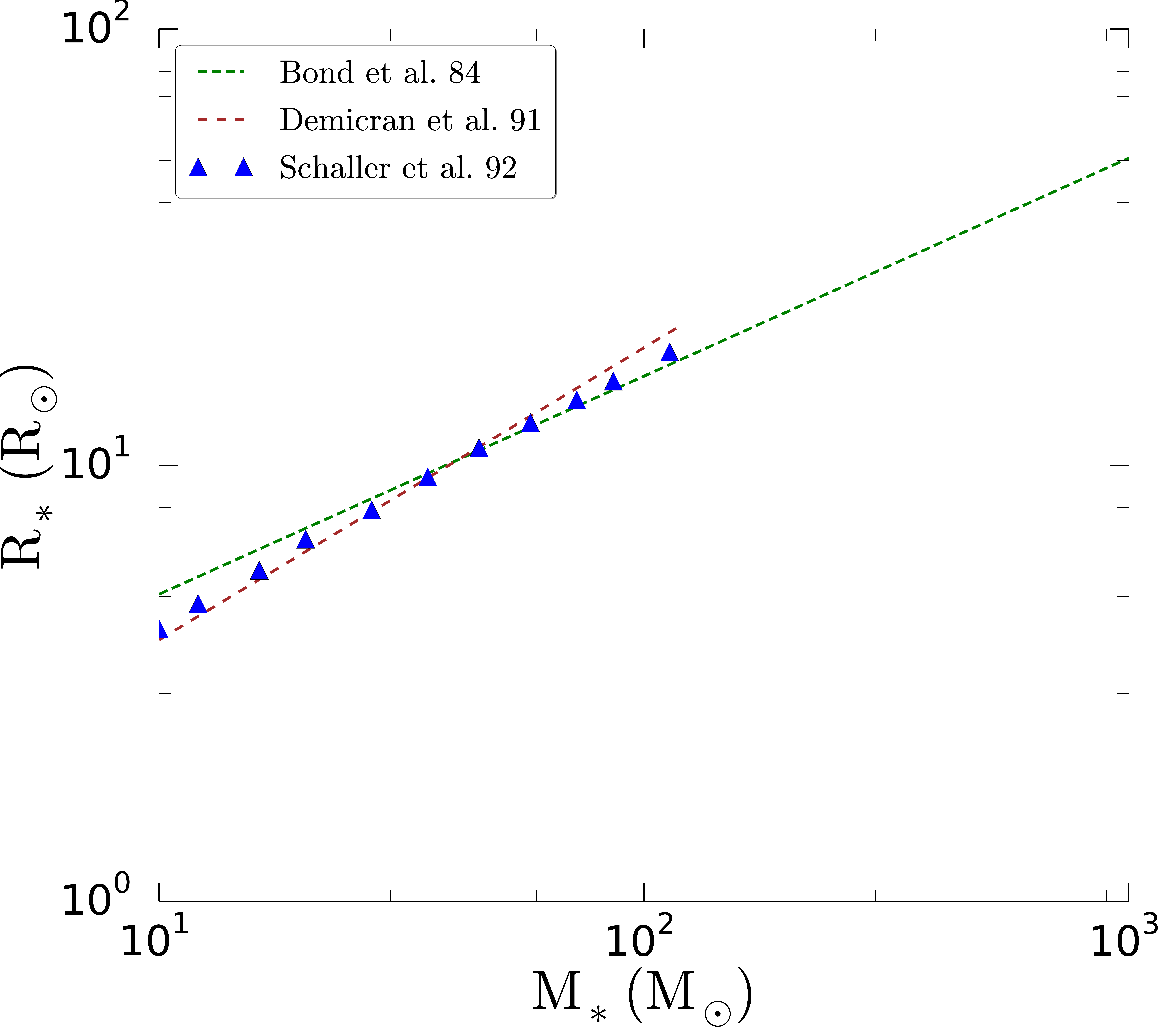}
	\caption{$M_\ast-R_\ast$ relations for MS stars from different studies. For masses in the range $10-120\MSun$ we adopted the \citet{Dem91,Sch92} data. The green dashed lines show the data from \citet{Bon84}. Even though the model is for really massive stars of mass $\gtrsim 10^4\MSun$, we can see it fits perfectly for masses $\gtrsim 50\MSun$.}
	\label{mass}
\end{figure}

The mass-radius ($M_\ast-R_\ast$) relation of the stars will play an important role in determining the number of collisions via the collisional cross section. Since the runaway collisions have to start before the most massive stars in the cluster turn into compact remnants, it is justified to consider all stars to be initially on the MS. In Fig. \ref{mass} we show the $M_\ast-R_\ast$ relations obtained from different studies \citep{Bon84,Dem91,Sch92}. The \citet{Dem91} and \citet{Sch92} data points match quite well below $M_\ast\lesssim 50\MSun$, while the \citet{Bon84} and \citet{Sch92} data points match quite well above $M_\ast\gtrsim 50\MSun$. We therefore combine these relations and adopt a $M_\ast-R_\ast$ relation as outlined below. The $M_\ast-R_\ast$ relation for stars more massive than $M_\ast\gtrsim 100\MSun$ is of crucial importance, as it determines the size of the runaway collision product. Unfortunately, the radius of stars with $M_\ast\gtrsim 100\MSun$ is poorly understood. SMSs may develop a very extended envelope of very low density \citep{Ish99,Bar01}. However, this extended envelope contains only a few percent of the total stellar mass, so it may play a negligible role in collisions. Moreover, such extended envelopes are not present in low metallicity main sequence stars \citep{Bar01}, which are our main focus in this paper.  Overall, the relation we employ is given as
\begin{eqnarray}
\label{less50}
\frac{R_\ast}{\Rsun}&=&1.60\times\left(\frac{M_\ast}{\MSun} \right)^{0.47}\,\,\,\, \mathrm{for}\,\, 10\MSun\lesssim M_\ast<50 \MSun,\\
\label{great50}
\frac{R_\ast}{\Rsun}&=&0.85\times\left(\frac{M_\ast}{\MSun} \right)^{0.67}\,\,\,\, \mathrm{for}\,\, 50\MSun\lesssim M_\ast,
\end{eqnarray}
where Eq. \ref{less50} is adopted from \citet{Bon84} and Eq. \ref{great50} is adopted from \citet{Dem91}.
One key quantity in our work is the stellar lifetime. The MS lifetime is given as \citep{Har88}
\beq
\tau_{\mathrm{MS}}=10^{10}\left(\frac{\mathrm{\MSun}}{M_\ast} \right)^{2.5}\,\text{yr}.
\eeq
For a $M_\ast\sim 20 \MSun$  it is $\sim$ 5.6 Myr. However, for $M_\ast\gtrsim 40\MSun$, the lifetime is almost constant ($\sim 3.5-5$ Myr) and independent of metallicity \citep{Hur00}. On the other hand, the SMS evolves over Kelvin-Helmholtz (KH) timescale \citep{Jan02}
\beq
t_{\mathrm{KH}}=6.34 \times 10^8 \left(\frac{M_\ast}{\mathrm{\MSun}} \right)^{-1} \,\text{yr}.
\label{lifetimeSMS}
\eeq
We stop our simulation at 5 Myr as we are  mostly interested in the early phases of cluster evolution corresponding to the most rapid rates of black hole seeds mass growth, in order to quantify the attainable final black hole masses for each of our simulation models. We do not take into account the change in stellar radius due to stellar evolution of MS stars despite the fact that stellar evolution could play an important role determining the final mass of the SMS \citep{Gle13,Kat19}. We also ignore the rotation of the stars which could significantly influence the evolution of the massive stars and hence the final mass of the SMS~\citep{Mae00,Leigh16}. Our main goal here is to build a simple model and to better understand the complicated physics involved in the study.

\subsection{Gas accretion}
Gas accretion plays a crucial role in our simulations as the stellar masses and radii can increase significantly depending on the accretion model, and hence the number of collisions will be increased. In our model the gas that is being accreted by the stars is assumed to be in a stationary state. We impose momentum conservation during the accretion process, which means the stars will slow down as they gain mass, and fall into the potential well of the cluster. The mass accreted onto the stars is removed from the (static) gas, leading to its depletion and eventually to a transition where the cluster evolution is determined only by N-body dynamics. However, gas motions do not follow the stellar motions and as a result sometimes accretion could add or remove momentum, less or more, respectively. In order to model the gas dynamics we need hydrodynamic simulations. 

We have considered different accretion scenarios in our work. As a first simplified choice we study the effect of a constant accretion rate three different accretion rates of $10^{-4},\,10^{-5}$ and $10^{-6} \MSun\mathrm{yr^{-1}}$, which roughly correspond to the Eddington and Bondi accretion rates under different assumptions for the mass of the central object. Even though it is highly unlikely that all stars in the cluster will accrete at the same constant rate, and individual star accretion might even be episodic \citep{Vor06,Vor15}, this ad-hoc approximation is a good starting point to understand how the interplay of accretion and collisions works in the cluster. 

Next we consider the Eddington accretion rate given as
\beq
\label{eddington}
\dot{M}_{\mathrm{Edd}}=\frac{4\pi G M_\ast}{\epsilon\kappa c_{\mathrm{s}}},
\eeq
where $c_{\mathrm{s}}$ is the speed of sound in the gas, $\epsilon$ is the radiative efficiency, i.e. the fraction of the rest mass energy of the gas that is radiated and $\kappa$ is the electron scattering opacity. Using $\kappa=0.4\,\mathrm{cm^2\,g^{-1}}$ and $\epsilon=0.1$, Eq. \ref{eddington} can be written as
\beq
\dot{M}_{\mathrm{Edd}}=2.20\times 10^{-8}\left(\frac{M_\ast}{\MSun} \right)\,\,\mathrm{\MSun yr^{-1}}.
\label{eddacc}
\eeq

Finally we consider the Bondi-Hoyle-Lyttleton accretion. Simulations \citep{Bon01} have shown that in larger clusters the stars accrete unequally, with the stars near the core accreting more than those near the outer envelopes of the cluster. This is due to the fact that the gas mass is accumulating near the core due to the cluster potential where it can be accreted by the stars. Even initially uniform clusters show a position dependent accretion as the gas and stars redistribute themselves in the host cluster potential. As a result of the position-dependent accretion, the final configurations show a significant amount of mass segregation. 

In principle the accretion rate of a star will depend on its cross section $\pi R_{\mathrm{acc}}^2$, where $R_{\mathrm{acc}}$ is the accretion radius, on the gas density $\rho_\infty$, and the relative velocity of the star with respect to the gas $v_\infty$, as
\beq
\label{accretion}
\dot{M}=\pi v_\infty \rho_\infty R_{\mathrm{acc}}^2.
\eeq

In the original Hoyle–Lyttleton (HL) problem the accretion radius in the supersonic regime is given as \citep{Hoy39,Hoy40,Hoy240}
\beq 
R_{\mathrm{HL}}=\frac{2GM_\ast}{v_\infty^2},
\eeq
which leads to the HL accretion rate:
\beq
\dot{M}_{\mathrm{HL}}=\frac{4\pi G^2M_\ast^2\rho_\infty}{v_\infty^3}.
\eeq
However, \citet{Bon52} defined the Bondi radius as
\beq
R_{\mathrm{B}}=\frac{GM_\ast}{c_{\mathrm{s}}^2}.
\eeq
The flow outside this radius is subsonic and the density is almost uniform, while inside the Bondi radius the gas becomes supersonic. This led \citet{Bon52} to propose an interpolated Bondi-Hoyle (BH) formula: 
\beq 
\dot{M}_{\mathrm{BH}}=\frac{2\pi G^2M_\ast^2\rho_\infty}{(v_\infty^2+c_{\mathrm{s}}^2)^{3/2}}.
\eeq
Studies have shown that there will be an extra factor of 2 which leads to the final accretion formula given as 
\beq 
\dot{M}_{\mathrm{BH}}=\frac{4\pi G^2M_\ast^2\rho_\infty}{(v_\infty^2+c_{\mathrm{s}}^2)^{3/2}}.
\eeq
This accretion rate can be written as in Eq. 2 of \citet{Macc12}:
\begin{equation}
\label{mainbondi}
\dot{M}_{\mathrm{BH}} = 7\times 10^{-9}\ \left(\frac{M_\ast}{\MSun}\right)^2\left(\frac{n}{10^{6}\, \mathrm{cm}^{-3}} \right)^2\left(\frac{\sqrt{c_{\mathrm{s}}^2+v_\infty^2}}{10^6\,\mathrm{cm\ s}^{-1}} \right)^{-3}\mathrm{\MSun yr^{-1}}.
\end{equation}
A recent study by \citet{Kaa19} has shown that the BH accretion rate will depend on the characteristic accretion radius of the cluster $R_{\mathrm{acc}}$ and  the mean separation between stars $R_{\bot}$, where $R_{\bot}=R_{\mathrm{cl}} N^{-1/3}$. The average accretion rate of an individual star is given as 
\begin{equation}
\label{kaaz}
  \langle \dot{M}_{\mathrm{BH}} \rangle=\left\{
  \begin{array}{@{}ll@{}}
    \dot{M}_{\mathrm{BH}}, & \text{when}\ R_{\bot} \gg R_{\mathrm{acc}}, \\
    N\times\dot{M}_{\mathrm{BH}}, & \text{when}\ R_{\bot} \leq R_{\mathrm{acc}}
  \end{array}\right.
\end{equation} 
In our work we consider a BH accretion rate given by Eq. \ref{kaaz}, where $\dot{M}_{\mathrm{BH}}$ is given by Eq.~\ref{mainbondi}.

\subsection{Handling collisions}
We adopt the sticky-sphere approximation to model collisions between the main sequence stars (see, for example, \citet{leigh12b, Lei17}. If the distance between the centers of two stars is less than the sum of their radii, we assume that the stars have merged and we replace them with a single object whose mass is equal to the sum of the masses of the colliding stars. We consider the collision product to be a MS star and the radius of the object to be determined by the $M_\ast-R_\ast$ relation described in Eqs. \ref{less50} and \ref{great50}. We assume that linear momentum is conserved during the collision.

However, studies have shown that the mass is not necessarily conserved during the collision of stars \citep{Sil02,Dal06,Tra07}. \citet{Ali20} have shown that the final mass of the colliding objects could change a lot depending on the mass loss rate. \citet{Gle13} have shown that the mass loss rate depends on the type of stars colliding. In our study, all the stars are MS stars and we have assumed both a mass loss recipe as given by Eq. 3 of \citet{Kat15}, 
\begin{equation}
    \Delta M=\text{min}\left[0.062\frac{M_2}{0.7M_1},0.062\right](M_1+M_2),
\label{katzmassloss}
\end{equation}
where $M_1$ and $M_2$ are the masses of the colliding stars, as well as constant mass loss rates of $3\%$ and $5\%$. Henceforth, we refer to Eq.~\ref{katzmassloss} as K15.

\subsection{Stability of numerical setup}
\begin{figure}
\includegraphics[width=\columnwidth]{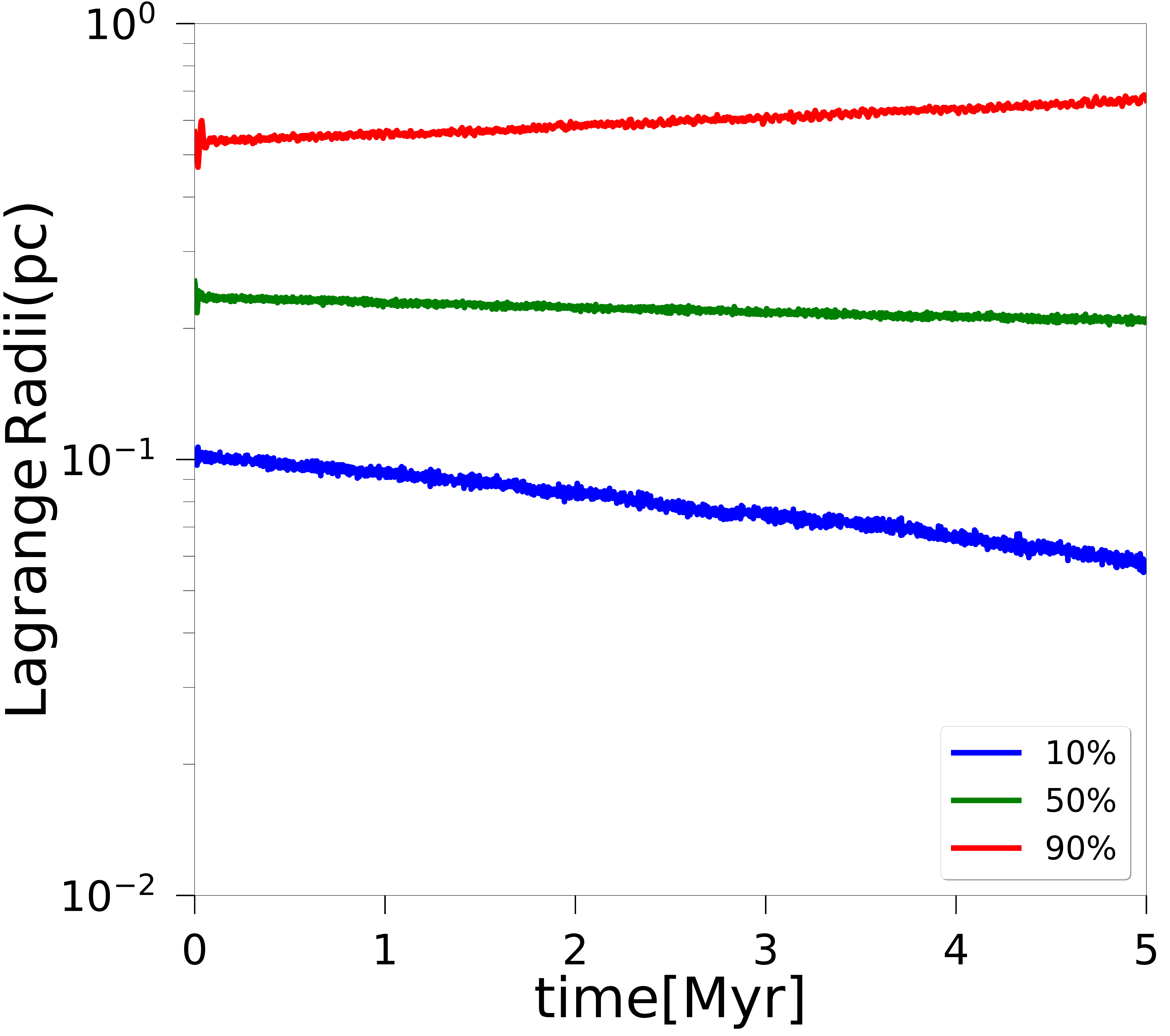}
\caption{Evolution of the Lagrange Radii for a test run without accretion with $N=5000$ particles, $M_{\mathrm{cl}}=M_{\mathrm{g}}=1.12\times 10^5 \MSun$, $R_{\mathrm{cl}}=R_{\mathrm{g}}=1$~pc, assuming a Salpeter IMF with mass range $10-100 \MSun$.}
\label{reference}
\end{figure}

While our work is based on the setup previously tested and employed by \citet{Boe18} and \citet{Ali20}, we test the stability pursuing a reference run where both the accretion and collisions are deactivated, to show that our setup corresponds to an overall stable initial condition. For this setup, we employ $N=5000$ particles, $M_{\mathrm{cl}}=M_{\mathrm{g}}=1.12\times 10^5 \MSun$, $R_{\mathrm{cl}}=R_{\mathrm{g}}=1$~pc, and we assume a Salpeter IMF with masses between $10-100 \MSun$. As accretion and collisions are switched off, we find the gas and stellar mass to be consistent. The time evolution of the Lagrangian radii which is the radius of an imaginary sphere around the centre of the stellar cluster containing a fixed fraction of its mass, is shown in Fig.~\ref{reference}.

\section{Results}\label{Results}
The main results of our simulations are presented in this section. We start by presenting our reference run in section~\ref{secreferencerun}. The dependence on the prescription for the accretion rate is explored in section~\ref{acc}, and an analysis in terms of collision and accretion time scales is provided in section~\ref{sectimescales}. The influence of the IMF is examined in section~\ref{imf}, the dependence on the cluster radius in section~\ref{radialdependence}, and  the implications of mass loss are finally examined in section~\ref{secmassloss}. The summary of the initial conditions and results is presented in table~\ref{tab:result}.

\subsection{Reference run}\label{secreferencerun}
\begin{figure*}
    \includegraphics[width=\columnwidth]{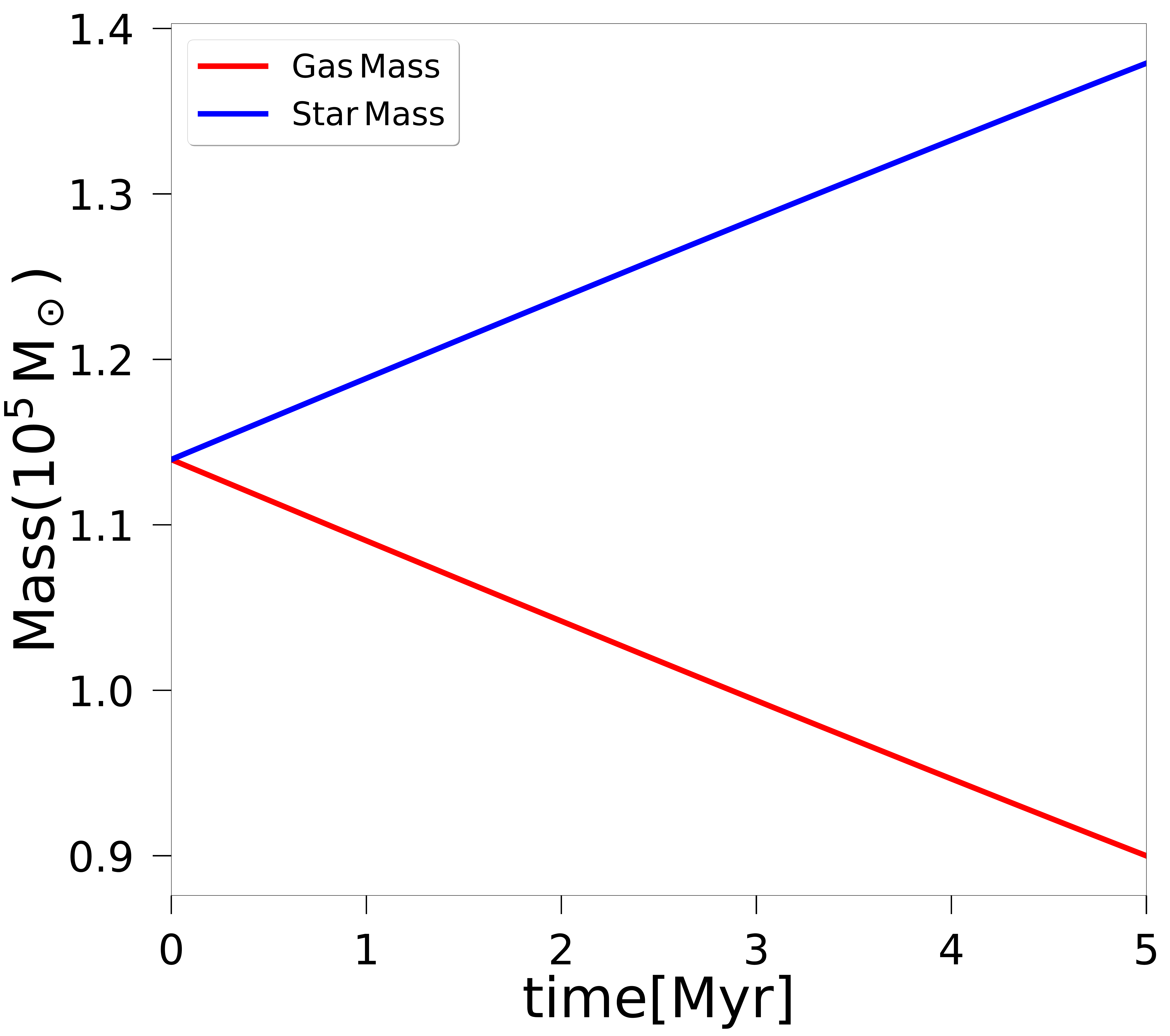}
    \includegraphics[width=\columnwidth]{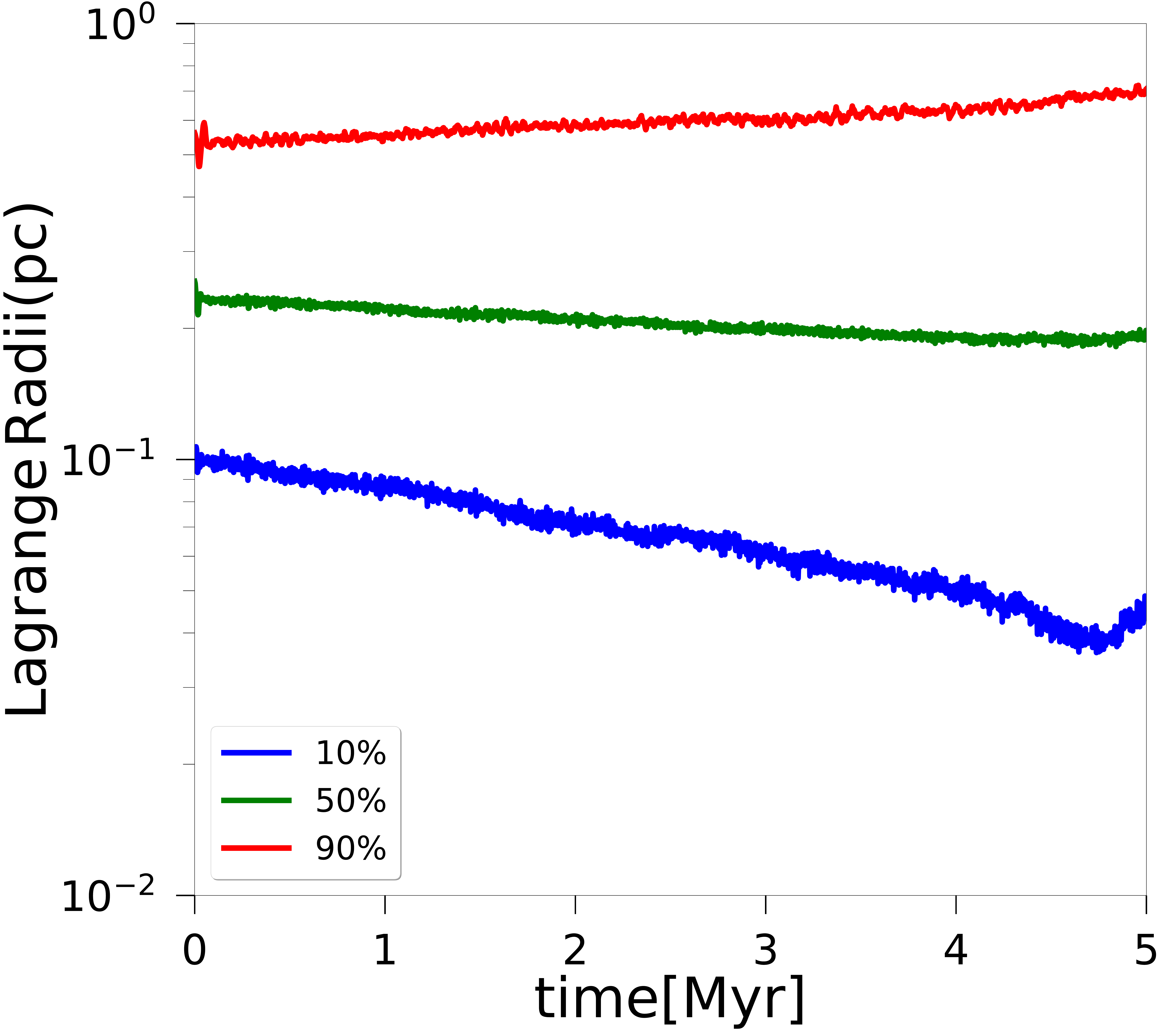}
    \includegraphics[width=\columnwidth]{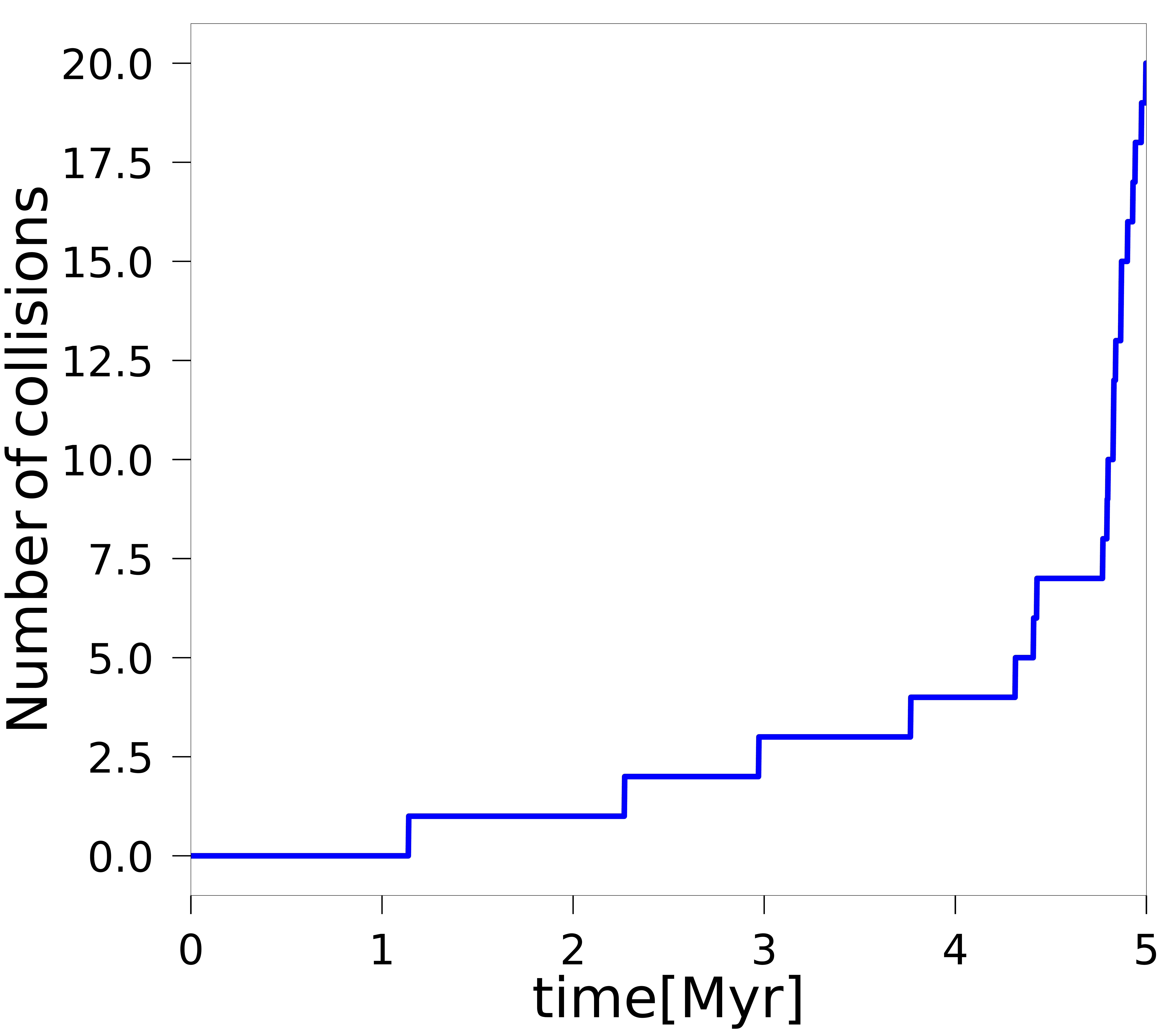}
    \includegraphics[width=\columnwidth]{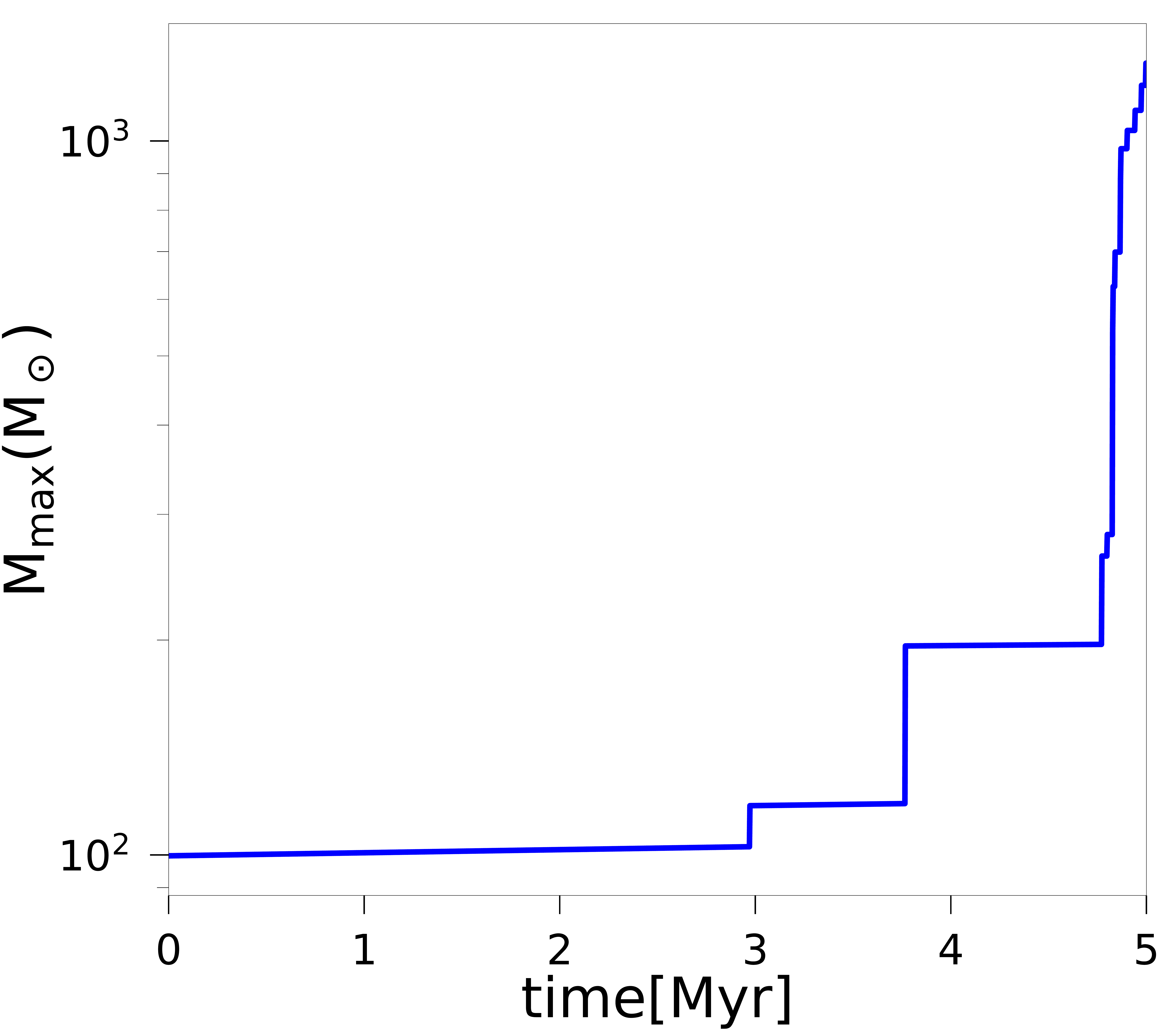}
    \caption{Simulation with $N=5000$, $M_{\mathrm{cl}}=M_{\mathrm{g}}=1.12\times 10^5 \MSun$, $R_{\mathrm{cl}}=R_{\mathrm{g}}=1$~pc, assuming a Salpeter IMF between $10-100 \MSun$ and an accretion rate of $\dot{m}=10^{-6} {\MSun\mathrm{yr^{-1}}}$. The top left panel shows the evolution of gas and stellar mass as a function of time, the top right panel the evolution of the Lagrangian radii, the bottom left panel the total number of collisions as a function of time, and the bottom right panel the evolution of the mass of the MMO as a function of time.}
    \label{acc-6}
\end{figure*}

We start by describing our reference run with $N=5000$, $M_{\mathrm{cl}}=M_{\mathrm{g}}=1.12\times 10^5 \MSun$, $R_{\mathrm{cl}}=R_{\mathrm{g}}=1$ pc, assuming a Salpeter IMF within a stellar mass interval of $10-100 \MSun$ and a constant accretion rate of $\dot{m}=10^{-6}\MSun\,\mathrm{yr}^{-1}$. The result of the simulation is shown in Fig.~\ref{acc-6}. In this configuration, accretion occurs rather gradually, and the gas mass decreases only by about $20\%$ over 5 Myr, while the stellar mass increases by the same amount. As shown in \citet{leigh14} and as reflected in the evolution of the Lagrangian radii, the inner part of the cluster, containing $10\%$ of the mass, goes through contraction and even the radius containing $50\%$ of the mass shows slight contraction, while the outer $90\%$ radius is moderately expanding. The first collision in the cluster occurs after about $1$ Myr and in total about $20$ collisions occur over 5 Myr, where the time interval between the collisions decreases at later times, potentially due to the larger cross section of the central massive object. It is important to say that not all these collisions are with the central object, but nevertheless they lead to the formation of more massive objects within the star cluster, and particularly in the last 1.5 Myr the collision products eventually merge with the central object, thereby contributing to its growth. 

\begin{figure*}
    \includegraphics[width=\columnwidth]{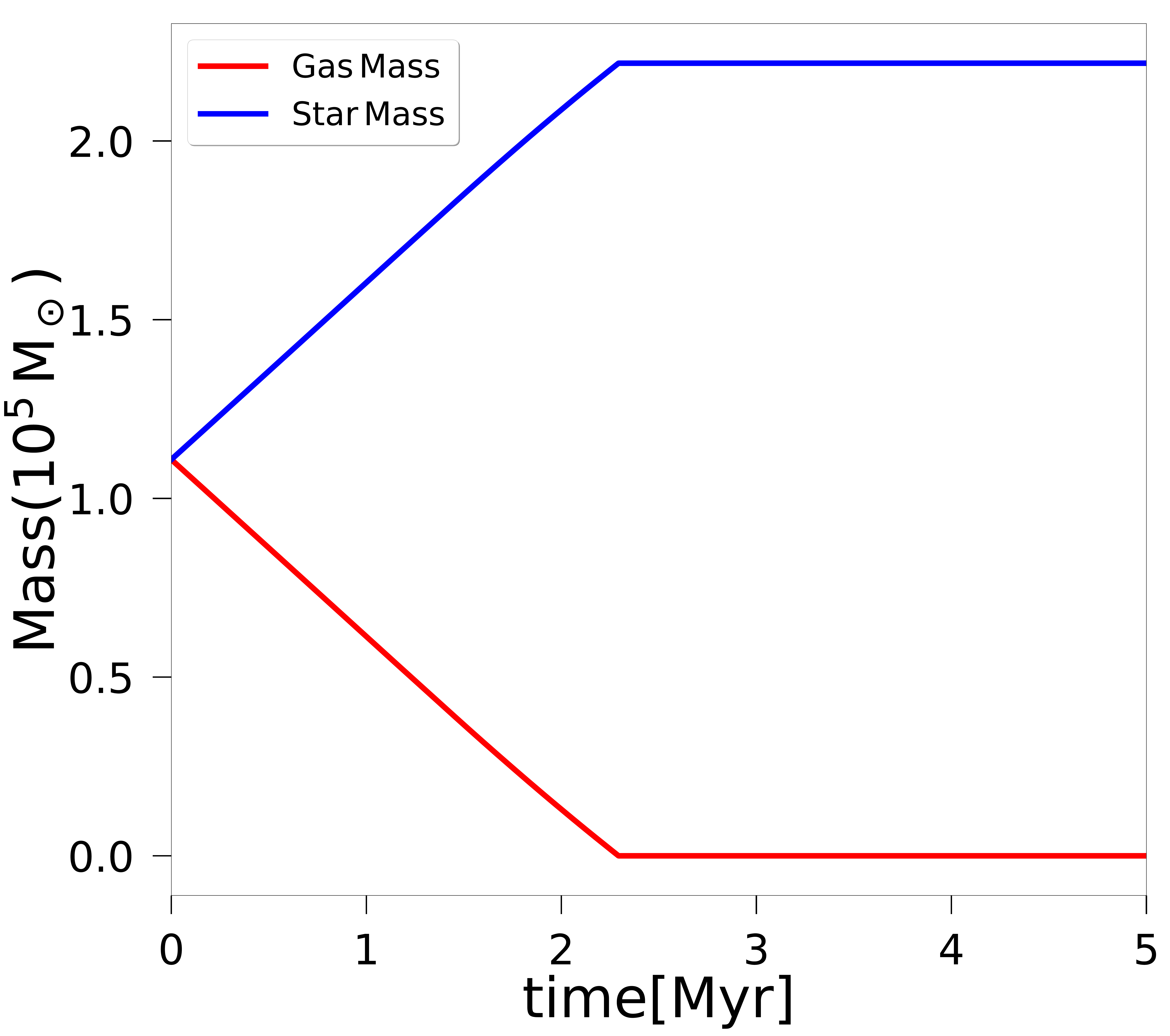}
    \includegraphics[width=\columnwidth]{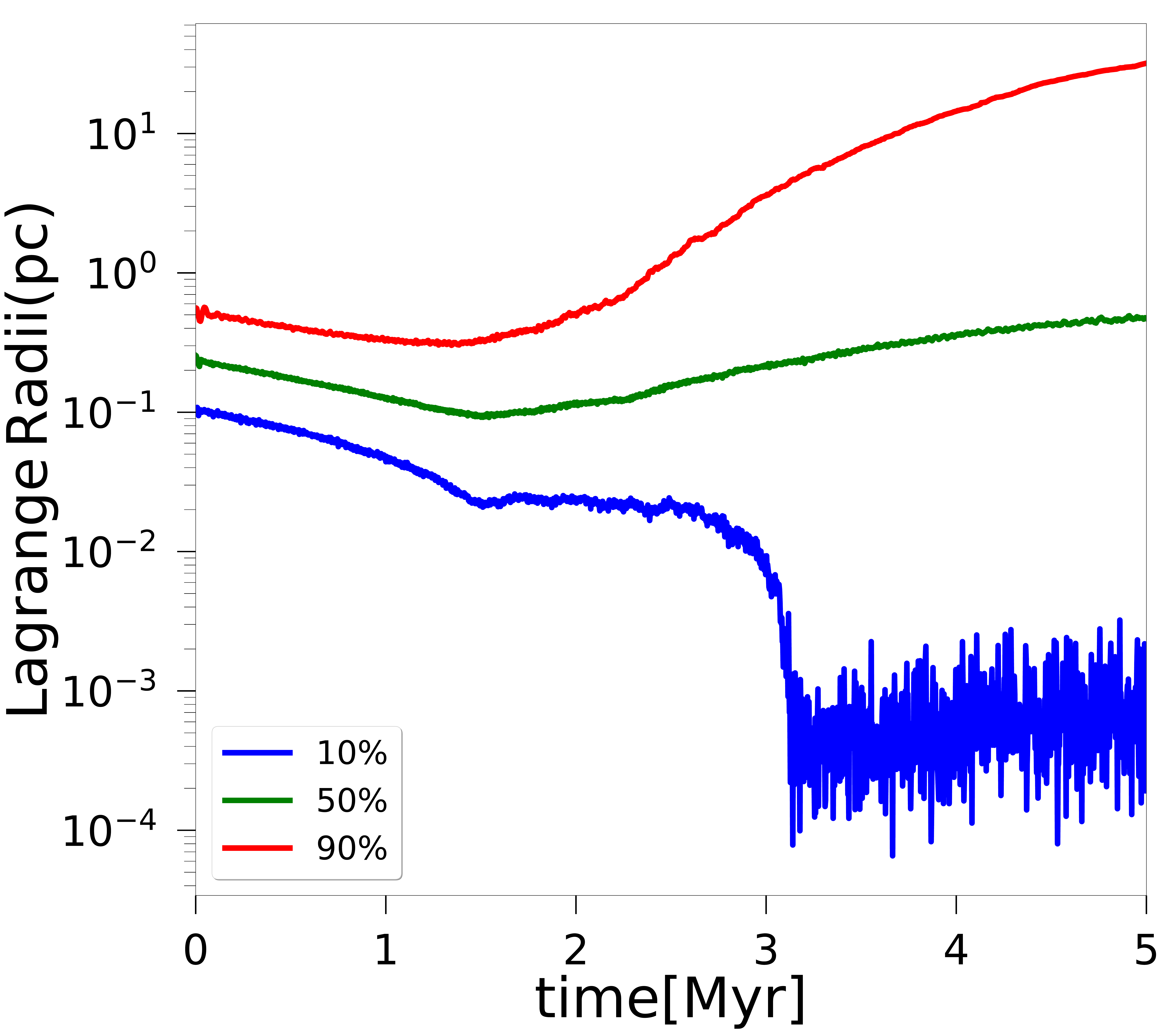}
    \includegraphics[width=\columnwidth]{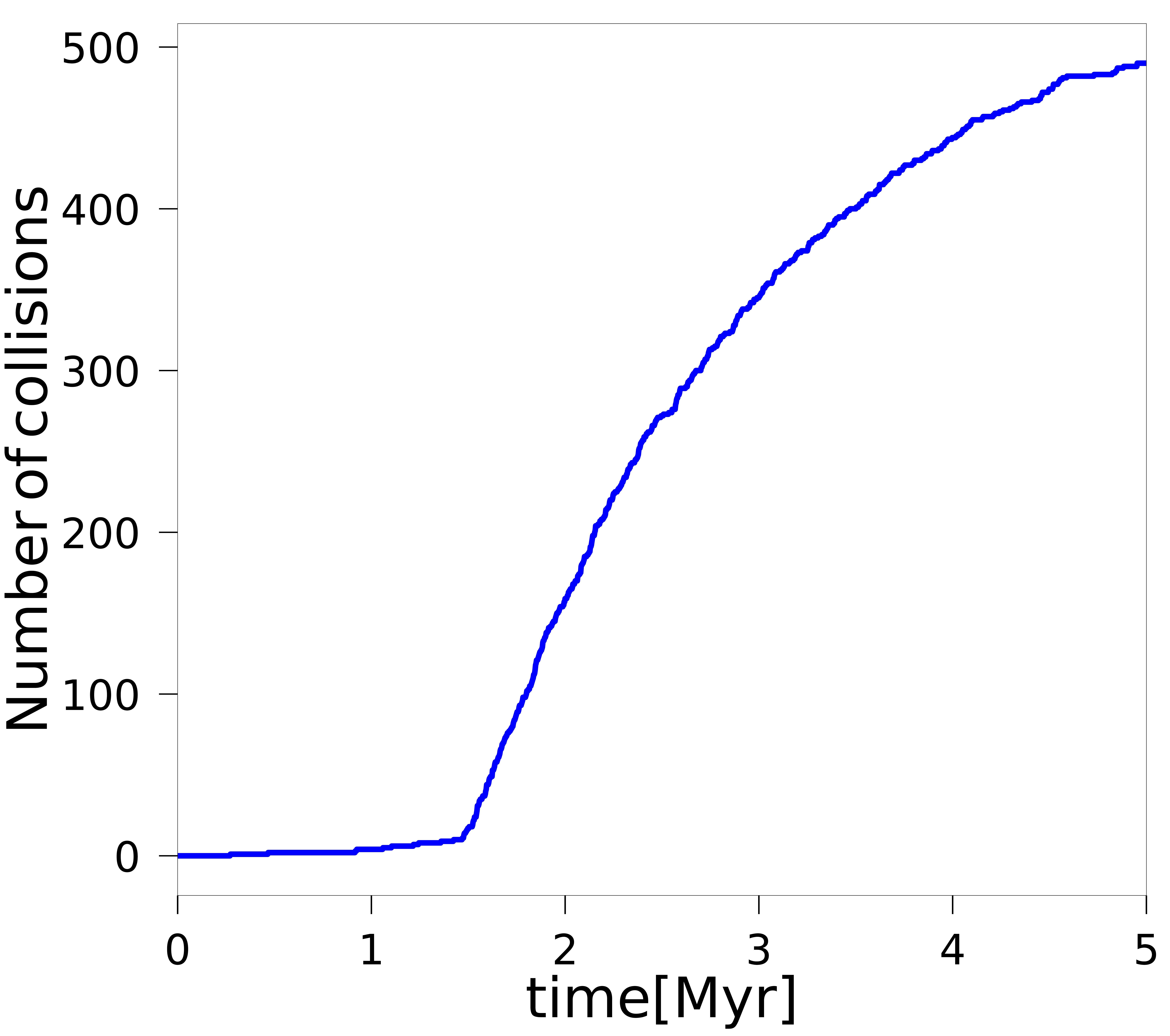}
    \includegraphics[width=\columnwidth]{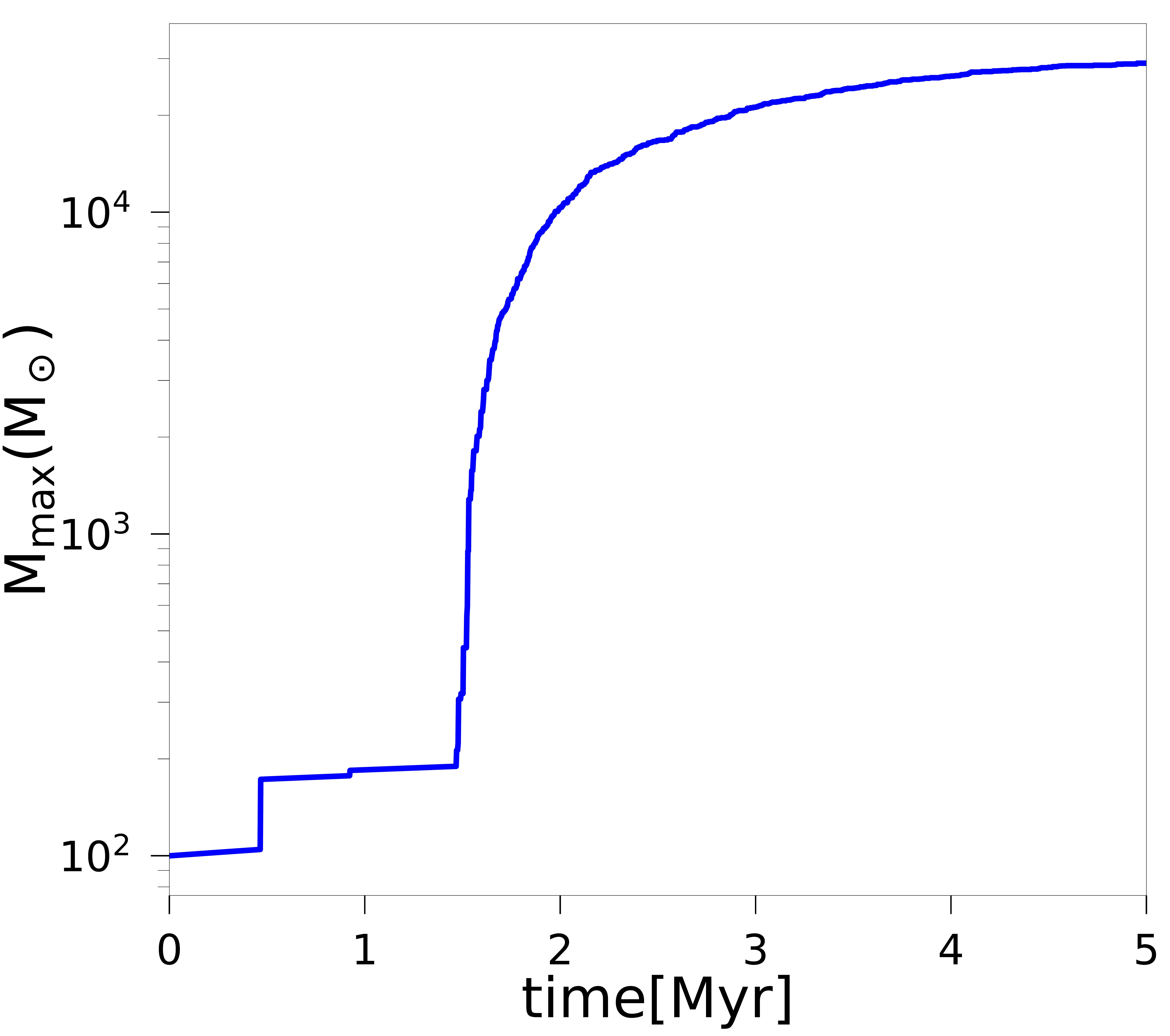}
    \caption{Same as in Figure \ref{acc-6}, but assuming instead $\dot{m}=10^{-5} {\MSun\mathrm{yr^{-1}}}$.}
    \label{acc-5}
\end{figure*}

We now explore how these results depend on the adopted value of the accretion rate. We first note that if we decrease the accretion rate by a factor of $5$, no collisions occur and the results are consistent with a simulation where the accretion and collisions are switched off. On the other hand, if the accretion rate is increased to a value of $\dot{m}=10^{-5}\MSun\,\mathrm{yr}^{-1}$, it has a considerable impact on the evolution of the cluster (see Fig.~\ref{acc-5}). First, we note in this case that the gas mass will be fully depleted after slightly more than 2 Myr, and the stellar mass therefore reaches a total value of $2.24\times10^5 \MSun$. Stellar collisions are considerably enhanced. They occur relatively early on and increase more rapidly after about 1.5 Myr, reaching a total of about $500$ collisions within 5 Myr. The mass of the MMO reaches about $3\times10^4 \MSun$ in the same time, predominantly driven by the merger of collision products. The $10\%$ Lagrangian radius undergoes initially moderate contraction as before. It however decreases rapidly after $3$ Myr, as the accretion rapidly accelerates core collapse; our highest accretion rates achieve core collapse within one million years (see below).  The $50\%$ Lagrangian radius slightly decreases for the first $1.3$ Myr and subsequently shows moderate expansion. The trend is very similar for the $90\%$ Lagrangian radius. This trend has been seen in previous simulations \citep[e.g.][]{leigh14}, especially after the formation of a massive object in the center \citep{Boe18, Ali20}.

\begin{figure*}
    \includegraphics[width=\columnwidth]{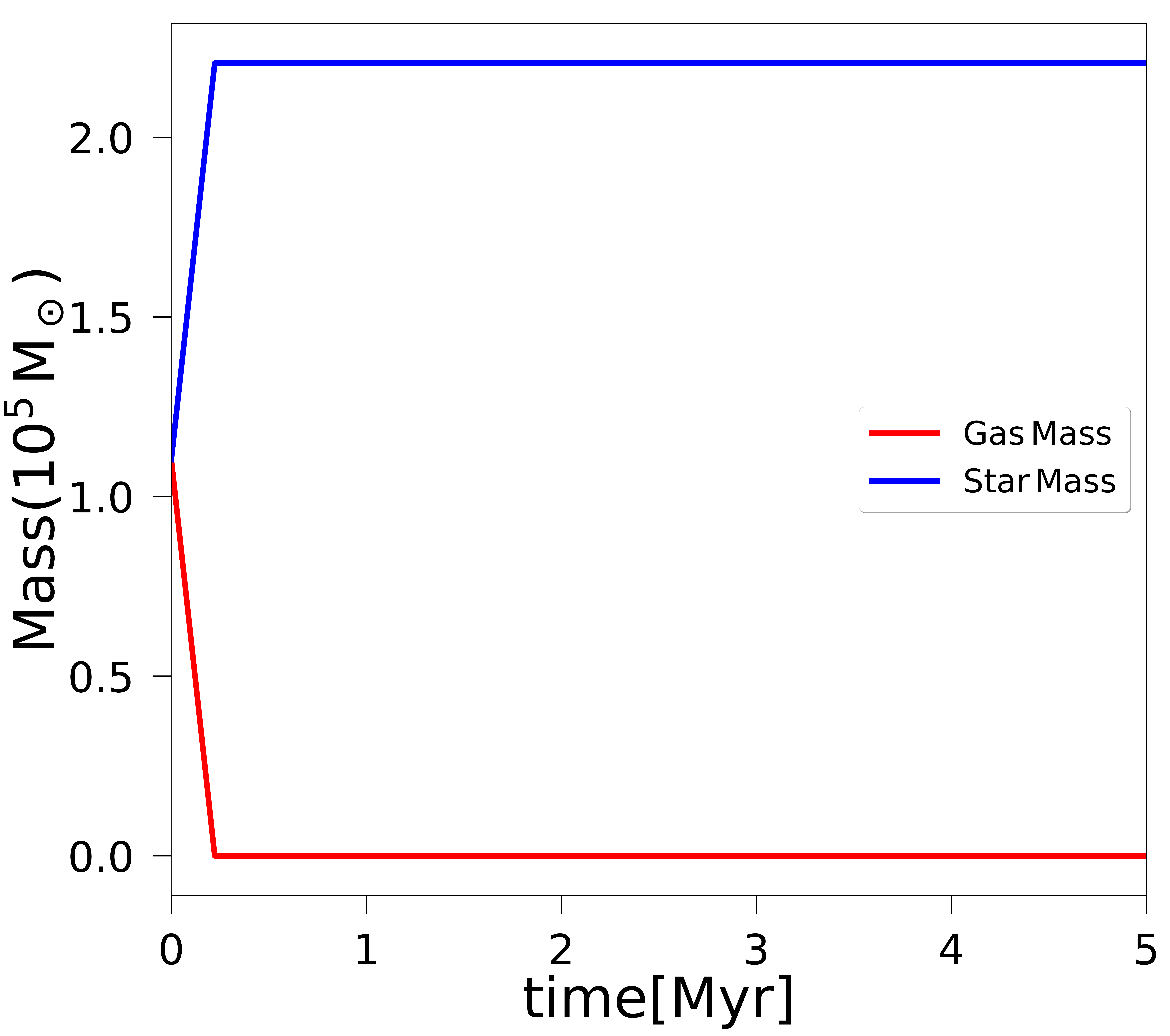}
    \includegraphics[width=\columnwidth]{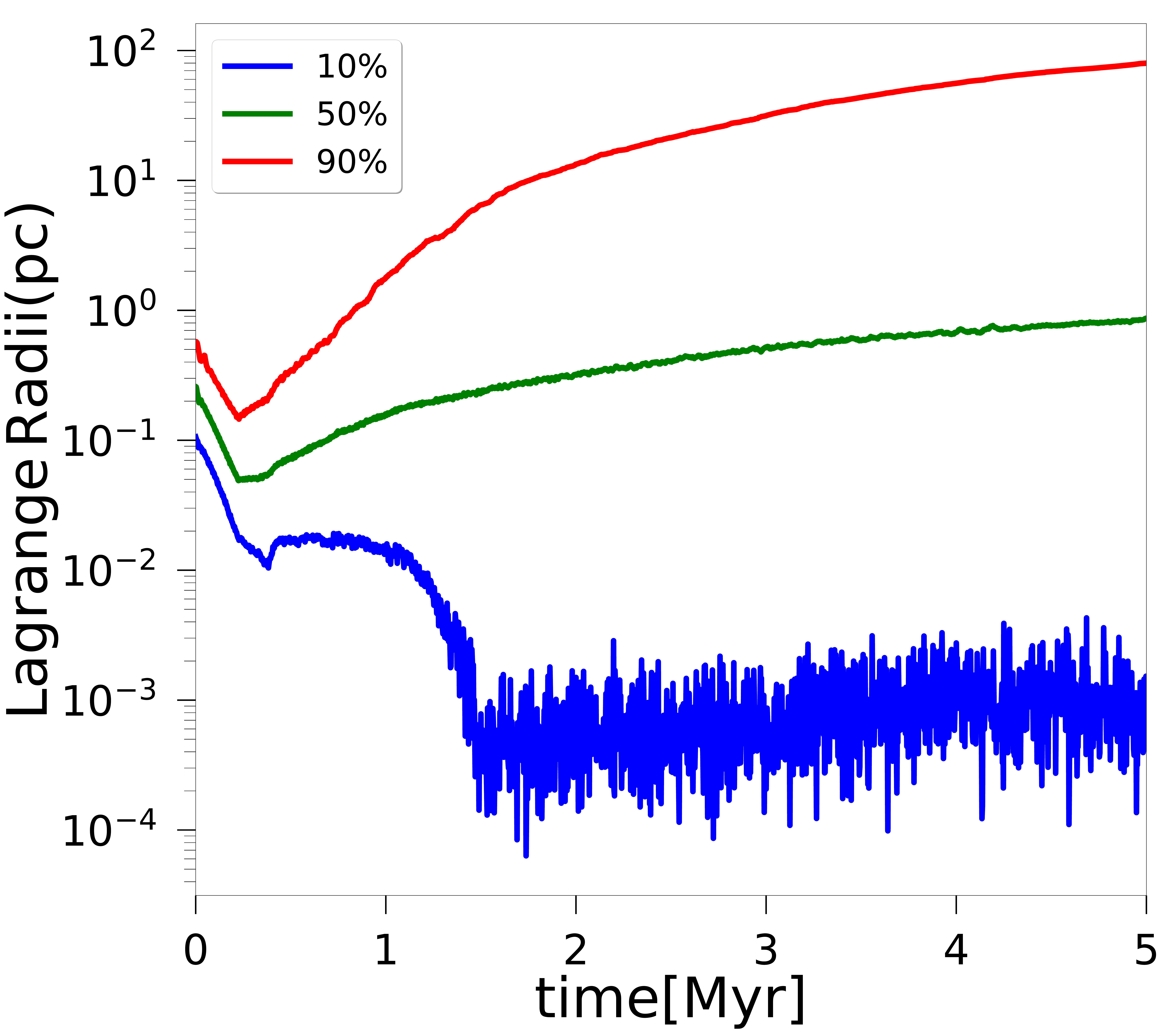}
    \includegraphics[width=\columnwidth]{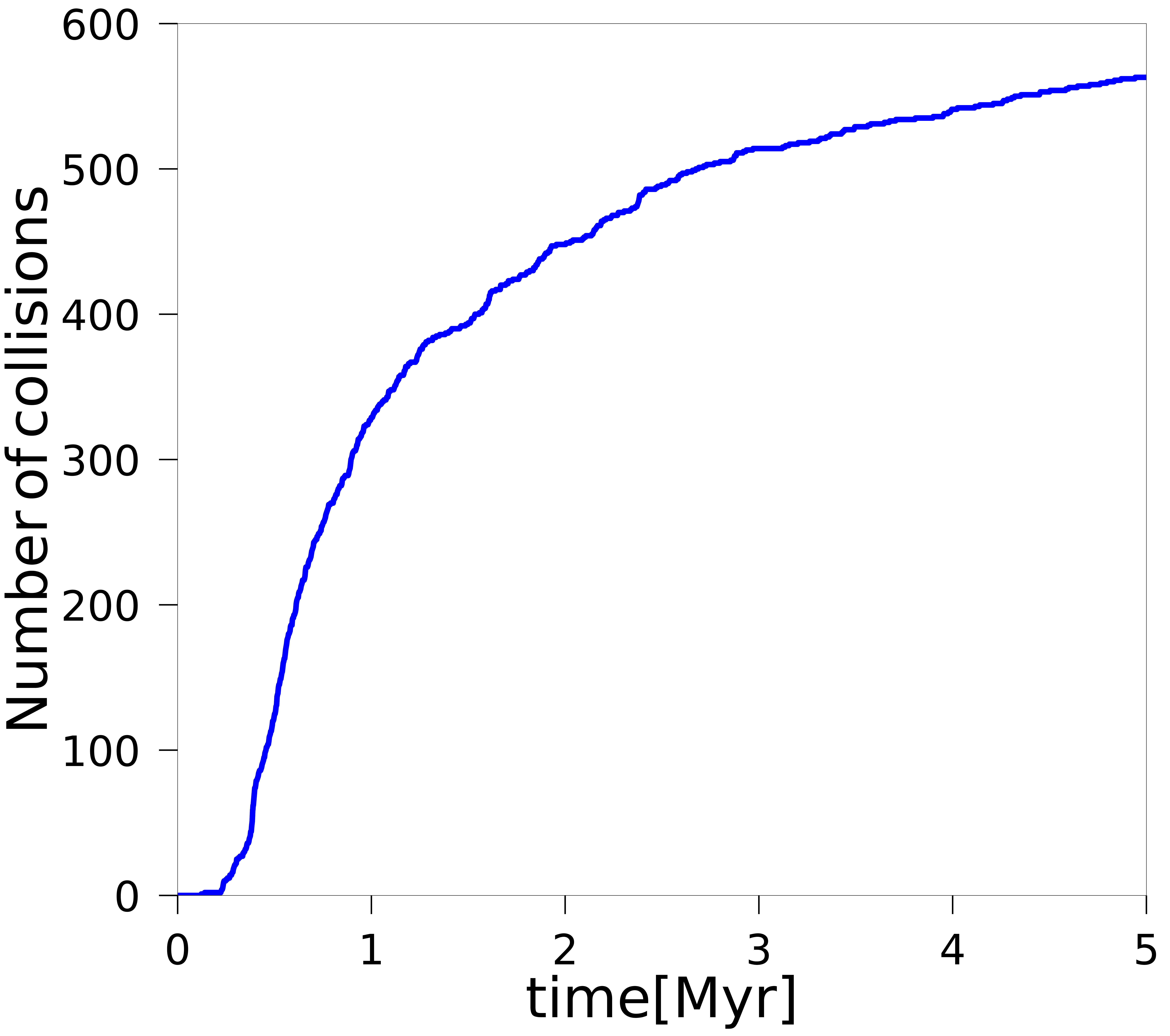}
    \includegraphics[width=\columnwidth]{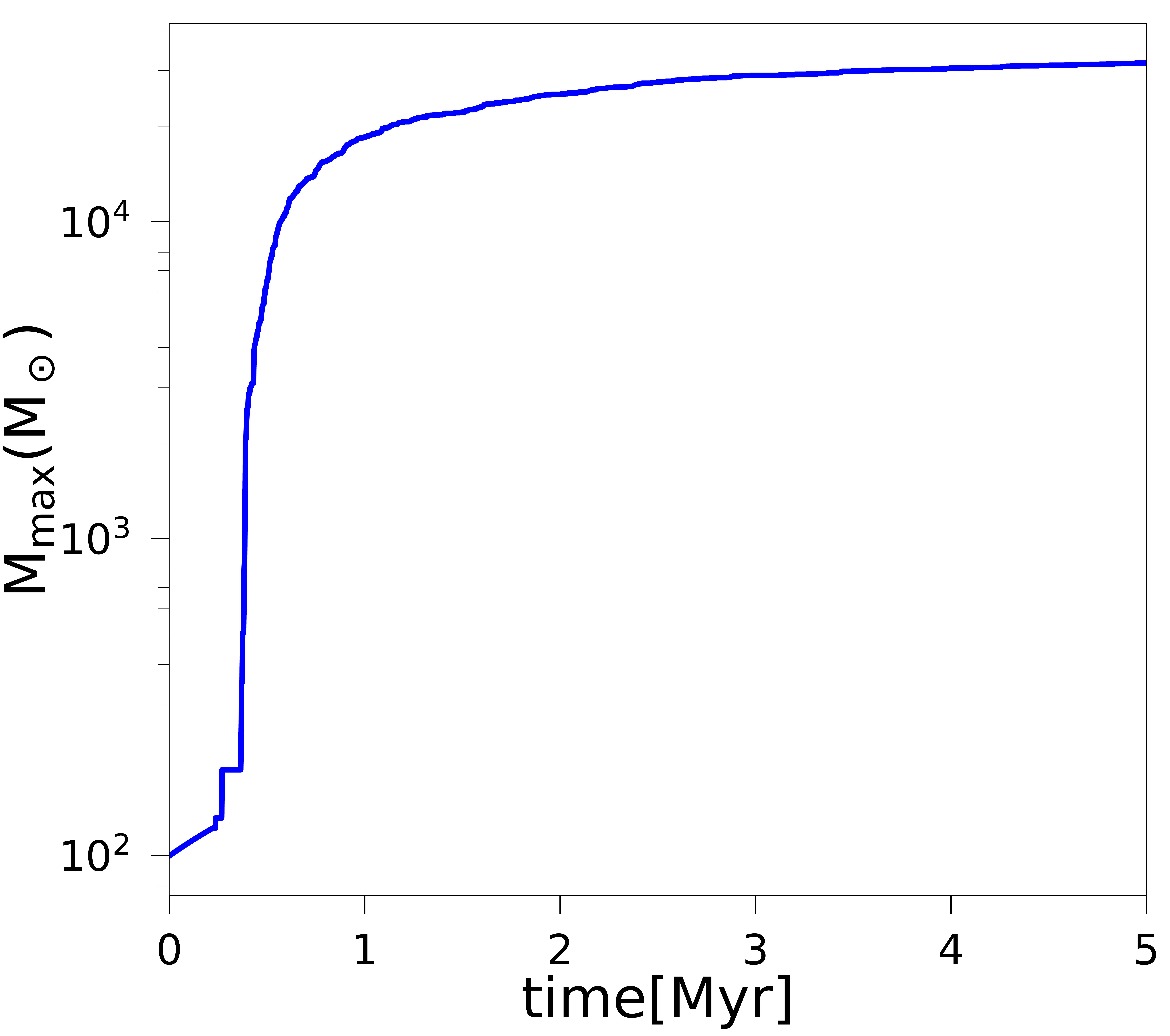}
    \caption{Same as in Figure \ref{acc-6}, but assuming instead $\dot{m}=10^{-4} {\MSun\mathrm{yr^{-1}}}$.}
    \label{acc-4}
\end{figure*}

As an even more extreme case, we consider the evolution for an accretion rate of $\dot{m}=10^{-4}\MSun\,\mathrm{yr}^{-1}$, as shown in Fig.~\ref{acc-4}. In this case, the gas mass is fully depleted after $\sim 0.2$ Myr.
Collisions are considerably enhanced and grow rapidly from about $0.2$ Myr onwards, reaching a total number of about 550 collisions after 5 Myr. The mass of the MMO increases very rapidly due to mergers of collision products, reaching $10^4 \MSun$ already after $0.8$ Myr, and about $3\times10^4 \MSun$ after 5 Myr. The final mass of the MMO is thus very similar to what is found in the case of an accretion rate of $\dot{m}=10^{-5}\MSun\,\mathrm{yr}^{-1}$, however the evolution is considerably accelerated, thus reaching the final mass earlier. The result indicates that the mass of the MMO does not become  much larger than about $10\%$ of the initial cluster mass. However, this not a fundamental limit. If all the gas gets accreted (which was seen in some of our simulations), the MMO can gain more mass via accretion and then collisions will only increase it more.  We verified that for all our models the free-fall time of the gas is shorter than the gas depletion time, i.e., all the gas will be replenished and stars will keep accreting.\footnote{If the free-fall time is longer than the gas depletion time, then there will be a rapid initial phase of accretion, followed by very little accretion while the gas slowly sinks down the cluster potential, eventually increasing again once the gas density gets high enough.} Dynamically high accretion rates will push clusters into core collapse on very short timescales compared to the rate from two-body relaxation. 

\subsection{Dependence on the physical recipe for the accretion rate}\label{acc}
\begin{figure*}
    \includegraphics[width=\columnwidth]{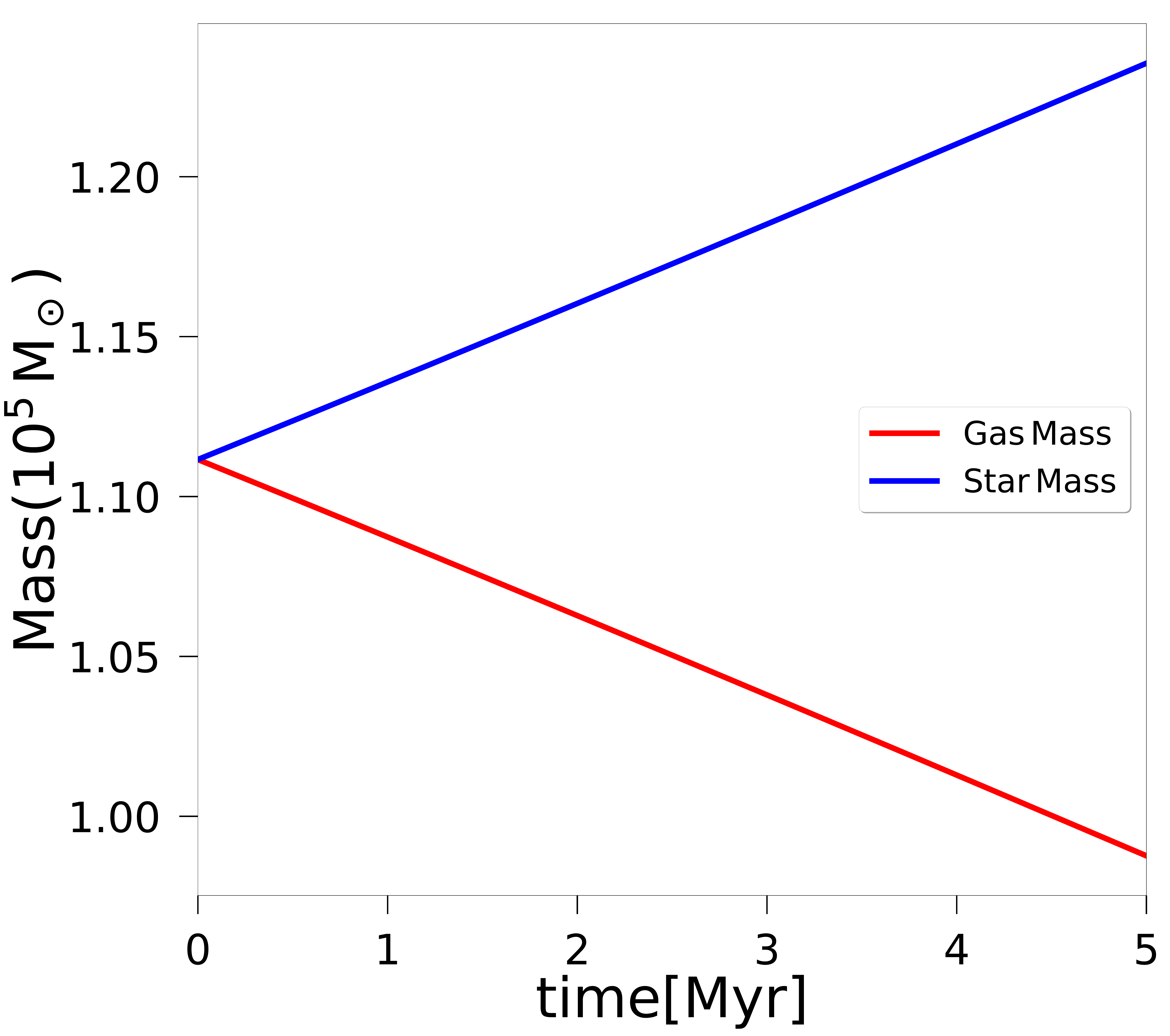}
    \includegraphics[width=\columnwidth]{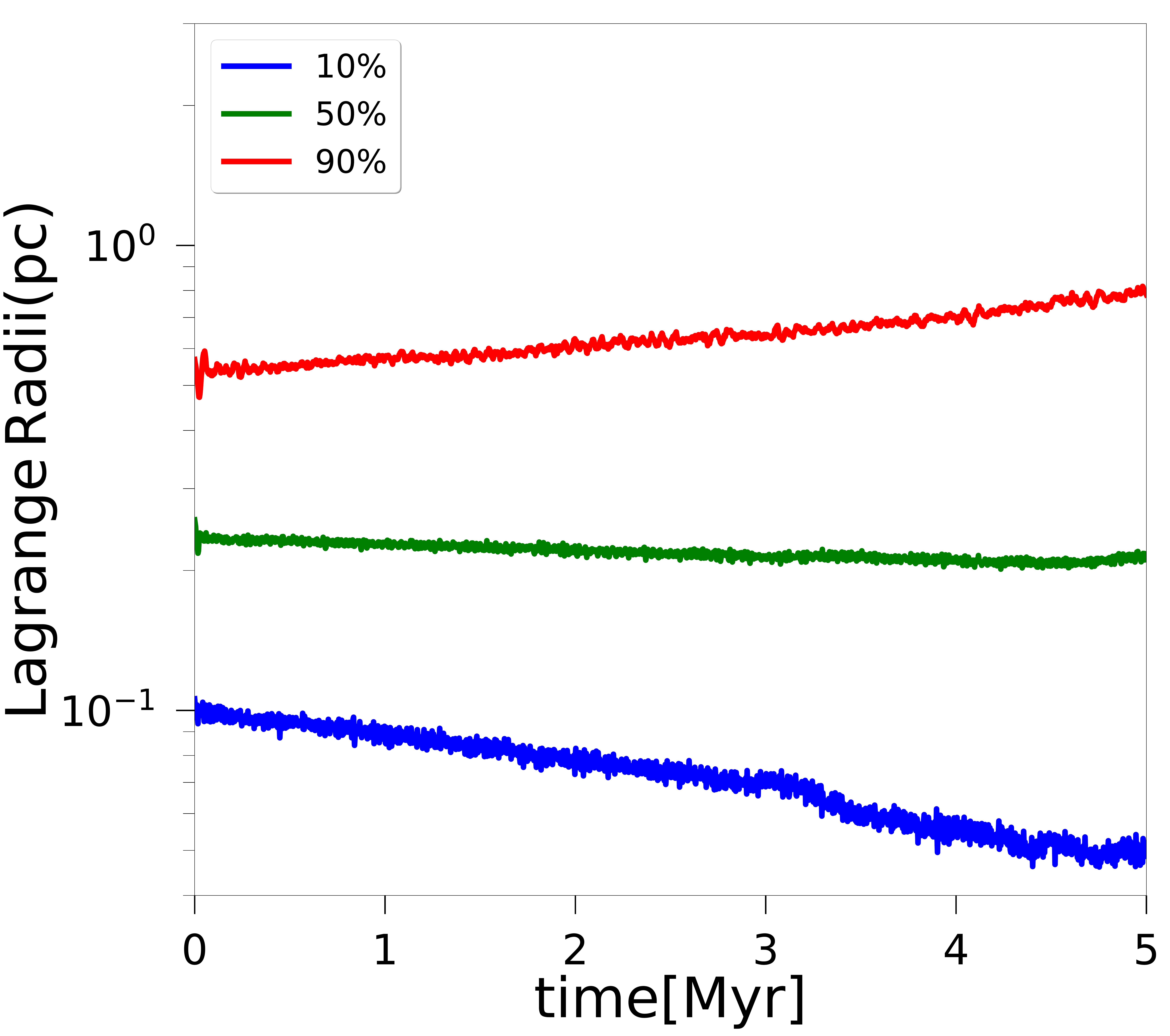}
    \includegraphics[width=\columnwidth]{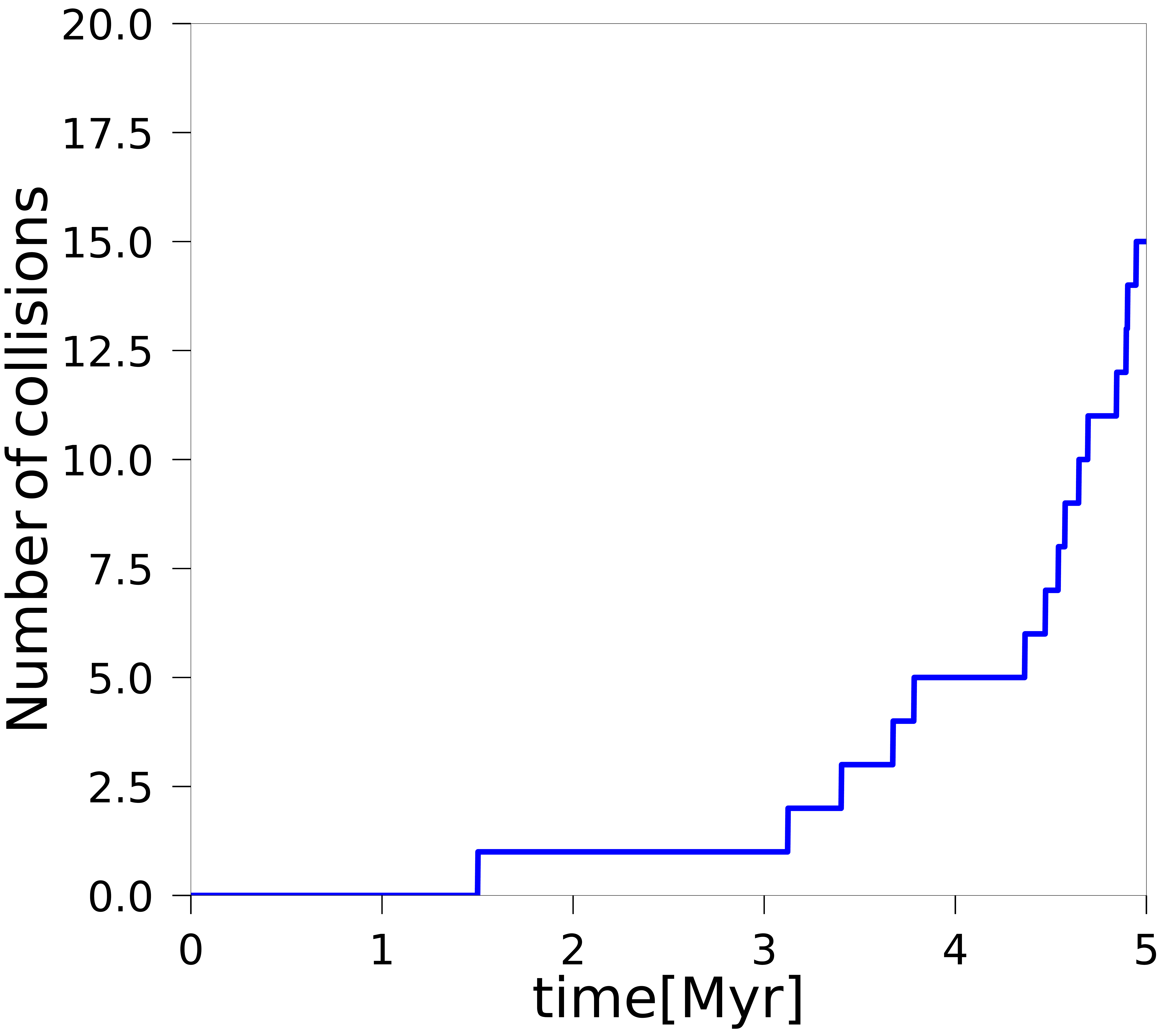}
    \includegraphics[width=\columnwidth]{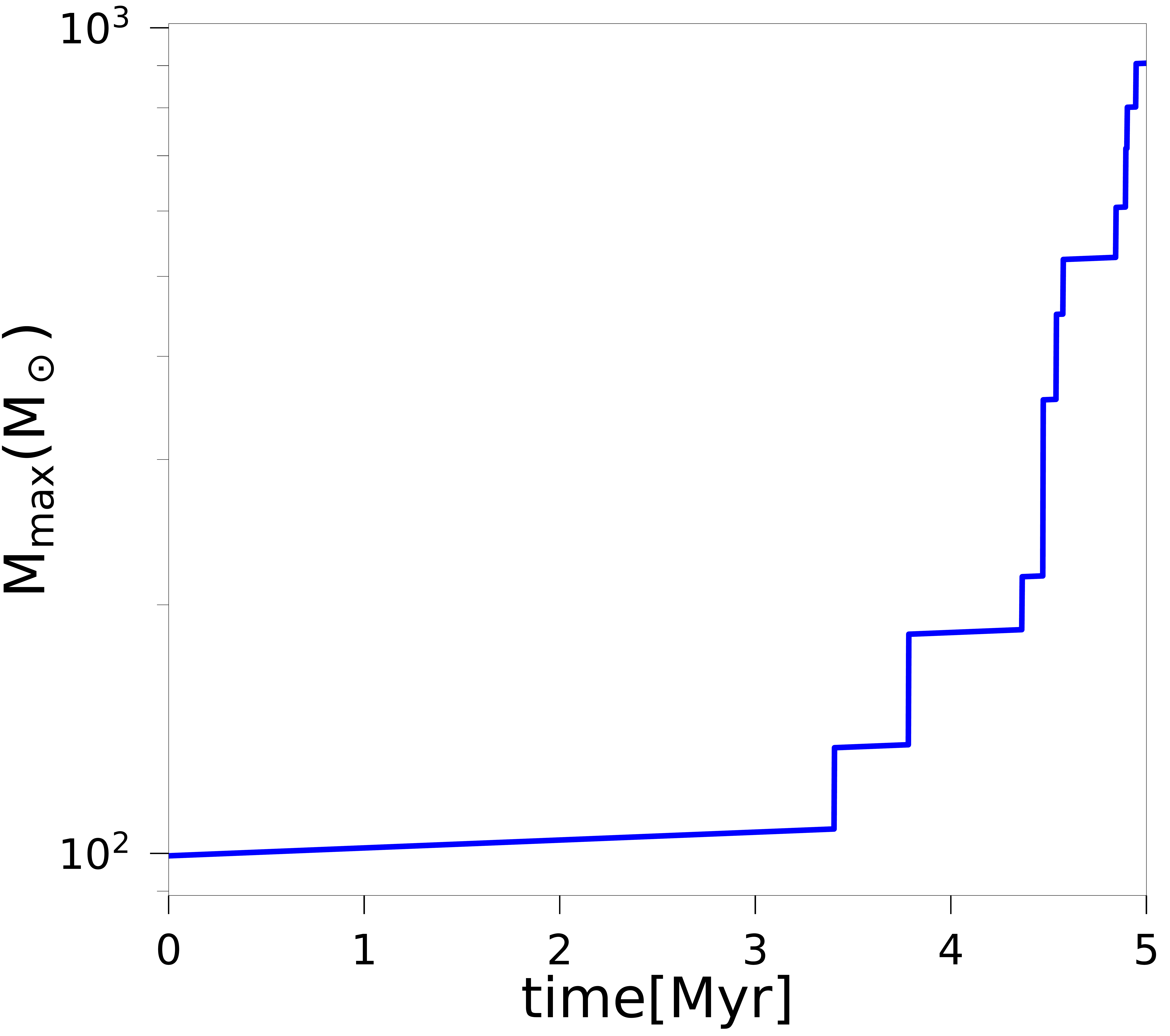}
    \caption{Same as in Figure \ref{acc-6}, but assuming instead Eddington accretion.}
    \label{Eddington}
\end{figure*}

As a next step, we explore the dependence of the evolution in the NSCs for different physical assumptions regarding the accretion rate, i.e. going beyond the simplified assumption of a constant accretion rate. In Fig.~\ref{Eddington}, we show the expected time evolution for the case of Eddington accretion scenario given by Eq.~\ref{eddacc}. In this case, we find that the evolution is comparable to what we found for a constant accretion rate of $\dot{m}=10^{-6}\MSun\,\mathrm{yr}^{-1}$. The gas mass decreases by about $15\%$ over 5 Myr, while the stellar mass increases by the same amount. The evolution of the Lagrangian radii is rather stable, with the $10\%$ and $50\%$ radii slightly contracting, and the $90\%$ radius moderately expanding. The first collision occurs after about $1.5$ Myr, and about $15$ collisions are reached after $5$ Myr. The growth of the MMO occurs most rapidly after about $3.5$ Myr, and reaches about $900 \MSun$. The results show that relevant numbers of collisions could occur, which can lead to a SMS mass of $\sim 10^3\MSun$ in the Eddington case.

\begin{figure*}
    \includegraphics[width=\columnwidth]{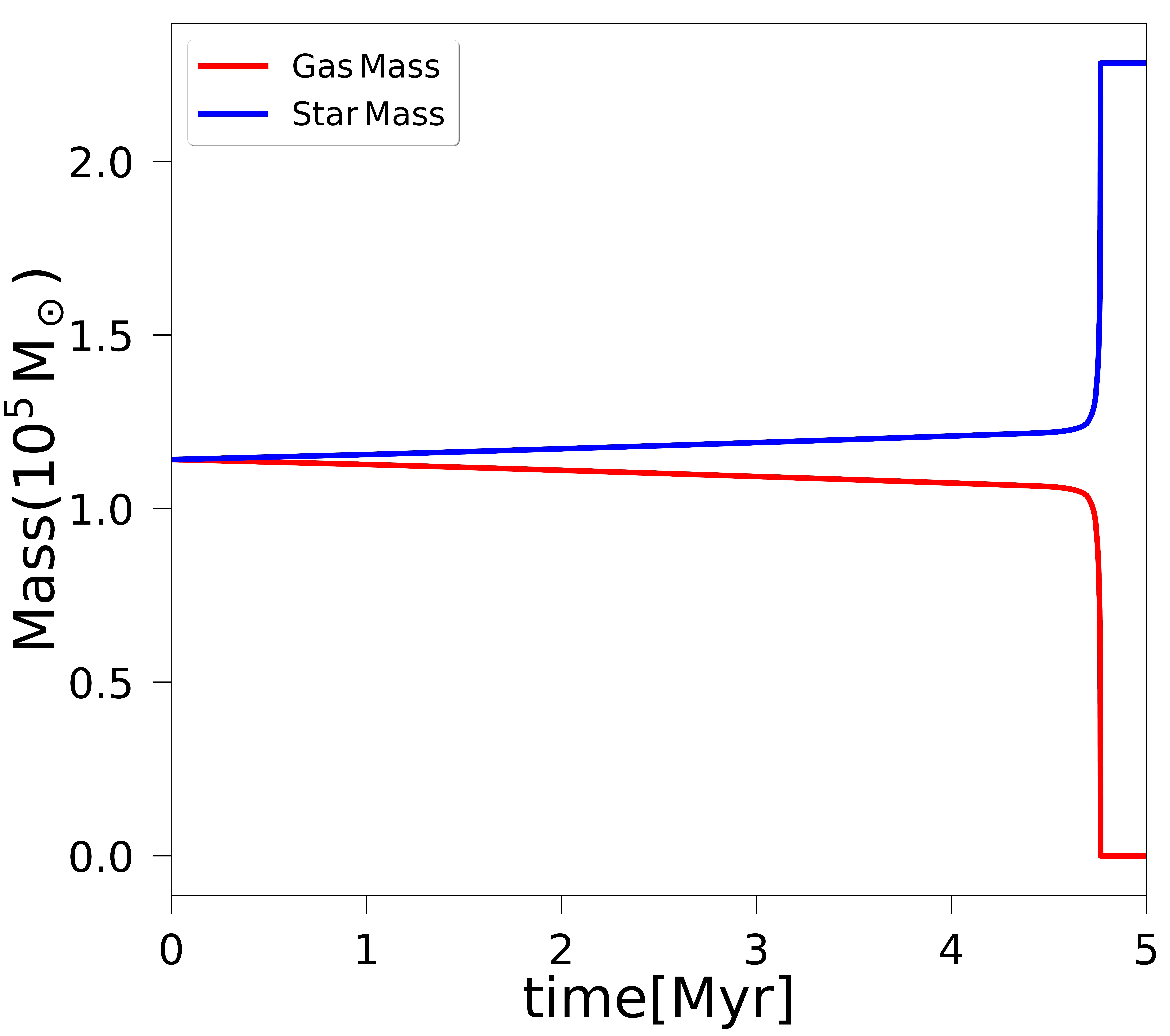}
    \includegraphics[width=\columnwidth]{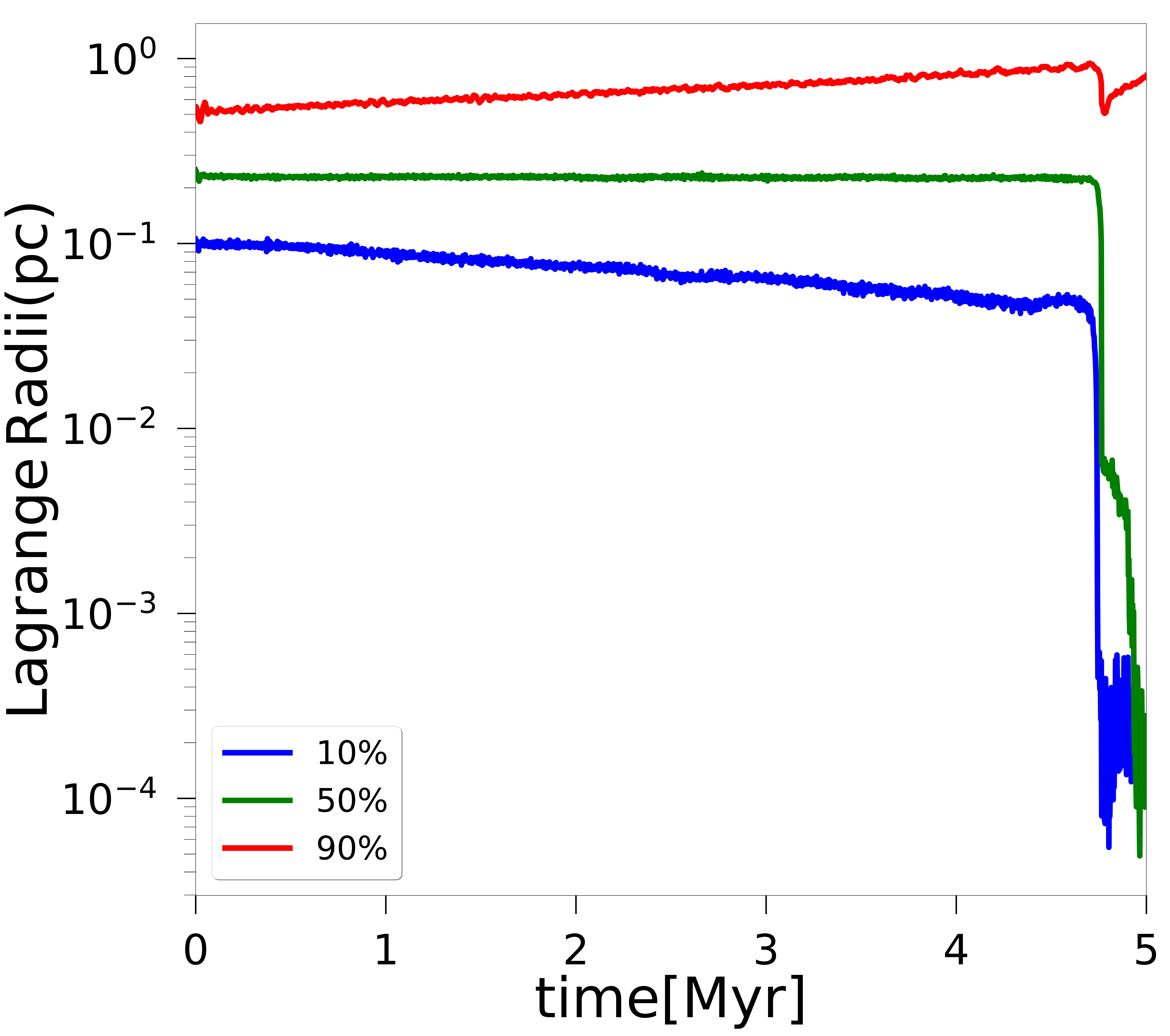}
    \includegraphics[width=\columnwidth]{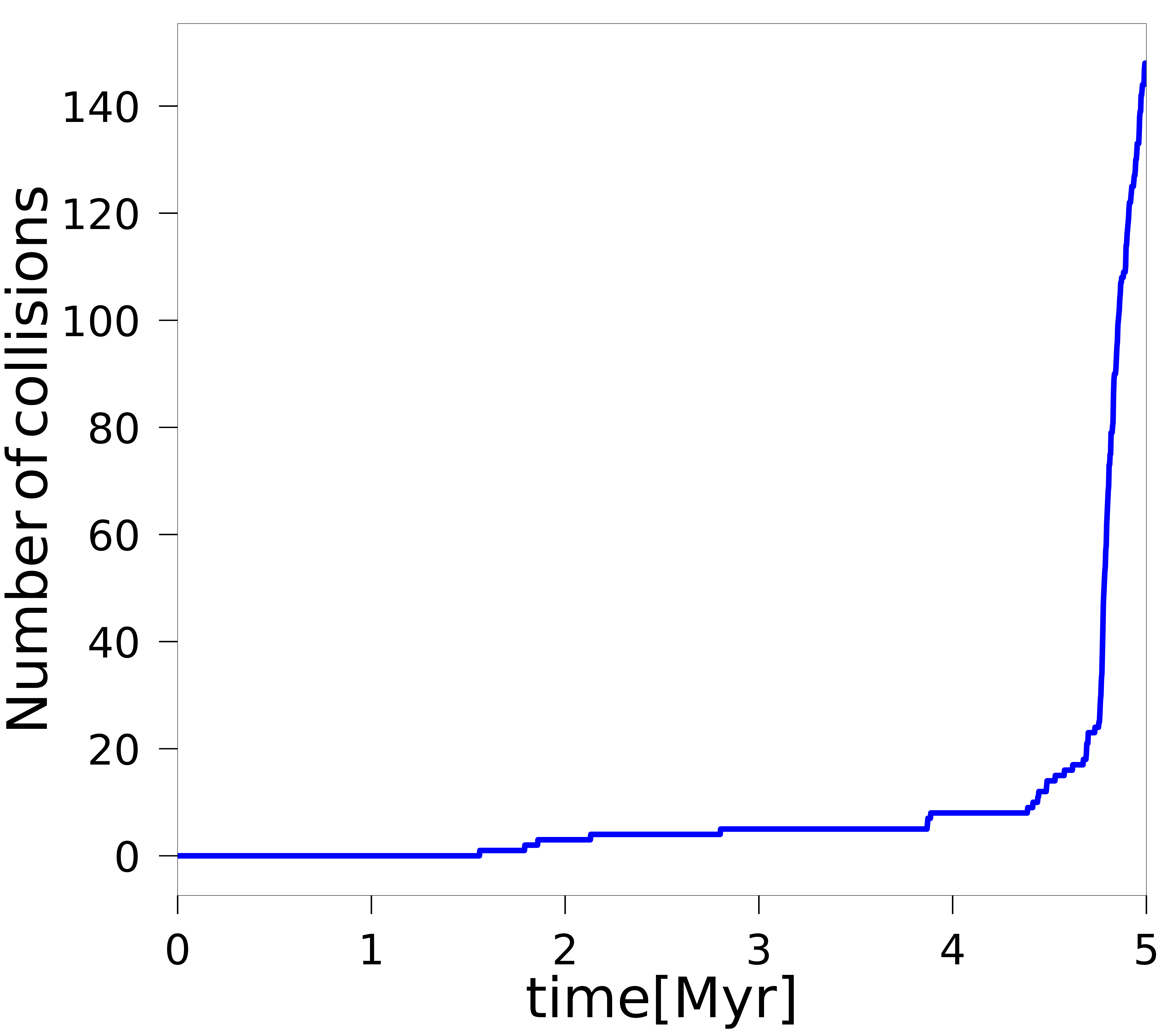}
    \includegraphics[width=\columnwidth]{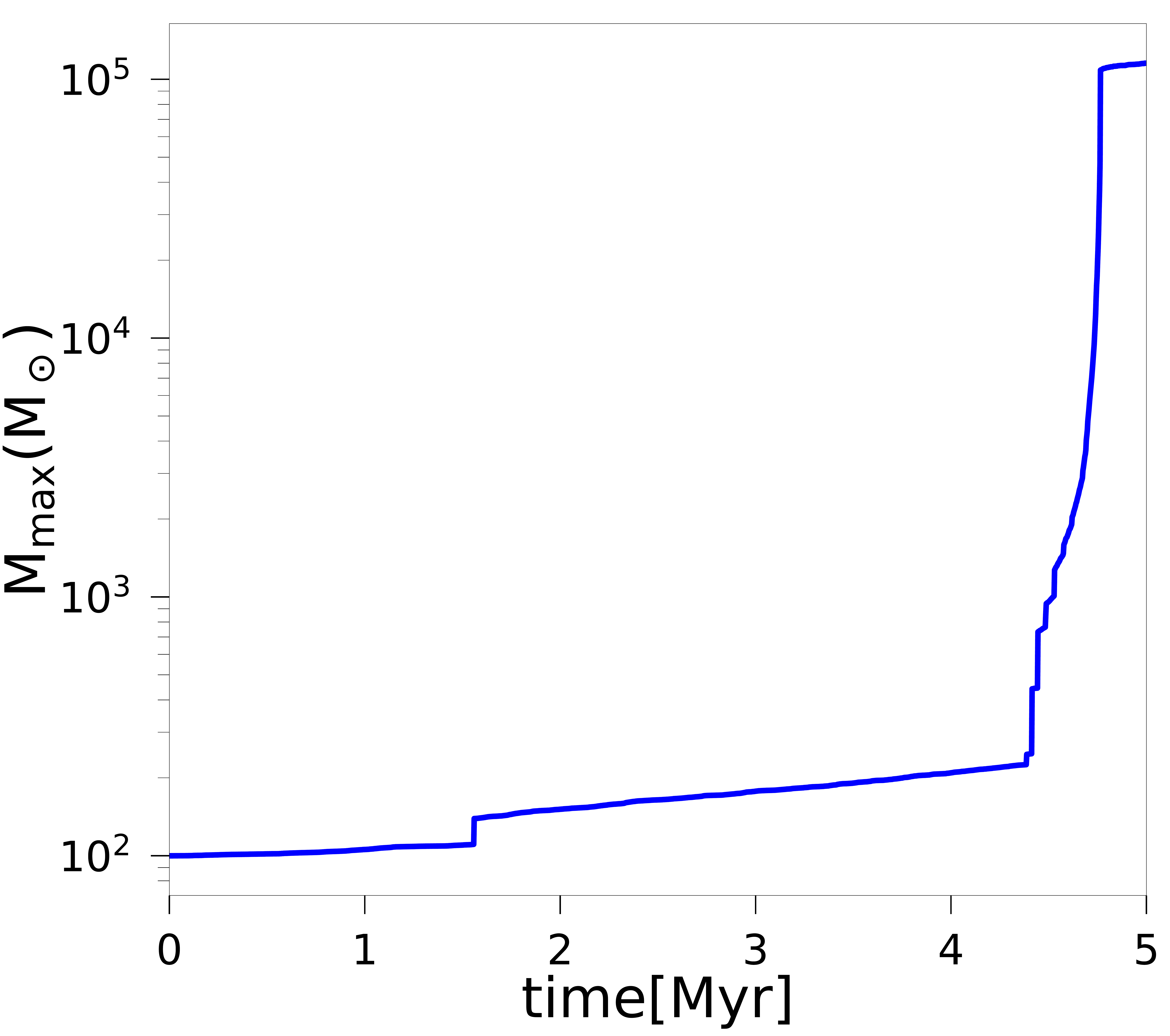}
    \caption{Same as in Figure \ref{acc-6}, but assuming instead Bondi accretion.}
    \label{bondi}
\end{figure*}

As a more optimistic scenario, we also consider the case of Bondi accretion given by Eq.~\ref{mainbondi}. The results are shown in Fig.~\ref{bondi}. As the Bondi accretion rate $\dot{M}_{\mathrm{BH}}\propto M_\ast^2$, this recipe has a strong impact on the evolution of the MMO. Initially the evolution takes place more slowly, due to an initially lower accretion rate, but is rapidly enhanced at late times as the mass of the accretor grows. As seen in the evolution of the gas and stellar mass, for a long time the accretion rates are lower than in the Eddington scenario, and only after about $4.7$ Myr the evolution becomes very rapid, and accelerates so much that the gas becomes fully depleted within a short time, while the stellar mass increases in the same way. The first collision occurs after about $1.5$ Myr and in total about 20 collisions occur before the accretion becomes strongly accelerated, with the MMO reaching about $200 \MSun$ within the first 4 Myr. During the accelerated phase of the evolution, more than 120 additional collisions occur, and the final mass reaches about $10^5 \MSun$. While the Lagrangian radii are very stable for the first $4.7$ Myr, they react to the extreme evolution occurring subsequently, with the $10\%$ Lagrangian radius strongly decreasing due to the formation of the MMO, and the same slightly later even for the $50\%$ radius. The $90\%$ radius shows first a minor decrease at the time when the evolution accelerates and subsequently expands. The evolution in this scenario is sufficiently extreme that our model assumptions will break down early on due to a lack of gas replenishment, so this part of our results needs to be regarded with caution. 

\begin{figure}
	\includegraphics[width=\columnwidth]{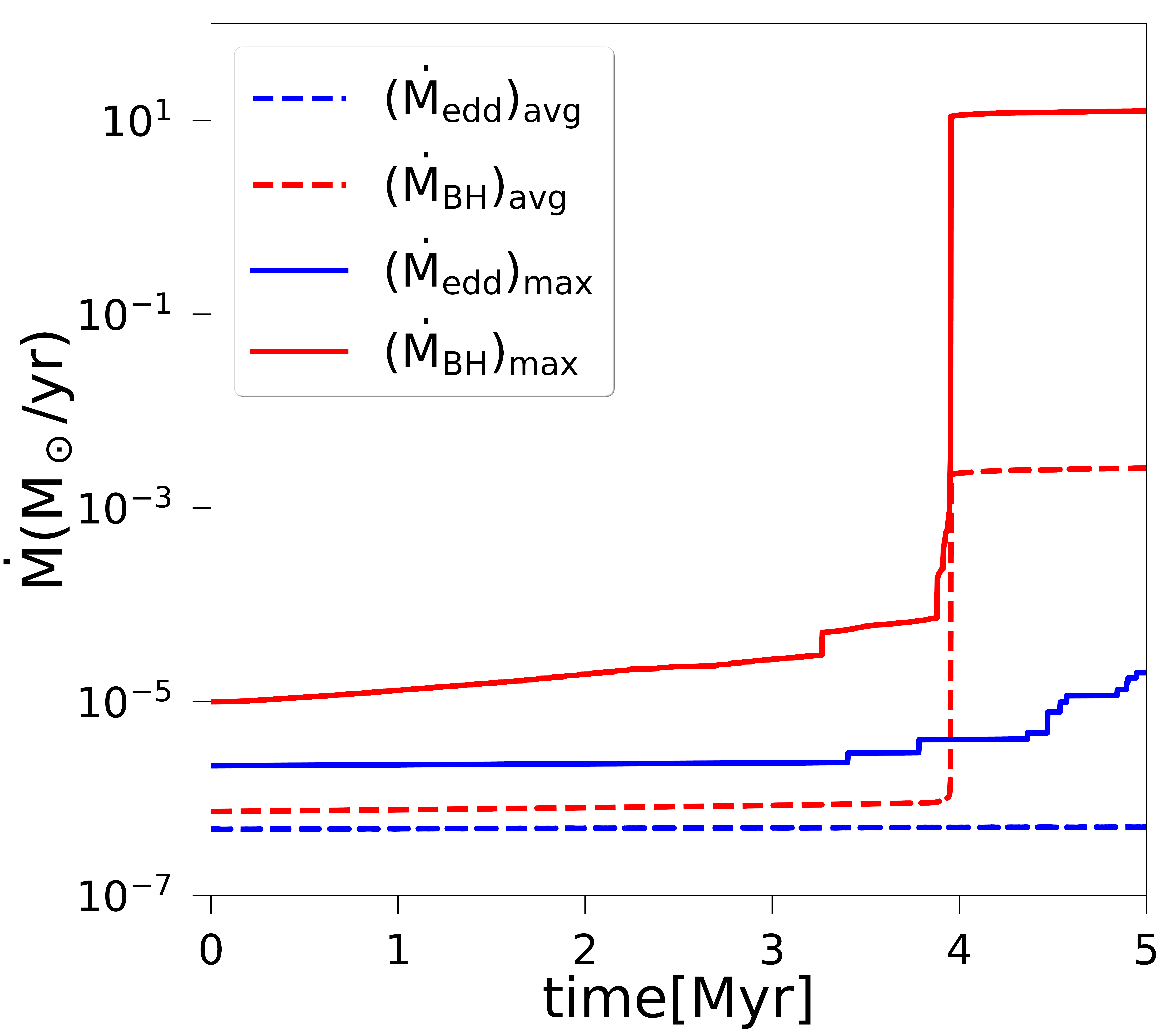}
	\caption{Average accretion rate and maximum accretion rate in the cluster as a function of time, for the models assuming Eddington or Bondi accretion. We assume an IMF with stellar masses of 10-100 $\MSun$.}
	\label{averageaccretion}
\end{figure}

The evolution of both the average accretion rate in the cluster and the maximum accretion rate is shown for the models with Eddington and Bondi accretion in Fig.~\ref{averageaccretion} (in the case of constant accretion rates, the plots would be trivial). The average accretion rates are almost constant as a function of time and increase only at late times. In the Eddington scenario, this increase is hardly visible by eye, while it is much more pronounced in the case of Bondi accretion, due to the steep increase of the accretion rate of the MMO, which then affects also the calculation of the average. We indeed find that the Bondi accretion rate for the MMO increases by more than two orders of magnitude, while it is about an order of magnitude in the case of Eddington accretion.

\subsection{Collision vs accretion time scale}\label{sectimescales}

To determine the respective relevance of collisions vs accretion for a single object, we aim in the following at a systematic comparison of the collision timescale ($t_{\mathrm{coll}}$) and accretion timescale ($t_{\mathrm{acc}}$). To evaluate collision timescale for the MMO, we adopt the formulation of \citet{Lei17} for a system consisting of heavy particles with mass $m_{\mathrm{A}}$ and light particles with mass $m_{\mathrm{B}}$:
\beq
t_{\mathrm{coll}}=\frac{G^3 m_{\mathrm{A}}^{1/2}\bar{m}M_{\mathrm{cl}}^{13/2}}{12\sqrt{2}m_{\mathrm{B}}^{3/2}N_{\mathrm{B}}(R_{\mathrm{A}}+R_{\mathrm{B}})^2|E|^{7/2}},\label{timecoll}
\eeq
where $\bar{m}$ is the average mass of all particles, $R_{\mathrm{A}}$ and $R_{\mathrm{B}}$ the radii of particles of type A and B, $N_{\mathrm{B}}$ is the number of particles of type B, and $|E|$ the total energy of the stars in the cluster, which we evaluate using the virial theorem and taking into account the gravitational potential from gas and stars via
\beq
|E|=\frac{GM_{\mathrm{tot}}M_{\mathrm{cl}}}{2R_{\mathrm{cl}}},
\eeq
with $M_{\mathrm{tot}}=M_{\mathrm{cl}}+M_{\mathrm{g}}$ being the total mass. It is important to note that eq.~\ref{timecoll} does not incorporate the effect of accretion on the collisions probability, while our numerical experiments in section~\ref{secreferencerun} establish that accretion enhances collisions. The collision timescale employed here for collisions with the MMO can in this sense be regarded as an upper limit. As demonstrated by \citet{Barrera2020}, in the absence of accretion, the prescription works best for lower particle numbers and less extreme mass ratios. Once one object becomes more massive than the rest, quasi-Keplerian motions and a loss cone formalism start to become more relevant. We do not account for that here and neglect these effects in the context of the collision timescale. The accretion timescale of the MMO is given by
\beq
t_{\mathrm{acc}}=\frac{M_{\mathrm{max}}}{\dot{m}},
\eeq
with $M_{\mathrm{max}}$ being the mass of the MMO and $\dot{m}$ is given by constant, Eddington and Bondi accretion rates.

\begin{figure*}
    \includegraphics[width=0.65\columnwidth]{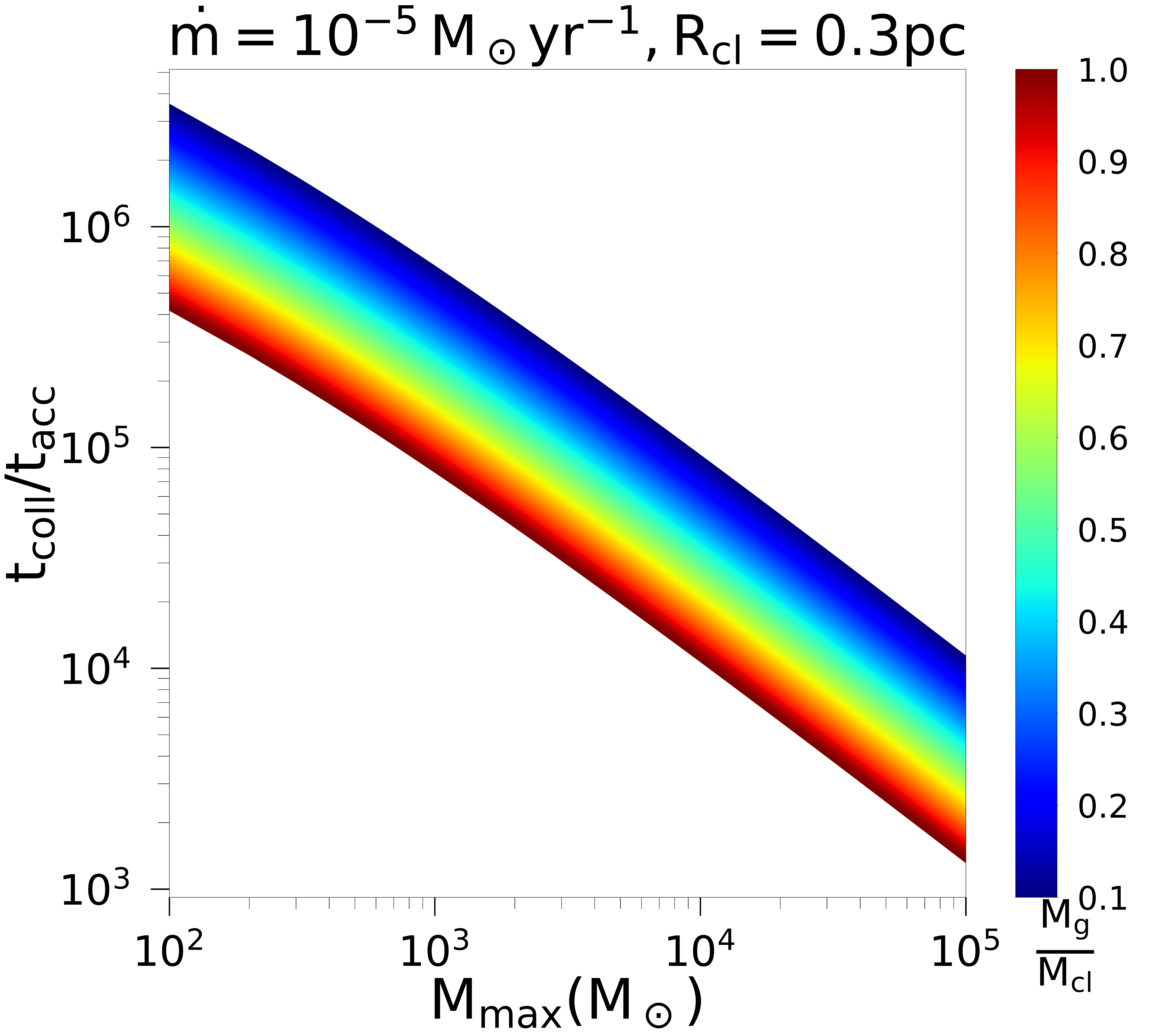}
    \includegraphics[width=0.65\columnwidth]{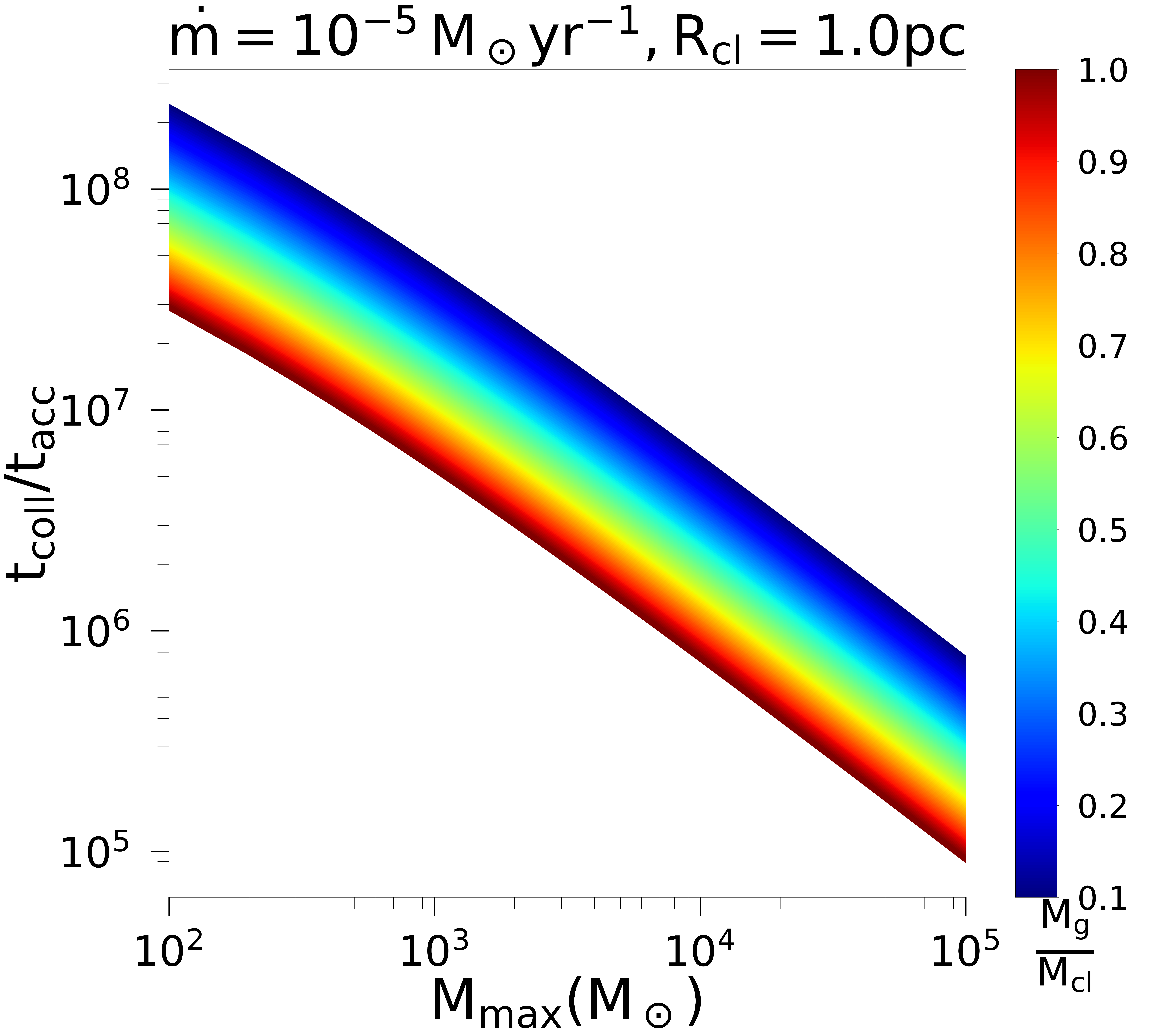}
    \includegraphics[width=0.65\columnwidth]{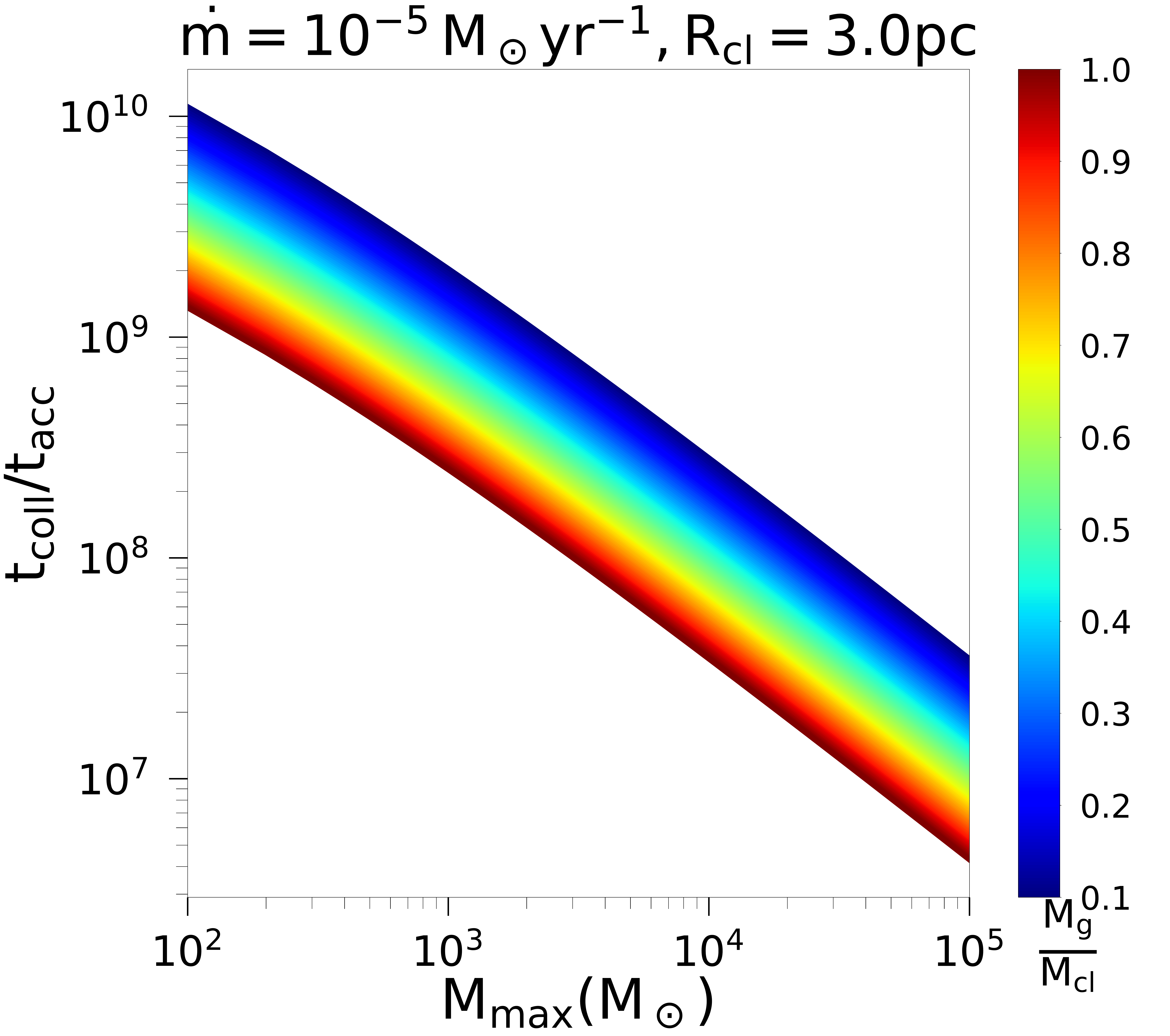}
    \includegraphics[width=0.65\columnwidth]{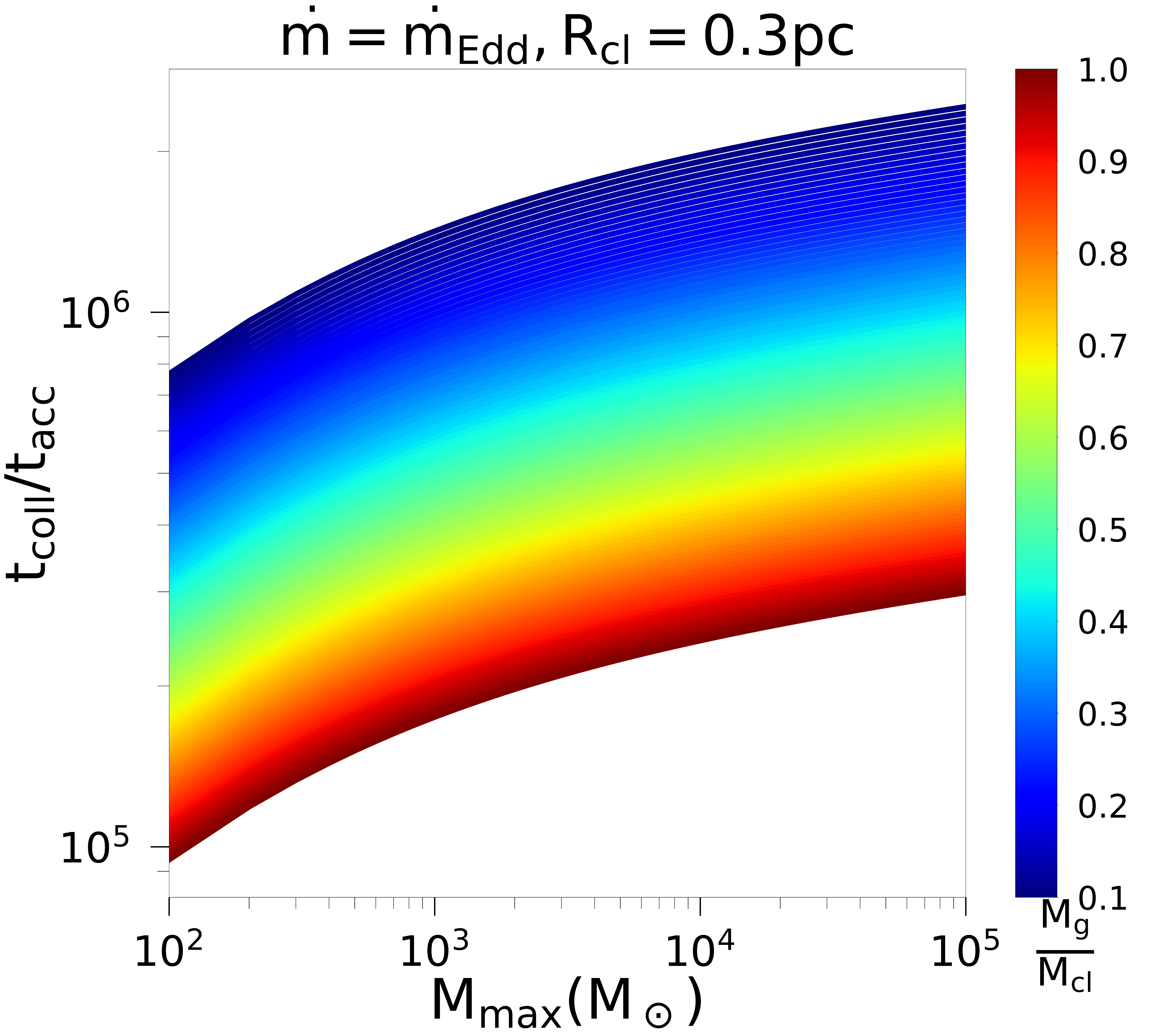}
    \includegraphics[width=0.65\columnwidth]{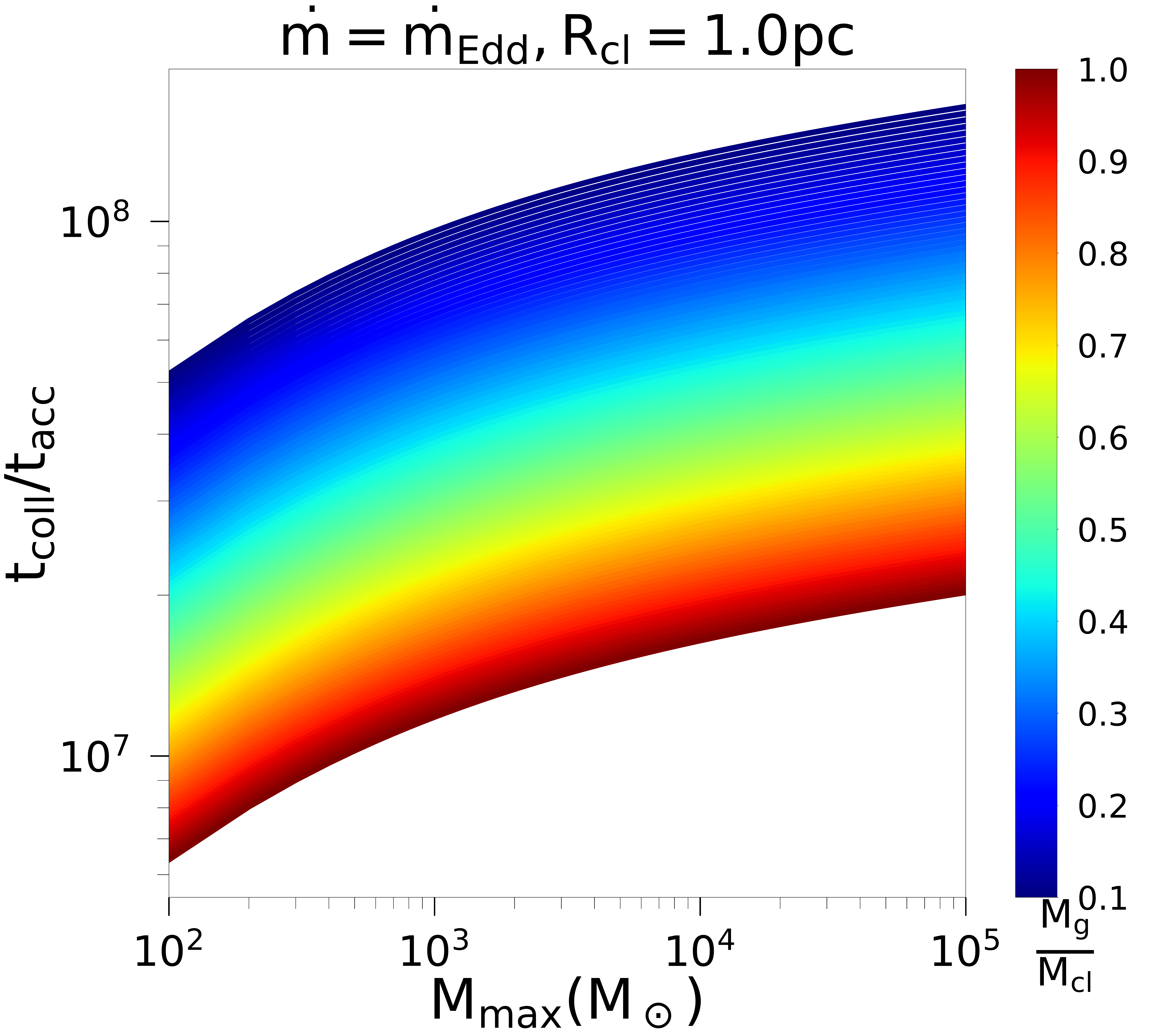}
    \includegraphics[width=0.65\columnwidth]{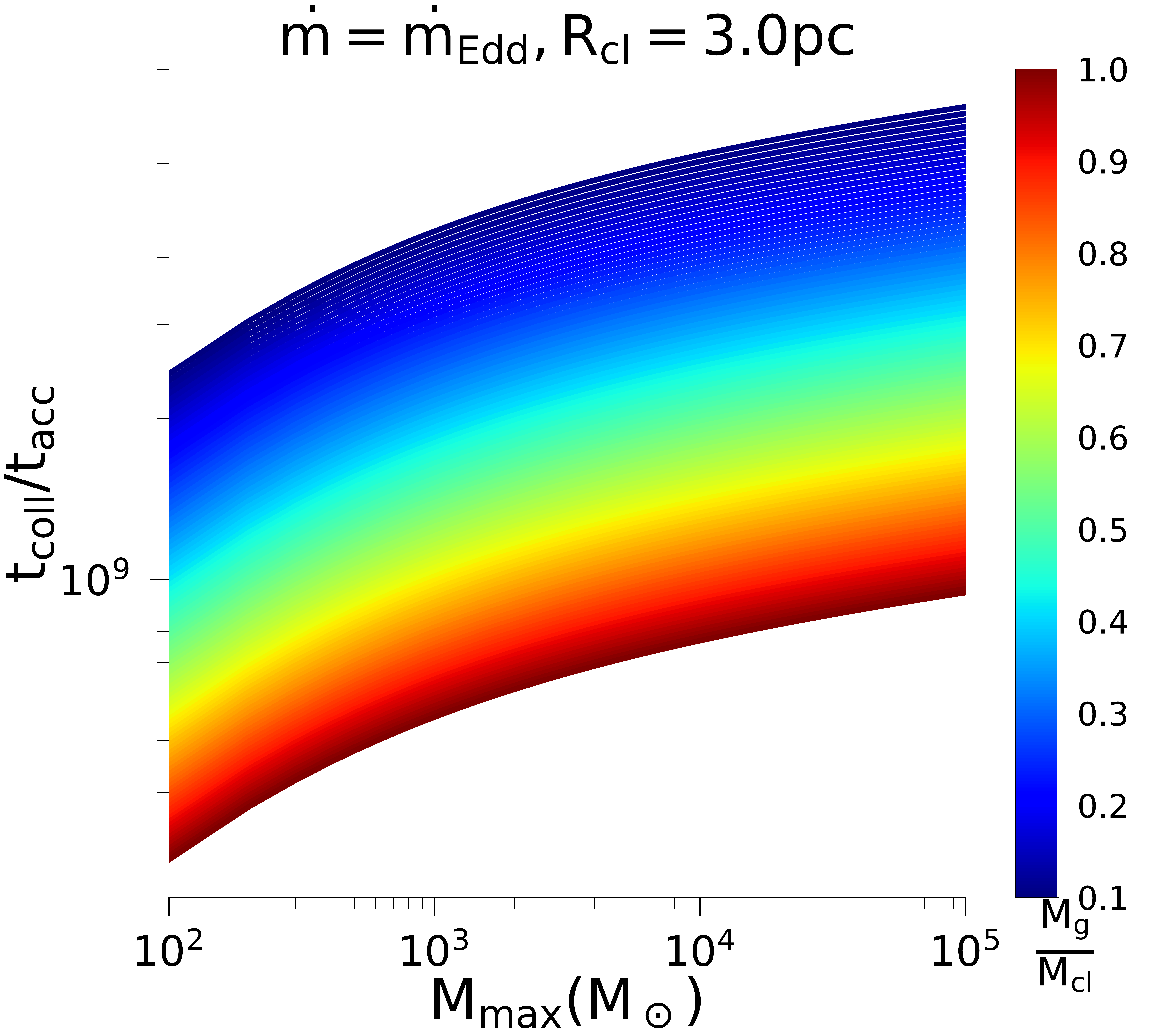}
    \includegraphics[width=0.65\columnwidth]{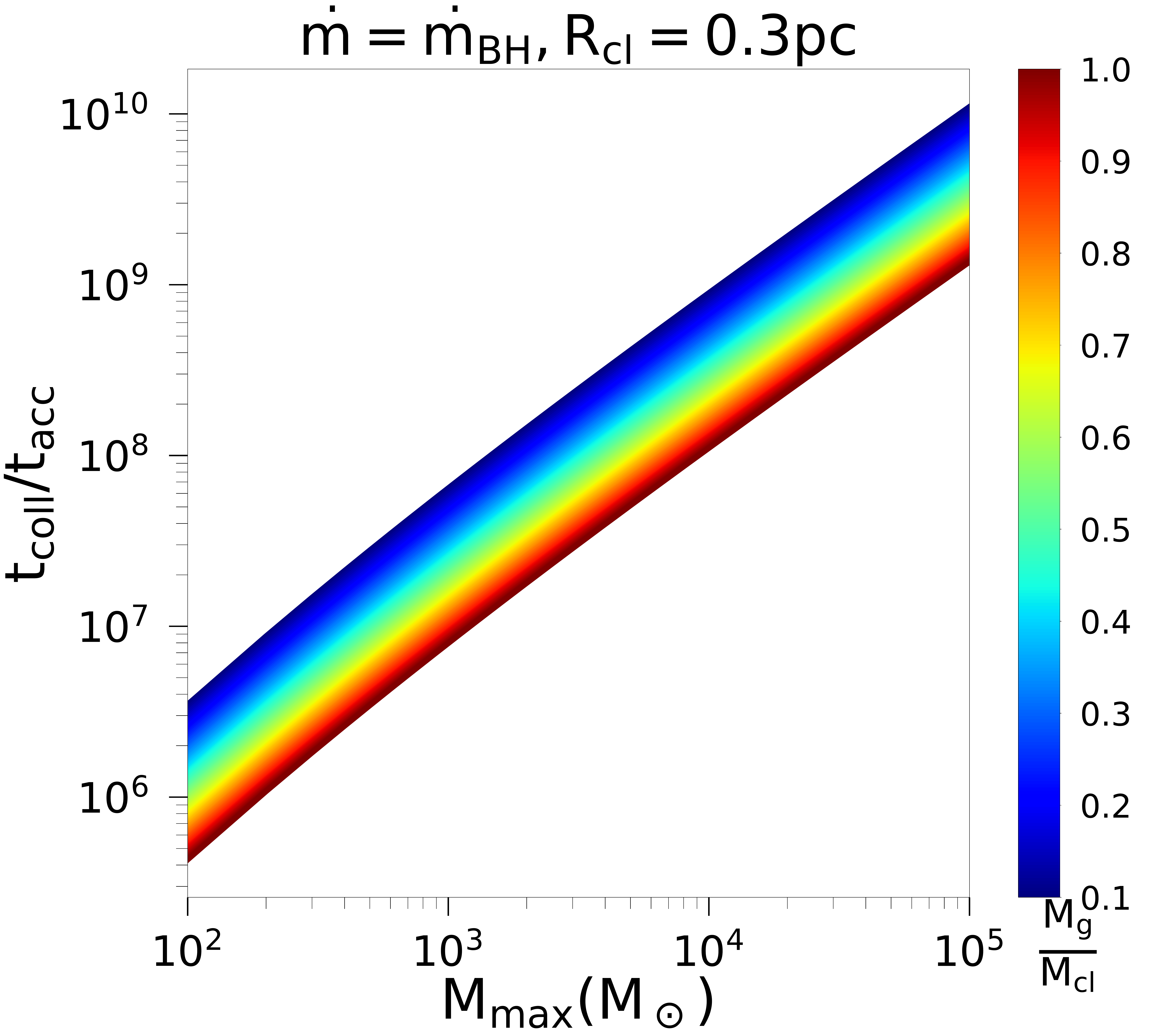}
    \includegraphics[width=0.65\columnwidth]{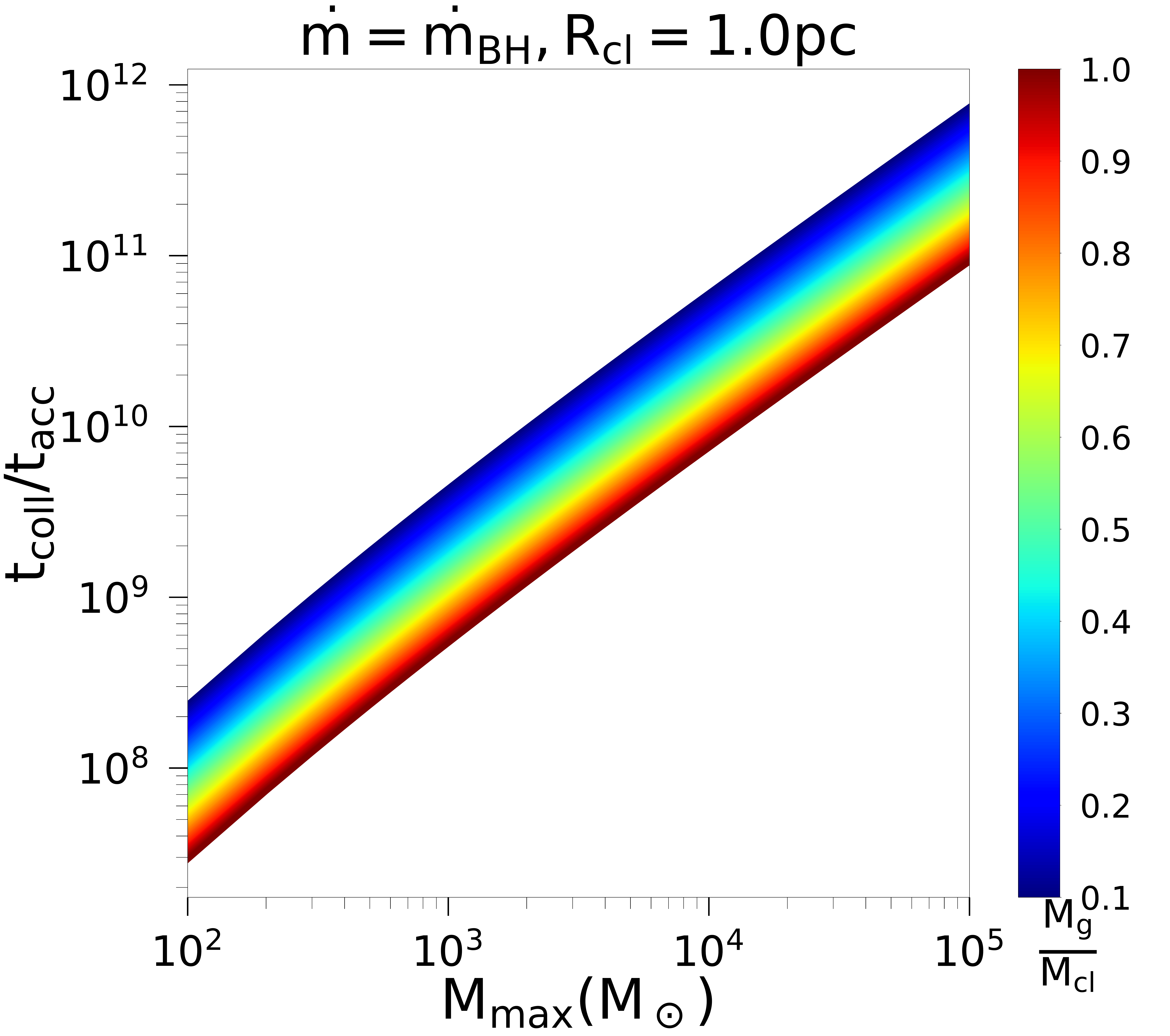}
    \includegraphics[width=0.65\columnwidth]{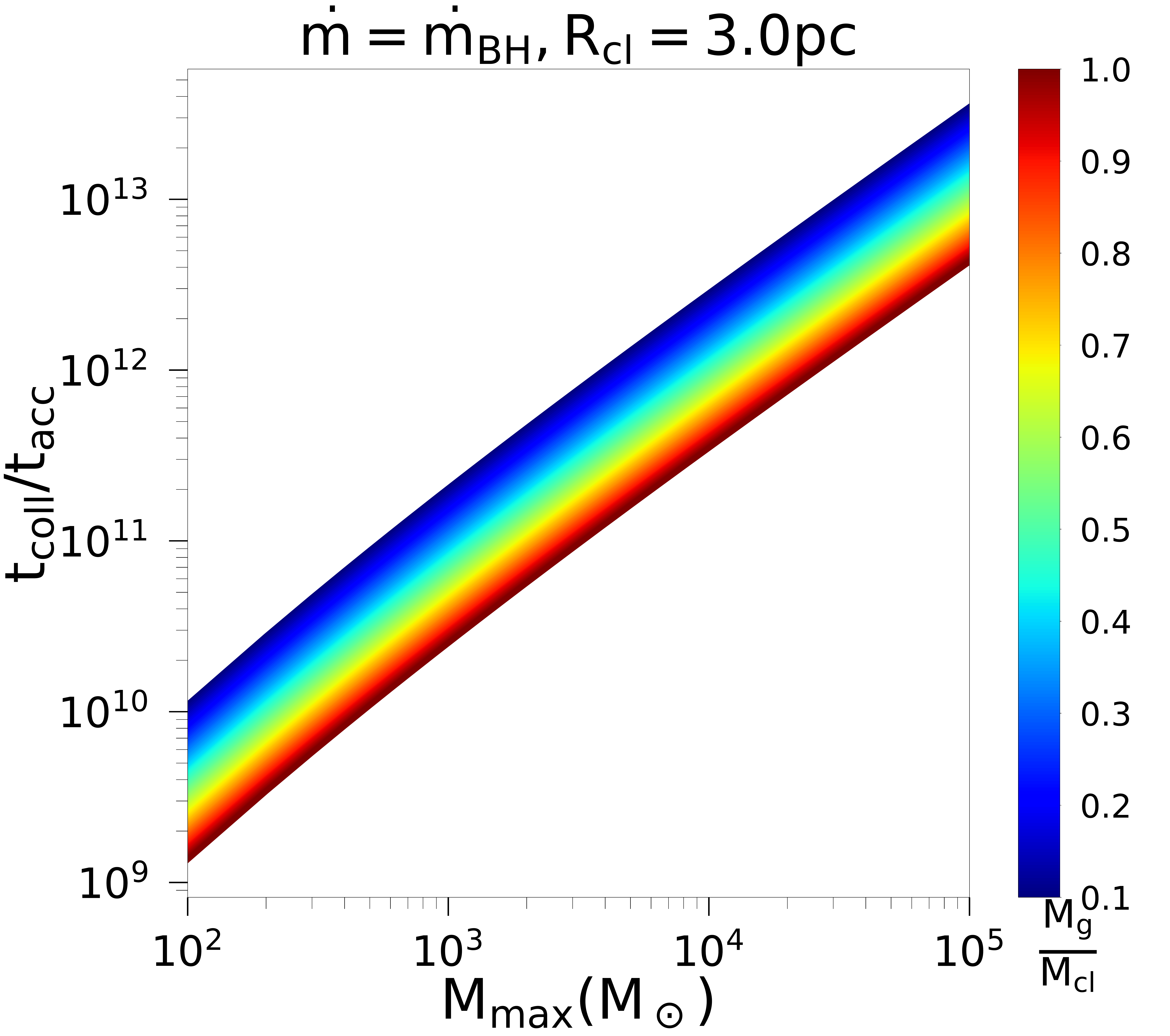}
    \caption{The ratio of collision over accretion timescales, $t_{\mathrm{coll}}/t_{\mathrm{acc}}$, as a function of stellar mass for the case of a constant accretion rate of $10^{-5}\acc$ (top panel), Eddington accretion rate (mid panel) and Bondi accretion rate (bottom panel). We show the results for clusters sizes of $0.3$~pc (left), $1$~pc (middle) and $3$~pc (right), assuming different ratios of gas to cluster mass. The expected ratio $t_{\mathrm{coll}}/t_{\mathrm{acc}}\gg 1$, suggesting that accretion dominates over collisions in this regime.}
    \label{timescales}
\end{figure*}

We plot the theoretically expected ratio of these timescales in Fig.~\ref{timescales}, considering the cases of a constant accretion rate of $10^{-5}\acc$, as well as Eddington accretion and Bondi accretion, as a function of the mass of the MMO. The color bar represents different ratios of gas to cluster mass. We consider $m_{\mathrm{A}}=20\MSun$, $\bar{m}=22\MSun$ and $N_{\mathrm{B}}=5000$ (motivated from the initial conditions of our simulations). The radii $R_{\mathrm{A}}$ and $R_{\mathrm{B}}$ are calculated using eq. \ref{less50} and \ref{great50}. 

\begin{figure}
    \includegraphics[width=\columnwidth]{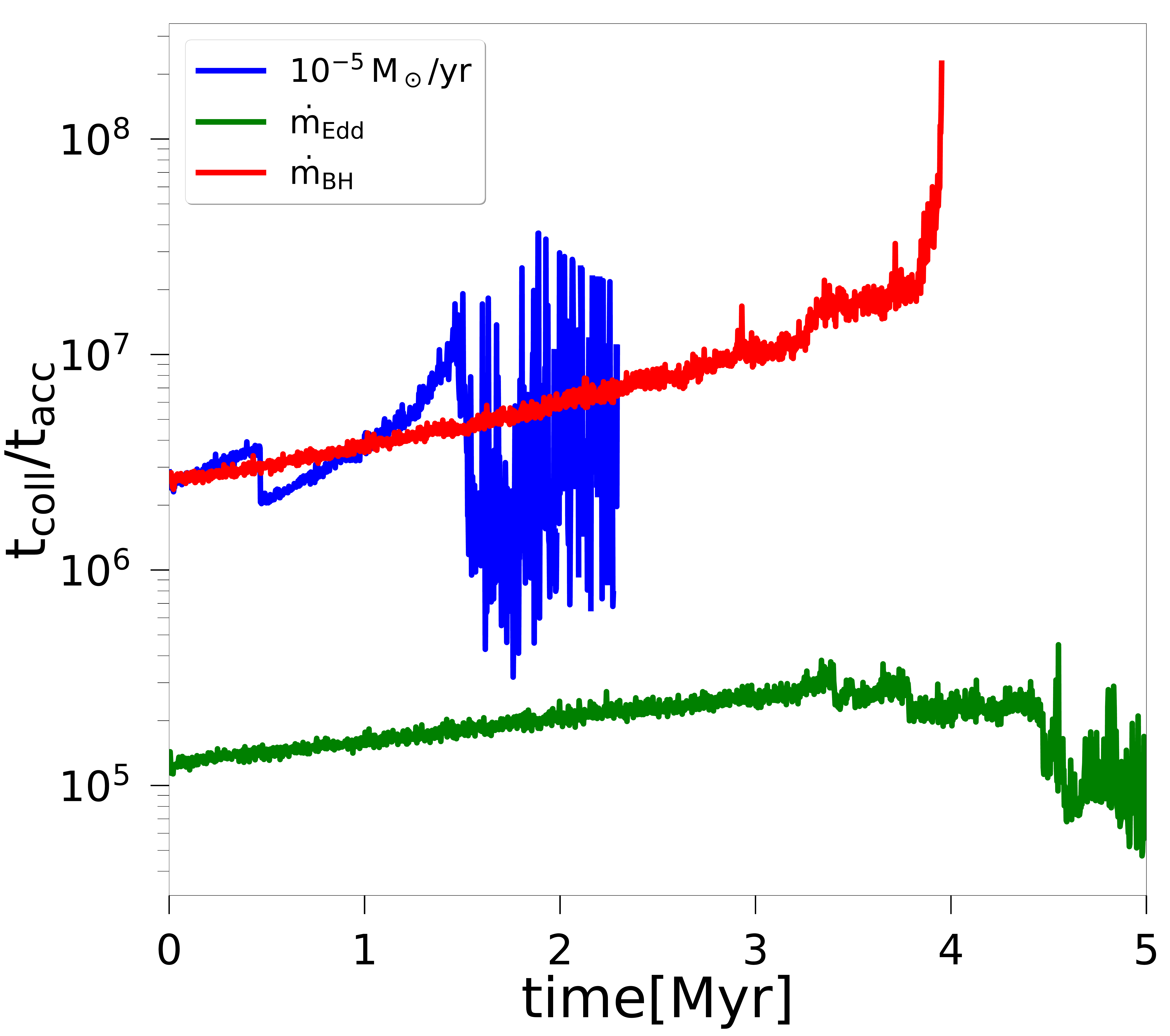}
    \caption{The ratio $t_{\mathrm{coll}}/t_{\mathrm{acc}}$, evaluated from simulations with N=5000, $M_{\mathrm{cl}}=M_{\mathrm{g}}=10^{5}\MSun$, $R_{\mathrm{cl}}=R_{\mathrm{g}}=1$pc as initial conditions, plotted for a constant accretion rate of $10^{-5}\acc$ (blue line), Eddington accretion rate (green line) and Bondi accretion rate (green line). The ratio $t_{\mathrm{coll}}/t_{\mathrm{acc}}\gg 1$, showing that accretion dominates over collisions in the simulations. }
    \label{timescalessim}
\end{figure}

In the considered parameter space covering several orders of magnitude, the ratio $\frac{t_{\mathrm{coll}}}{t_{\mathrm{acc}}}\gg 1$ , suggesting that accretion dominates over collisions in this regime. For $t_{\mathrm{coll}}\approx t_{\mathrm{acc}}$, we would require extremely compact clusters with $R_{\mathrm{cl}}\ll0.1$~pc or the mass of the MMO would have to be much less than $100\MSun$. So in principle a situation can occur for which collisions will dominate over accretion, but it will be for a small part of the parameter space. We find a very similar result if we plot the ratio of these timescales directly from the conditions obtained in the simulations, which we plot in Fig.~\ref{timescalessim}. Since our simulations have an initial range of masses between 10-100 $\MSun$ we assumed the most massive particle to be the type A and consider the rest of the particles as type B. The ratio of the timescales is again considerably larger than $1$. 

In principle this suggests that accretion should be more important than collisions for the clusters considered here. However, it is important to note from the results in section~\ref{secreferencerun} that only a moderate number of collisions occurs for an accretion rate of accretion rate $\sim 10^{-6}\acc$ (and we checked that no collisions occur  for a low accretion rate $\sim 2\times10^{-7}\acc$), but a relevant number of collisions occurs for a high accretion rate of $\sim 10^{-5}\acc$ or $10^{-4}\acc$. It is thus important to realize that the collision probability is affected by accretion, an effect which is not incorporated into the formula above for the collision timescale. This is compatible with the work of ~\citet{Dav10}, who have shown that the number of stellar collisions should be $\propto N^{5/3}\dot{M} ^{2/3}$, where in their work $\dot{M}$ is the accretion rate onto the cluster, which is however reflected in accretion onto individual stars.

We also emphasize that collision timescale in Eq.~\ref{timecoll} is for a single object. To obtain the timescale for the collision of any two objects in the cluster, the collision timescale adopted here should be multiplied with $2/(N-1)$ \citep{Lei17}, bringing both timescales much closer together, or allowing collision timescale to be shorter for part of the parameter space.  Hence, stellar collisions will be strongly favoured in clusters with high $N$ such as in all our models. However, it is important to note that this collision probability was computed using a simplified model where all the stars in the cluster are of the same mass, in contrast to a realistic IMF adopted in our models. The presence of an IMF will lead to mass segregation as massive stars tend to sink to the core through dynamical friction where they may eventually decouple from the remainder of the cluster. This is also known as ``Spitzer Instability''~\citep{Spi69}. The Spitzer instability will lead to a shorter relaxation time in the core and an increased collisional cross section, both of which will increase the collision rate. What we can derive from the considerations in this subsection, is that collisions overall will be relevant and frequently occur, which may in turn may even accelerate the accretion process, due to the mass dependence in the Eddington and Bondi rate, and in particular early collisions may then determine the point when accretion starts to become more efficient.

\subsection{Effect of the IMF}\label{imf}
\begin{figure*}
    \includegraphics[width=0.65\columnwidth]{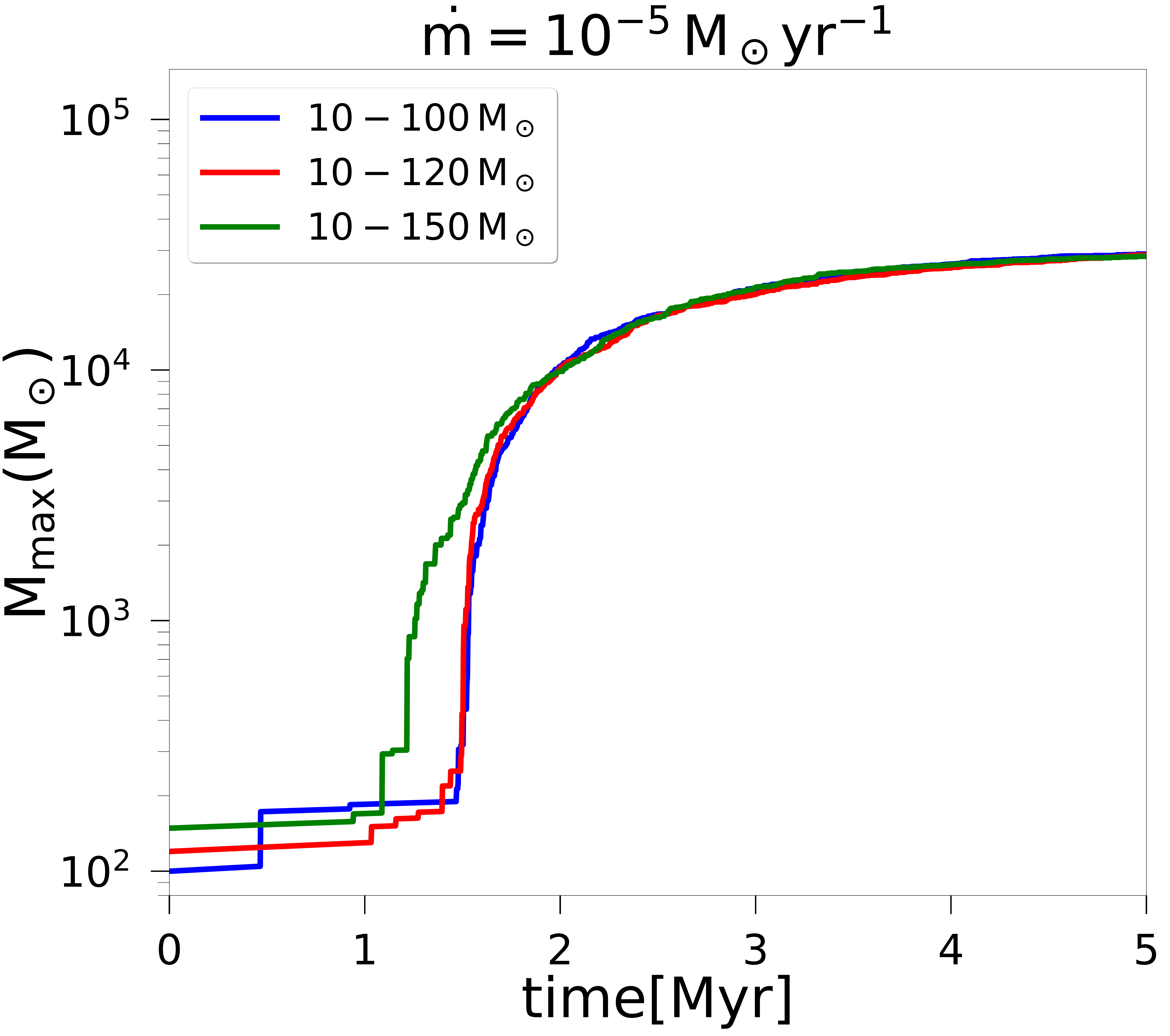}
    \includegraphics[width=0.65\columnwidth]{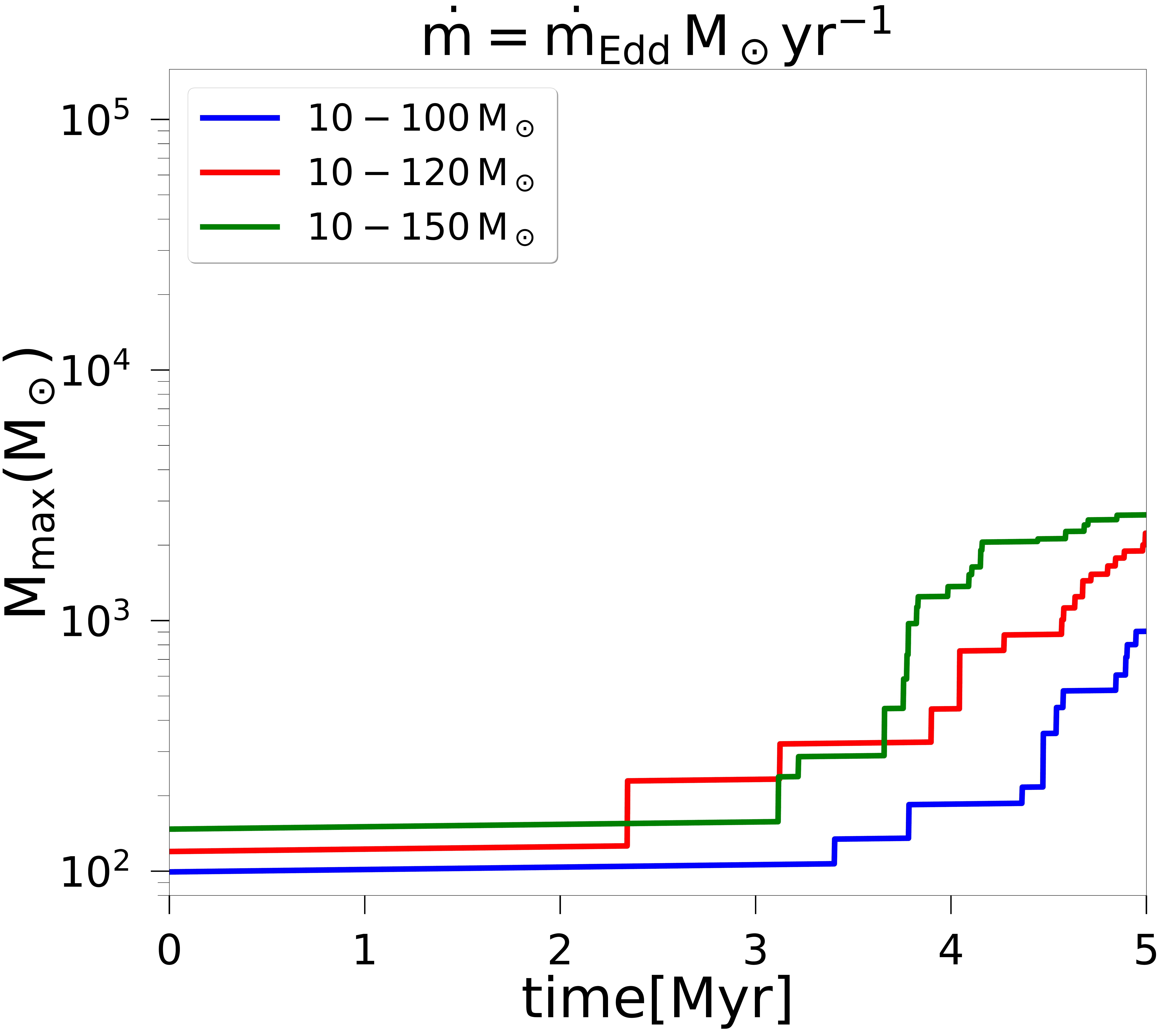}
    \includegraphics[width=0.65\columnwidth]{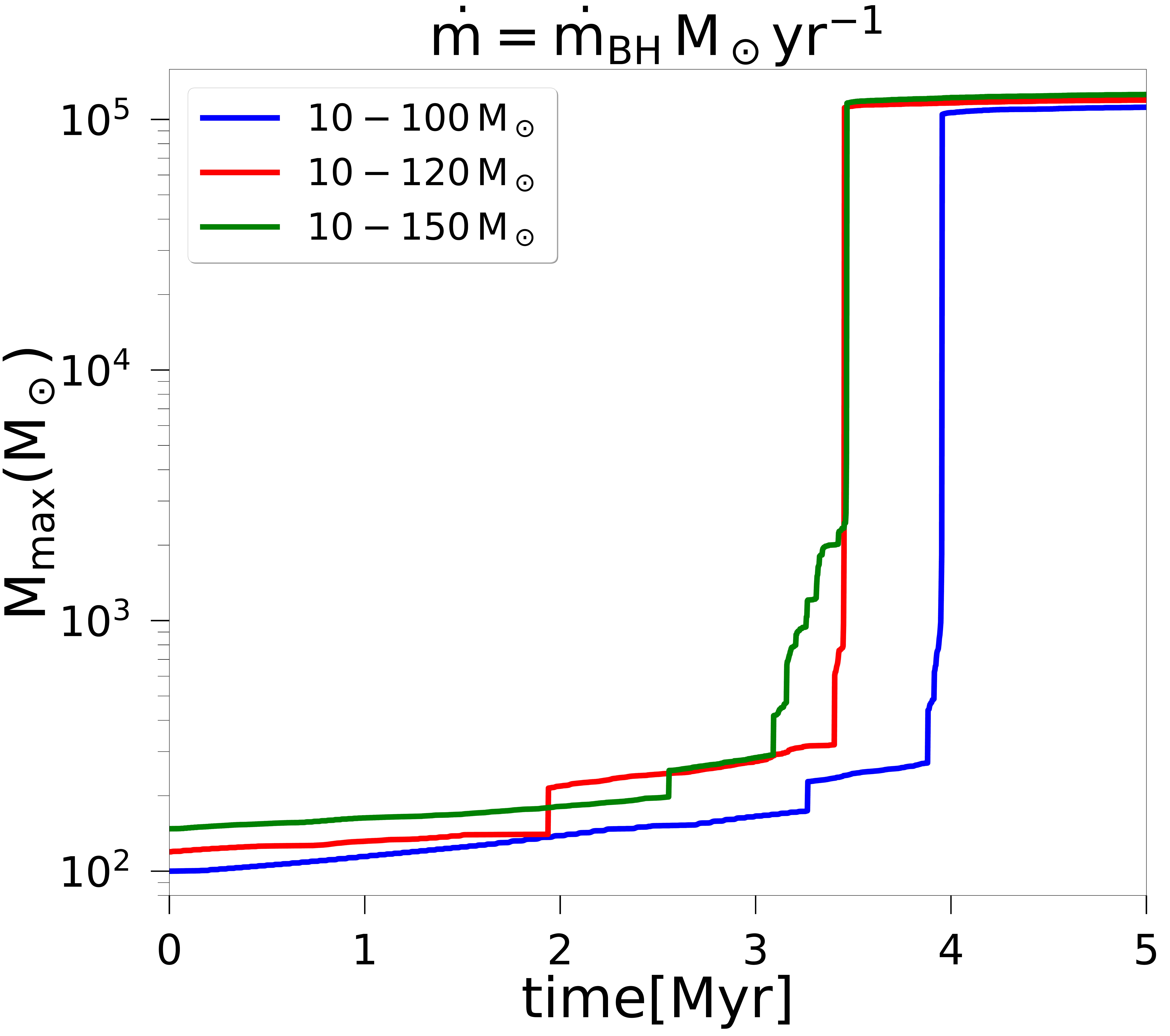}
\caption{Effect of varying the upper-mass end of the IMF on the mass of the MMO. The left panel shows the time evolution in case of a constant accretion rate of $\dot{m}=10^{-5}\acc$, the mid panel for the case of Eddington accretion, and the right panel for the case of Bondi accretion.   }
\label{IMF}
\end{figure*}

To understand how the results so far depend on additional assumptions, we explore how the results for the different accretion rates change using different assumptions regarding the IMF. We generally assume a Salpeter IMF with a lower-mass cutoff of $10 \MSun$, and vary the upper mass limit in the initial stellar mass distribution, considering $100$, $120$ and $150 \MSun$. In the first case with a constant accretion rate of $\dot{m}=10^{-5}\MSun\,\mathrm{yr}^{-1}$, Fig.~\ref{IMF} shows that such a variation has only a very moderate impact. The intrinsic scatter, when varying the initial conditions, means that the evolution at late times  basically cannot be uniquely distinguished between the different cases, but it is likely that overlap in the parameter space of the results will occur when considering a larger set of initial conditions.

For the case of Eddington accretion, we see very clearly that the MMO in our simulations grows more rapidly where the IMF extends to a higher mass. This is expected due to the dependence of the Eddington accretion rate on the mass of the accretor, accelerating both the accretion of the MMO as well as the conversion of gas mass into stellar mass in general. We note that the first collisions tend to occur earlier in the case of a larger initial mass, and we find a difference in the final mass by about a factor of 3 comparing the most extreme cases. In case of mass-dependent accretion recipes, the growth of the MMO is thus further enhanced if the initial IMF already extends to higher masses.

Finally, in the case of Bondi accretion, the dependence on the initial mass and the resulting behavior is even more extreme. Due to the steep dependence of the rate on the accretor mass, the central mass remains low for extended periods of time, then suddenly accelerates in a run-away fashion, as seen in the previous subsection. When the upper-mass cutoff of the stellar mass is higher, this rapid acceleration occurs earlier. For the cases considered here, a final mass of $10^5 \MSun$ is reached in all cases, only at different times. We note again that the latter corresponds to a rather extreme assumption, so this part of the results needs to be regarded with caution. It is interesting to note that a more moderate increase of the mass of the MMO is found at earlier times when the Bondi accretion rate is low and a larger higher-mass cutoff in the IMF will favor this growth. Similar results were also found by \citet{leigh13}.

\subsection{Dependence on cluster radius}\label{radialdependence}

\begin{figure*}
\includegraphics[width=0.65\columnwidth]{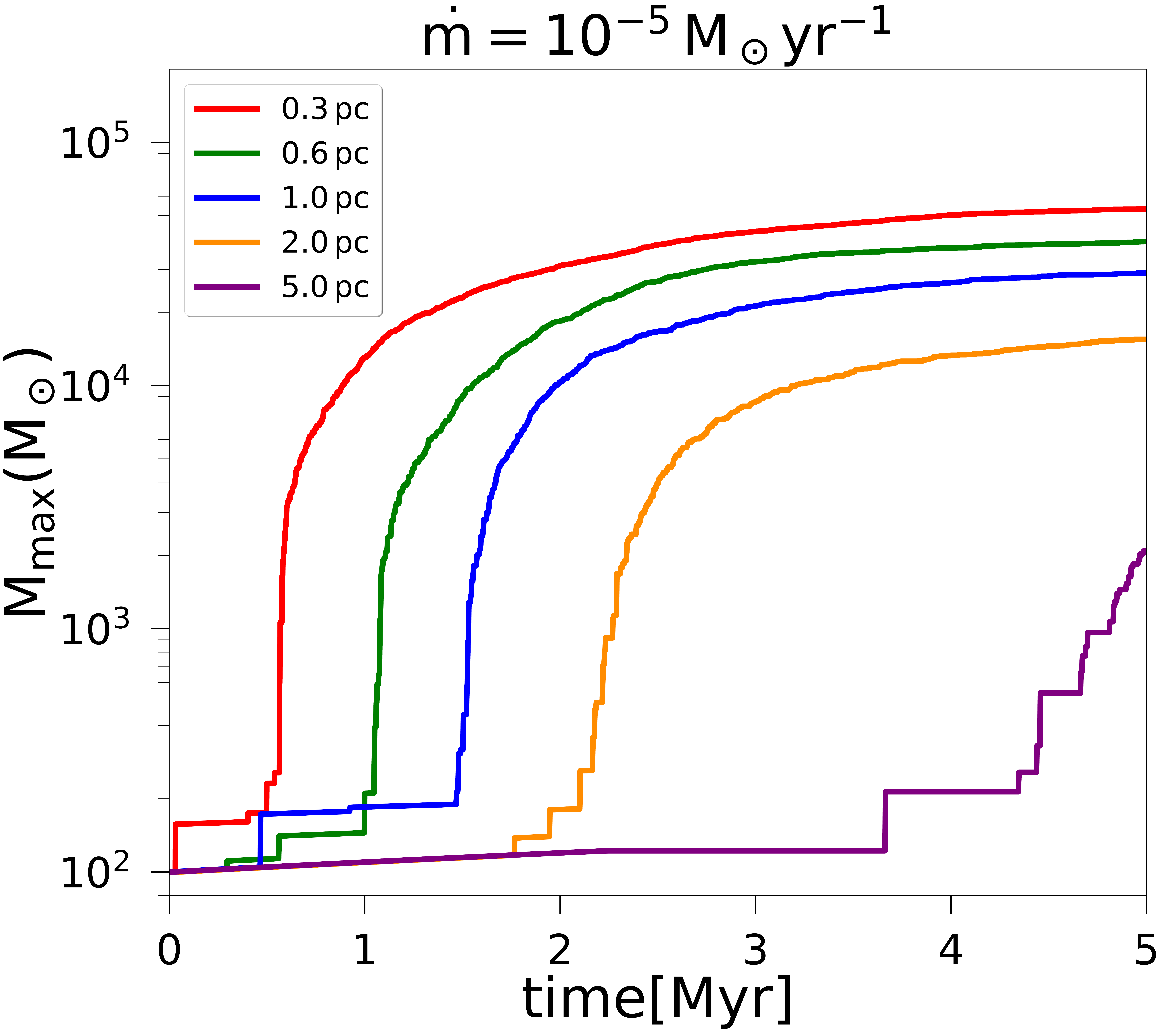}
\includegraphics[width=0.65\columnwidth]{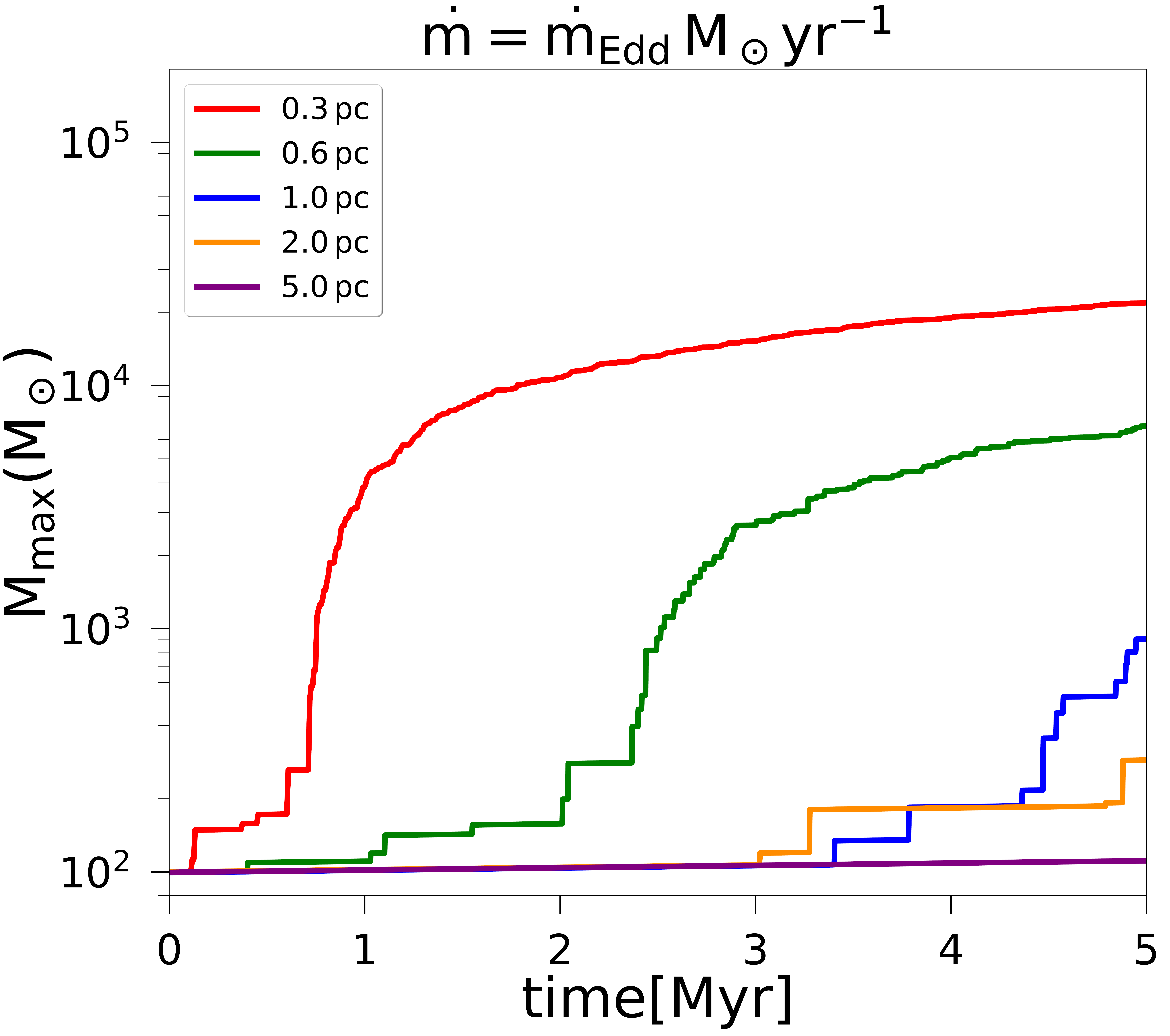}
\includegraphics[width=0.65\columnwidth]{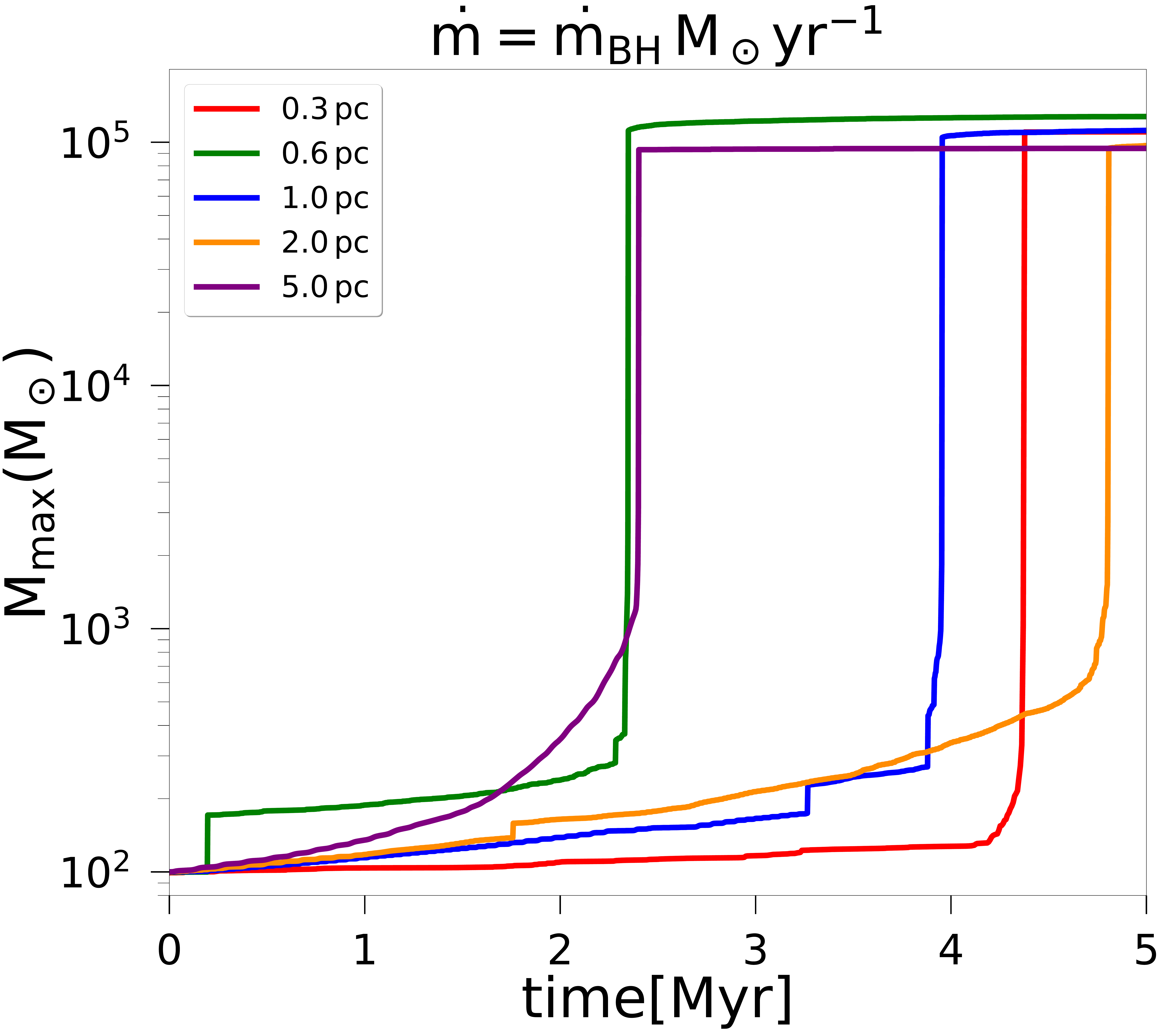}
\includegraphics[width=0.65\columnwidth]{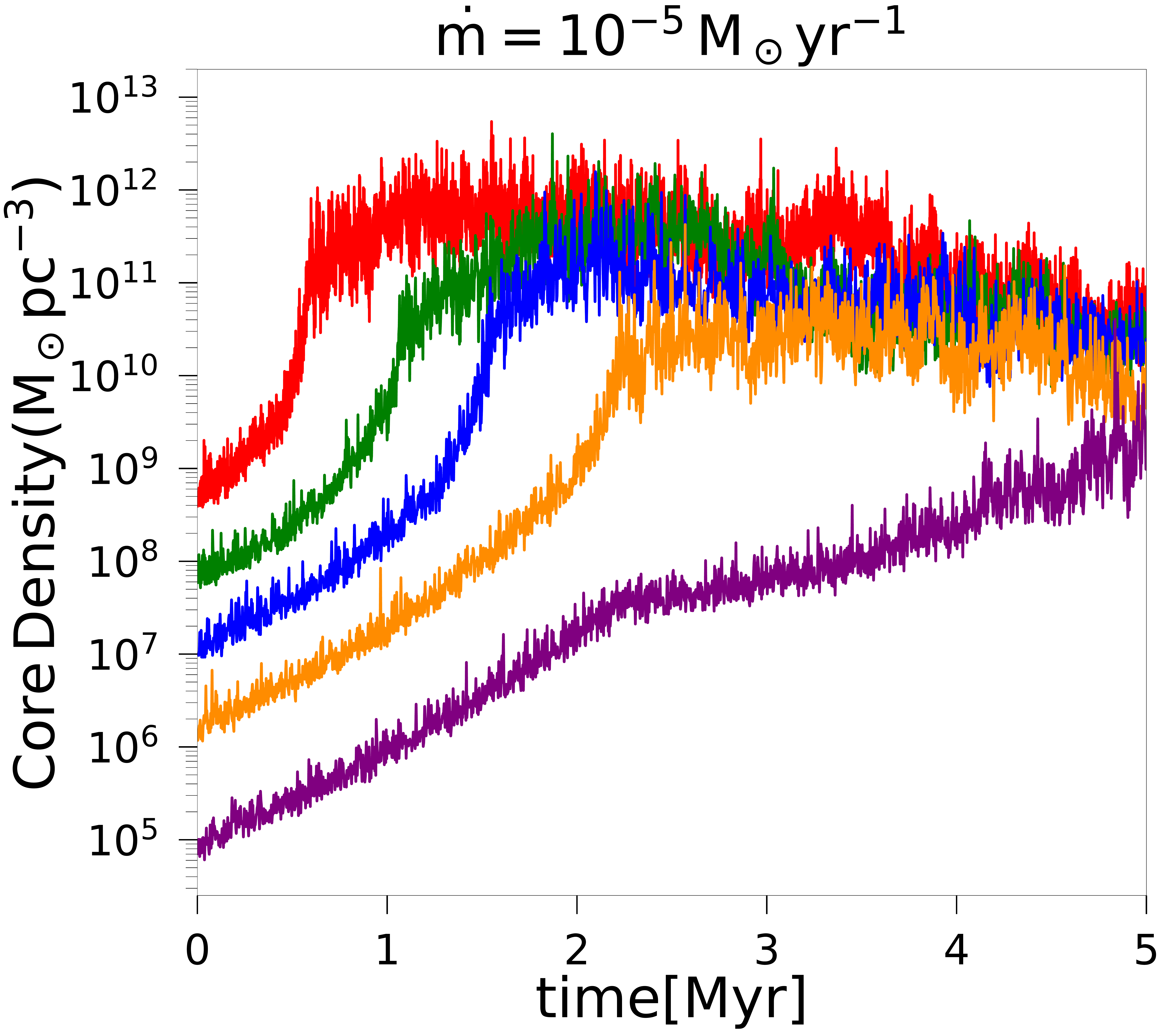}
\includegraphics[width=0.65\columnwidth]{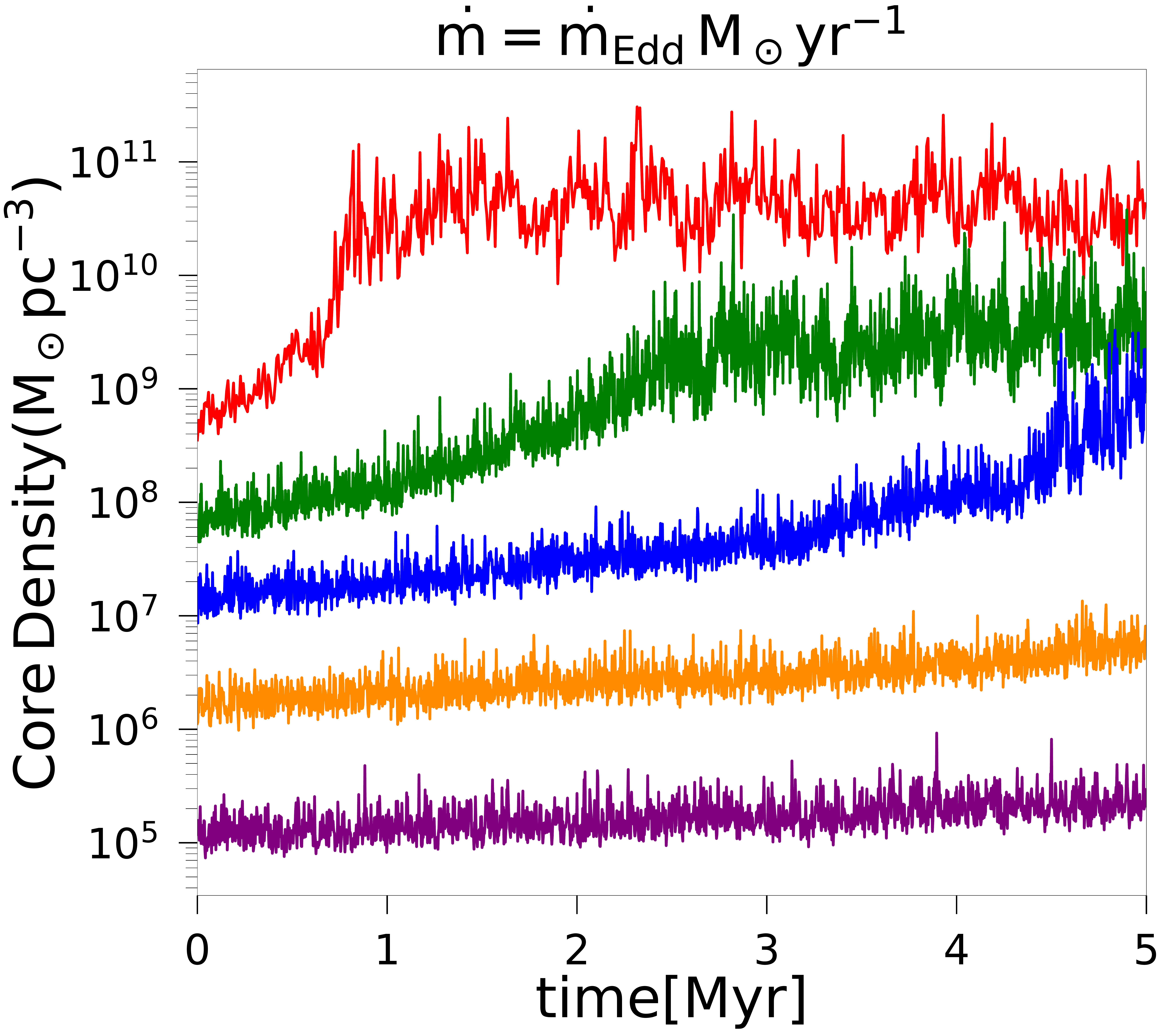}
\includegraphics[width=0.65\columnwidth]{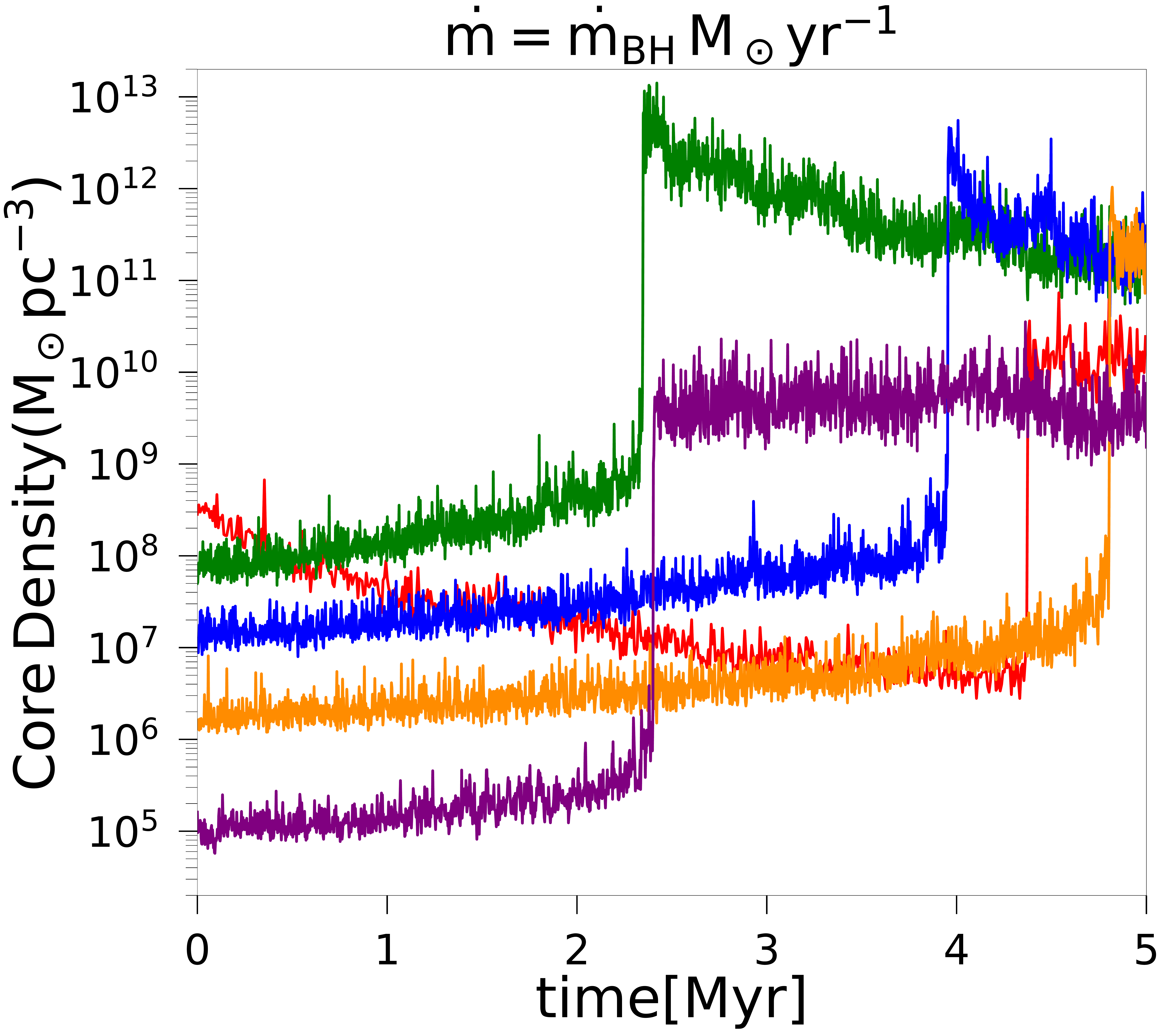}
\caption{Effect of $R_{\mathrm{cl}}$ (or the density) on the mass of the MMO. The left panel shows the time evolution in case of a constant accretion rate of $\dot{m}=10^{-5}\, \mathrm{\MSun yr^{-1}}$, the mid panel for the case of Eddington accretion, and the right panel for the case of Bondi accretion. In the bottom panel we show the evolution of the core density as a function of time, the color represents the same as the top panel. We adopt here our reference IMF between $10$ and $100\MSun$.}
\label{radius}
\end{figure*}

To also understand how the different accretion mechanisms operate for different central densities of the cluster, we vary $R_{\mathrm{cl}}$ and thus the central density of the cluster. Specifically, $R_{\mathrm{cl}}$ has been varied between $0.3$ and $5$~pc, leading to an evolution as shown in Fig.~\ref{radius} for the different scenarios. Depending on $R_{\mathrm{cl}}$, the initial central stellar densities range from $10^5\MSun\,\mathrm{pc}^{-3}$ up to $10^9\MSun\,\mathrm{pc}^{-3}$, and due to subsequent evolution and depending on accretion scenarios, even core densities of $10^8\MSun\,\mathrm{pc}^{-3}$ to $10^{12}\MSun\,\mathrm{pc}^{-3}$ can be reached.

For a constant accretion rate of $\dot{m}=10^{-5}\MSun\,\mathrm{yr}^{-1}$, we find a considerable dependence on $R_{\mathrm{cl}}$, with collisions in principle occurring in all cases, but their number depending significantly on the size of the cluster (see table \ref{tab:result}). For a large cluster with $R_{\mathrm{cl}}=5$~pc, the MMO still grows to about $2000 \MSun$, with most of the growth occurring after $3.5$ Myr due to the collisions. Even for  $R_{\mathrm{cl}}=2$~pc, the growth starts considerably earlier, with a steep rise in the mass of the MMO after $2$ Myr, and reaching a final mass of more than $10^4 \MSun$. For the most compact cluster with $R_{\mathrm{cl}}=0.3$~pc, rapid growth starts after $0.5$ Myr and the final mass reaches about $5\times10^4 \MSun$. The growth of the central object in these cases is mostly due to collisions, driven by the increase of the central density in the cluster, driven by contraction to the increase in mass. The mass of the MMO correlates well with the compactness of the cluster in the cases considered here.

In the Eddington accretion scenario where at least initially the accretion rate tends to be lower, the evolution depends more strongly on the size of the cluster, and for $R_{\mathrm{cl}}=2$~pc or more, no significant growth of the mass of the MMO is found. In the case of $R_{\mathrm{cl}}=1$~pc, the central mass starts growing after about 3.5 Myr, reaching about $800 \MSun$. In the case of more compact clusters, the evolution starts earlier, even within the first Myr in case of a $0.3$~pc cluster, with the MMO reaching a final mass of about $2\times10^4 \MSun$. The growth overall is less steep in this scenario, as the Eddington accretion rate is initially lower, thus leading to a more gradual increase in total stellar mass and the contraction of the cluster occurs more gradually. Again, the mass of the MMO correlates well with the compactness of the cluster for these configurations.

In the case of Bondi accretion, the behavior becomes somewhat more extreme. It is not directly related to the initial $R_{\mathrm{cl}}$, but depends more on when the first collision happens to occur, thereby producing an object of larger mass, to which the Bondi rate reacts very sensitively. The occurrence of this first collision will statistically vary for different cluster sizes, and in fact Fig.~\ref{radius} shows no clear dependence of the overall evolution on the $R_{\mathrm{cl}}$. In all cases, the evolution initially occurs very slowly, though at different times for the different simulations, a sudden acceleration occurs once the Bondi rate becomes sufficiently large. The final mass is always about $10^5 \MSun$, forming in between 2 and 5 Myr for all cases considered here. In this case, the evolution is mostly driven by accretion, with collisions not playing a relevant role.

In the cases considered here, the cluster size has the lowest impact in the case of the Bondi accretion rate, due to the very steep dependence on the mass of the MMO in this scenario. For more moderate mass dependencies, the cluster size is however found to be highly relevant, and can strongly affect the resulting evolution. Understanding the physical mechanism of accretion along with its dependence of mass will thus be crucial to determine the relevance of this parameter. While a more compact cluster in principle shortens the timescale for collisions, we also note that the latter is not necessarily the dominant effect. Particularly in the case of Bondi accretion, early collisions in the cluster will rather accelerate the accretion process due to the steep dependence on mass, thereby leading to an even more rapid growth via accretion. For the Eddington scenario and in case of a constant accretion rate, the shortening of collision timescale in comparison is more relevant,  and in these cases there is a clear correlation between having a more compact cluster and a more massive central object.

\subsection{Effect of mass loss during collisions}\label{secmassloss}
\begin{figure*}
    \includegraphics[width=0.65\columnwidth]{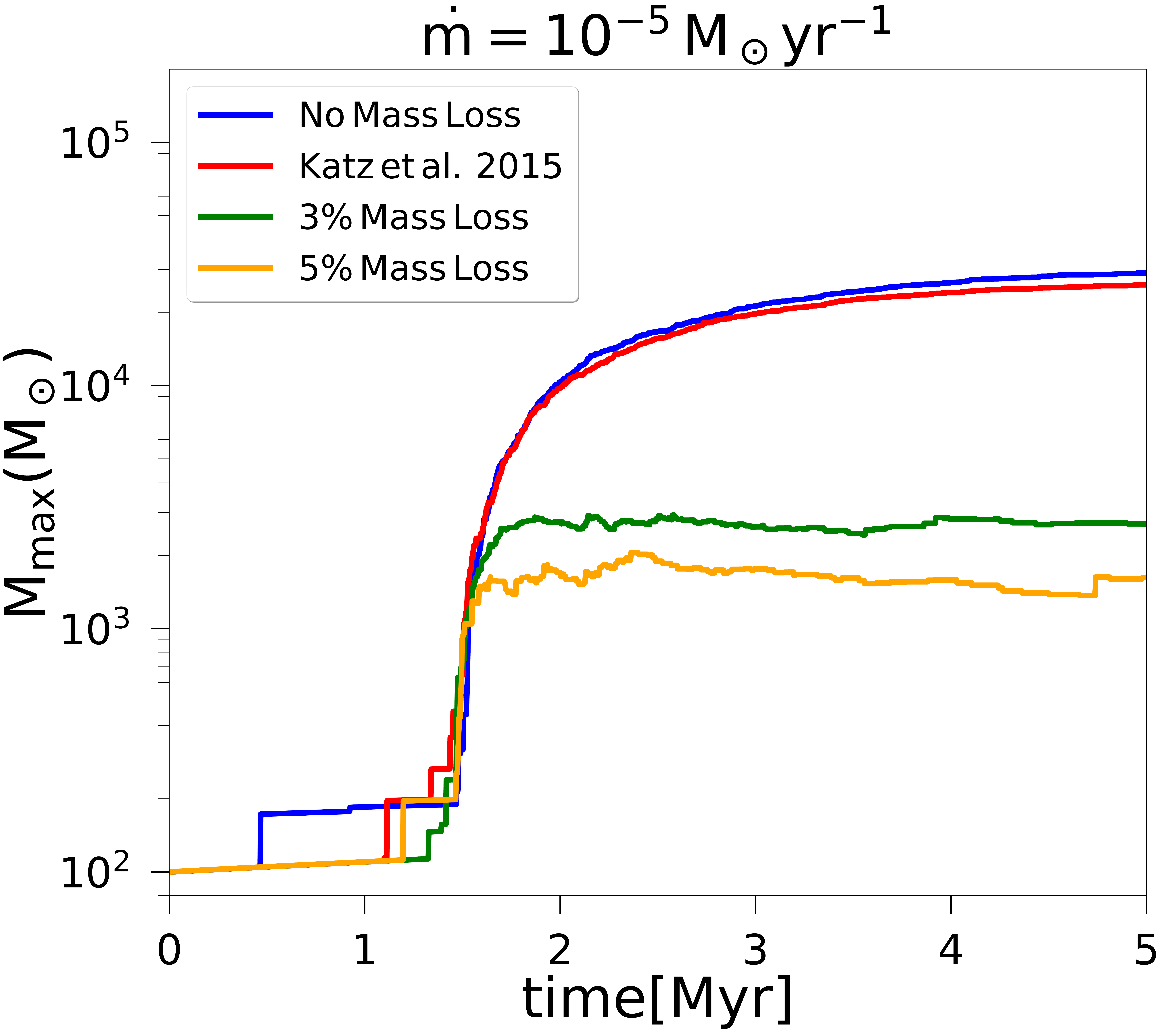}
    \includegraphics[width=0.65\columnwidth]{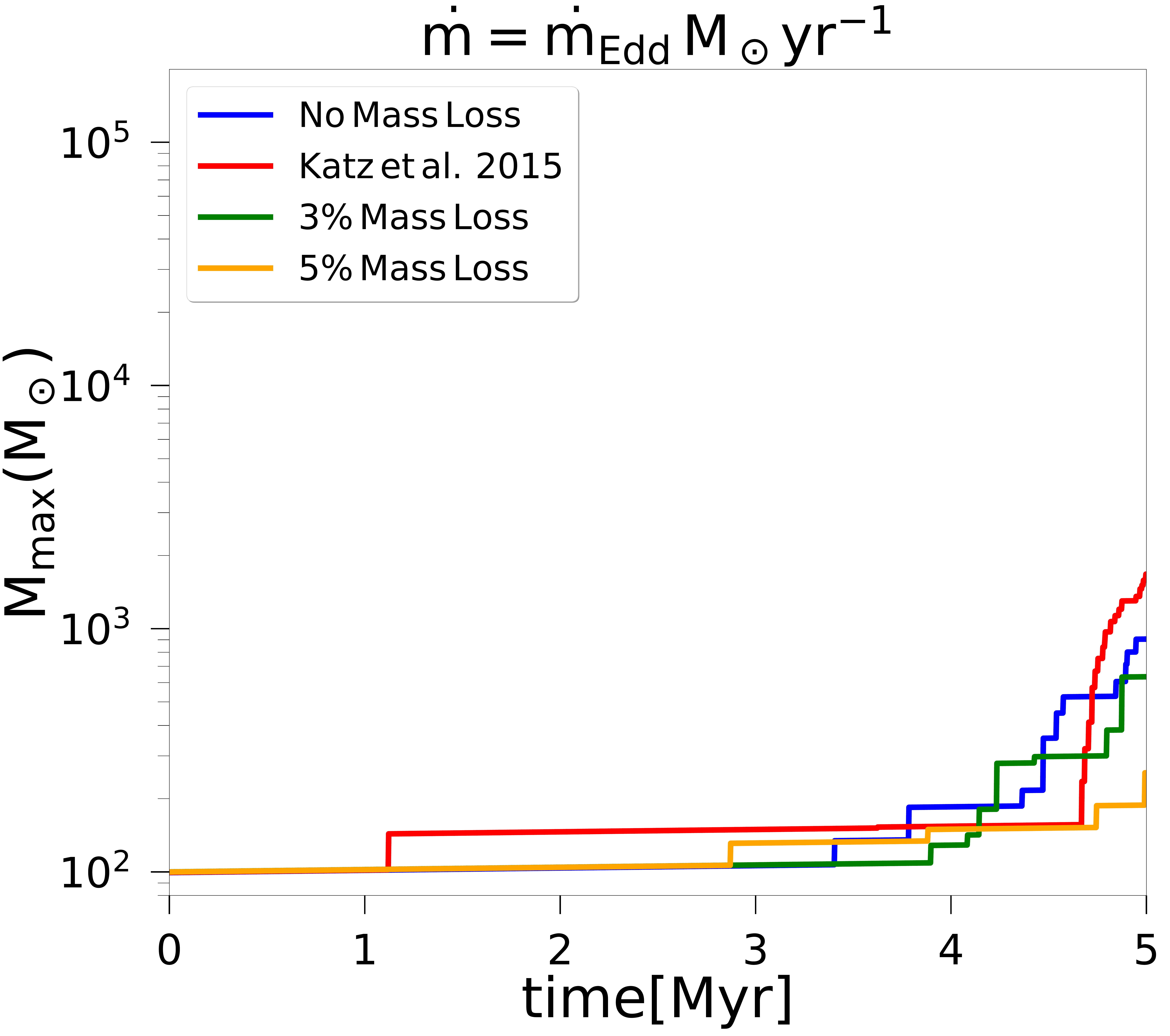}
    \includegraphics[width=0.65\columnwidth]{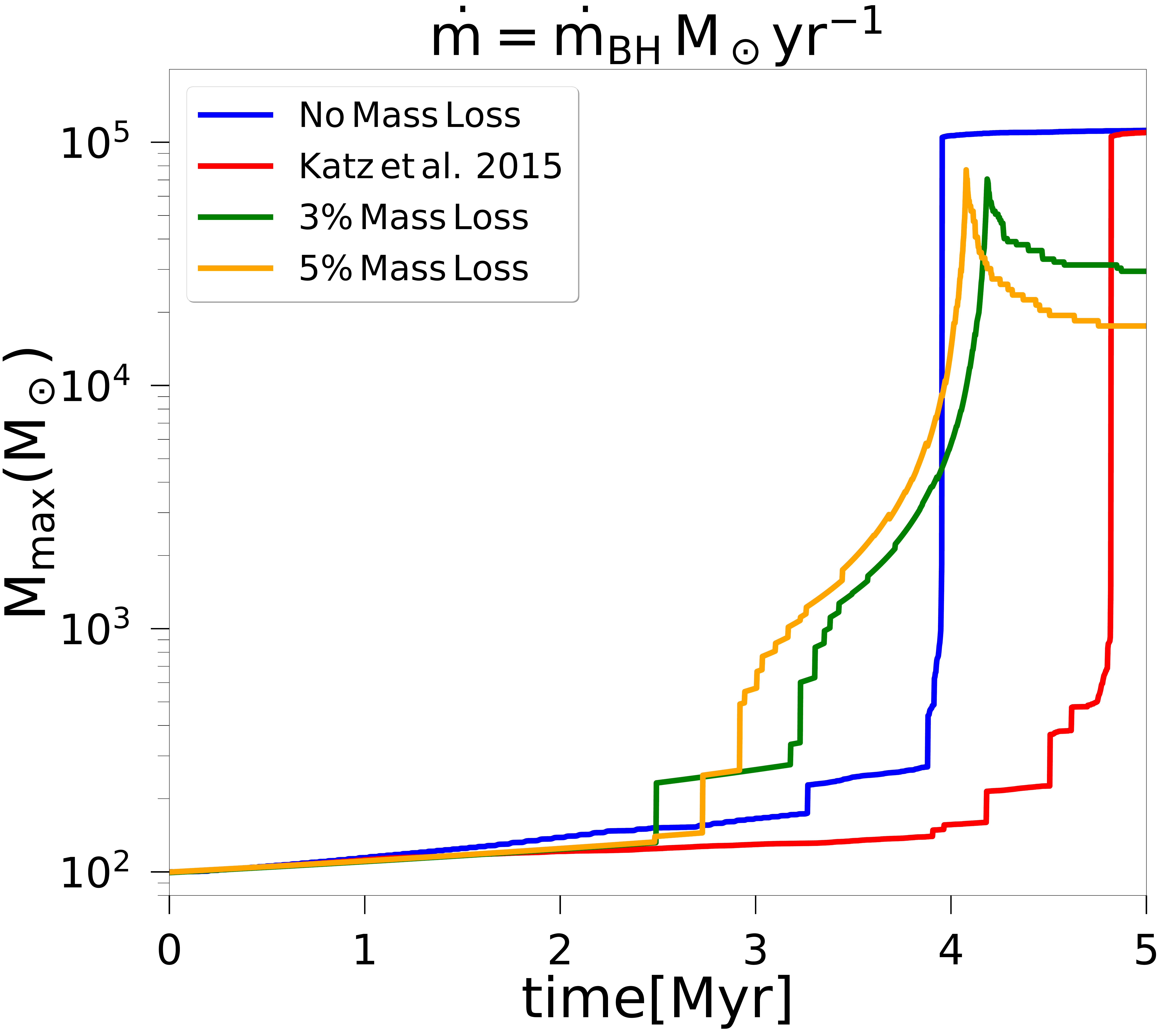}
    \caption{Effect of mass loss due to collisions on the mass of the MMO. The left panel shows the time evolution in case of a constant accretion rate of $\dot{m}=10^{-5}\, \mathrm{\MSun yr^{-1}}$, the mid panel for the case of Eddington accretion, and the right panel for the case of Bondi accretion. The blue line, the red line and the green line represent the cases of no mass loss, mass loss according to Eq. \ref{katzmassloss}, a constant mass loss fraction of  3$\%$  and a constant mass loss fraction of  5$\%$, respectively. We adopt here our reference IMF between $10$ and $100\MSun$.
    \label{massloss}
    }
\end{figure*}

Finally, the possible effect of mass loss during collisions is explored for different scenarios and the results are shown in Fig.~\ref{massloss}, considering different accretion scenarios, all for our reference IMF with masses between $10-100\MSun$.

The mass loss during collisions is particularly relevant in the case of a constant accretion rate, where we consider the reference case of $\dot{m}=10^{-5}\acc$. Both in the cases without mass loss as well as in K15, we find almost the same final mass of the MMO of about $3\times10^4\MSun$. This shows that most of the collisions occur at very high mass ratios, where K15 suggests the mass loss to have little impact. However, if this assumption is not correct or rather an underestimate, our calculation with a mass loss fraction of $3\%$ shows that the final mass could be considerably reduced by roughly a factor of 10, consistent with similar results by \citet{Ali20} for primordial protoclusters. The reason is that in this case, particularly for high mass ratios during the collisions, the mass loss can become comparable to the mass that is added during the collision, or potentially even larger, and then compensate for part of the accretion. The final mass is similar but a bit further reduced in case of a $5\%$ mass loss fraction, reflecting the flattening that was also found by \citet{Ali20} when plotting final masses as a function of the mass loss fraction.

In the case of Eddington accretion, we first have to note that here the results depend somewhat sensitively on when the first collision occurs, as subsequently the accretion rate will be enhanced, thereby providing a somewhat statistical element that depends on the nonlinear evolution. For this reason, the simulation with the K15 obtains the highest mass of a few times $10^3\MSun$, a very similar value as the simulation with no mass loss. In case of a $3\%$ mass loss fraction, the final mass is reduced to about $600\MSun$, and about $250\MSun$ for a $5\%$ fraction. The latter reflects roughly the expected trend regarding the final masses, with some super-imposed scatter due to the time of when the first collision occurs in the models.

Finally, in the case of Bondi accretion, in principle a similar behavior is found, only with an even steeper increase of the final masses at late times, and a more flat evolution initially, as also found before in our exploration of Bondi accretion. The highest final mass is attained here in the case with no mass loss, with about $10^5\MSun$ after about 4 Myr. Using K15, a very similar final mass is obtained after almost 5 Myr. The curves with $3\%$ and $5\%$ mass loss fractions behave similarly, starting to increase more significantly after about 3 Myr, reaching a maximum close to $10^5\MSun$ after about 4 Myr, while subsequently showing a decrease of their mass once  the gas reservoir is exhausted and accretion stops, while mass loss still continues due to collisions. The final mass after 5 Myr is thus between $2-4\times10^4\MSun$ in these cases.

\begin{table*}

    \begin{center}
        Summary of the results
    \end{center}

    \begin{tabular}{cccccccc}
    \hline
    \hline
    $\dot{m}_{\mathrm{acc}}(\MSun\mathrm{yr}^{-1})$ &$\Delta M(\MSun)$ & IMF($\MSun$) & $M_{\mathrm{cl}}=M_{\mathrm{g}}\,(\MSun)$ & $R_{\mathrm{cl}}=R_{\mathrm{g}}\,(\mathrm{pc})$ & $N$ & $N_{\mathrm{coll}}$ & $M_{\mathrm{max}}$\\
    \hline
    $10^{-4}$ & None & 10-100 & 1.10$\times 10^5$ & 1.0 & 5000 & 563 & 3.16$\times 10^4$ \\
    $10^{-5}$ & None & 10-100 & 1.10$\times 10^5$ & 1.0 & 5000 & 490 & 2.90$\times 10^4$ \\
    $10^{-6}$ & None & 10-100 & 1.10$\times 10^5$ & 1.0 & 5000 & 20 & 1.29$\times 10^4$ \\
    $10^{-5}$ & None & 10-120 & 1.14$\times 10^5$ & 1.0 & 5000 & 440 & 2.88$\times 10^4$ \\
    $10^{-5}$ & None & 10-150 & 1.22$\times 10^5$ & 1.0 & 5000 & 404 & 2.84$\times 10^4$ \\
    $\mbh$ & None & 10-100 & 1.10$\times 10^5$ & 1.0 & 5000 & 153 & 1.12$\times 10^5$ \\
    $\mbh$ & None & 10-120 & 1.14$\times 10^5$ & 1.0 & 5000 & 152 & 1.19$\times 10^5$ \\
    $\mbh$ & None & 10-150 & 1.22$\times 10^5$ & 1.0 & 5000 & 178 & 1.26$\times 10^5$ \\
    $\medd$ & None & 10-100 & 1.10$\times 10^5$ & 1.0 & 5000 & 15 & 9.06$\times 10^2$ \\
    $\medd$ & None & 10-120 & 1.14$\times10^5$ & 1.0 & 5000 & 31 & 2.23$\times 10^3$ \\
    $\medd$ & None & 10-150 & 1.22$\times 10^5$ & 1.0 & 5000 & 23 & 2.64$\times 10^3$ \\
    $10^{-5}$ & None & 10-100 & 1.10$\times 10^5$ & 0.3 & 5000 & 1018 & 5.33$\times 10^4$ \\
    $10^{-5}$ & None & 10-100 & 1.10$\times 10^5$ & 0.6 & 5000 & 691 & 3.91$\times 10^4$ \\
    $10^{-5}$ & None & 10-100 & 1.10$\times 10^5$ & 2.0 & 5000 & 246 & 1.55$\times 10^4$ \\
    $10^{-5}$ & None & 10-100 & 1.10$\times 10^5$ & 5.0 & 5000 & 28 & 2.08$\times 10^4$ \\
    $\mbh$ & None & 10-100 & 1.10$\times 10^5$ & 0.3 & 5000 & 454 & 1.10$\times 10^5$ \\
    $\mbh$ & None & 10-100 & 1.10$\times 10^5$ & 0.6 & 5000 & 426 & 1.28$\times 10^5$ \\
    $\mbh$ & None & 10-100 & 1.10$\times 10^5$ & 2.0 & 5000 & 28 & 9.68$\times 10^4$ \\
    $\mbh$ & None & 10-100 & 1.10$\times 10^5$ & 5.0 & 5000 & 24 & 9.44$\times 10^4$ \\
    $\medd$ & None & 10-100 & 1.10$\times 10^5$ & 0.3 & 5000 & 409 & 2.19$\times 10^4$ \\
    $\medd$ & None & 10-100 & 1.10$\times 10^5$ & 0.6 & 5000 & 107 & 6.83$\times 10^3$ \\
    $\medd$ & None & 10-100 & 1.10$\times 10^5$ & 2.0 & 5000 & 12 & 2.88$\times 10^2$ \\
    $\medd$ & None & 10-100 & 1.10$\times 10^5$ & 5.0 & 5000 & 0 & 1.11$\times 10^2$ \\
    $10^{-5}$ & K15 & 10-100 & 1.10$\times 10^5$ & 0.3 & 5000 & 471 & 2.59$\times 10^4$ \\
    $10^{-5}$ & $0.3\%$ & 10-100 & 1.10$\times 10^5$ & 0.6 & 5000 & 272 & 2.69$\times 10^4$ \\
    $10^{-5}$ & $0.5\%$ & 10-100 & 1.10$\times 10^5$ & 2.0 & 5000 & 234 & 1.62$\times 10^4$ \\
    $\mbh$ & K15 & 10-100 & 1.10$\times 10^5$ & 0.3 & 5000 & 86 & 1.10$\times 10^5$ \\
    $\mbh$ & $0.3\%$ & 10-100 & 1.10$\times 10^5$ & 0.6 & 5000 & 79 & 2.95$\times 10^4$ \\
    $\mbh$ & $0.5\%$ & 10-100 & 1.10$\times 10^5$ & 2.0 & 5000 & 76 & 1.76$\times 10^4$ \\
    $\medd$ & K15 & 10-100 & 1.10$\times 10^5$ & 0.3 & 5000 & 33 & 1.68$\times 10^3$ \\
    $\medd$ & $0.3\%$ & 10-100 & 1.10$\times 10^5$ & 0.6 & 5000 & 16 & 6.34$\times 10^2$ \\
    $\medd$ & $0.5\%$ & 10-100 & 1.10$\times 10^5$ & 2.0 & 5000 & 11 & 2.56$\times 10^2$ \\
    \hline
    \end{tabular}
    \caption{$\dot{m}_{\mathrm{acc}}(\MSun\mathrm{yr}^{-1}$): accretion model, $\Delta M(\MSun)$: Mass loss due to collision, IMF($\MSun$): Initial Mass Function used in the models, $M_{\mathrm{cl}}=M_{\mathrm{g}}\,(\MSun)$:Initial mass of the cluster (= mass of the gas), $R_{\mathrm{cl}}=R_{\mathrm{g}}\,(\mathrm{pc})$: Initial radius of the cluster (=radius of the gas), $N$:Initial number of particles, $N_{\mathrm{coll}}$: Total number of collisions, $M_{\mathrm{max}}$: Final mass of the MMO in the cluster \label{tab:result} }
    
\end{table*}

\subsection{Caveats}
One of the main caveats of this work is that the simulation setup we employed here is highly idealised and not based on cosmological initial conditions, and also not solving the hydrodynamic equations. Hence, the processes in real cosmological systems could be different if there are differences in the structure, or if the accretion recipes that we explored here do not represent the type of accretion that will occur in these halos. The main goal of this work was to build a simplified model that allows us to study a large part of the parameter space, and to study the evolution of the cluster in that parameter space over timescales of a few Myr, which would not be possible if we were to include the full hydrodynamics. If we were to do so, then in addition to the the cooling, the chemistry of the gas and  feedback processes due to the stars would need to be modeled in more detail. Recent studies by~\citet{Kro20} and ~\citet{Nat21} discuss in more detail about gas accretion onto black holes inside NSCs.

The results of this work motivate us to run more simulations and will be able to guide future work to explore parts of the parameter space that can be regarded as particularly promising. It will be particularly important to understand which of the accretion recipes employed here, if any, provides an accurate prescription that will be consistent with results from real hydrodynamical simulations. We want to point out that sticking to only the first few Myr of evolution of the cluster minimizes concerns related to numerical issues i.e. if we were to re-run the simulations, we are confident that we would get the same results over the relevant timescales we consider. Over a longer timescale we would expect more divergence in the results of identical simulation setups. In addition, we note that other complicating factors such as mass loss due to stellar winds, stellar evolution, rotation etc. are currently not included in our simulations. Particularly, it is very uncertain how collisions affect stellar evolution, and the mass loss through winds for stars more massive than $1000$~M$_\odot$ is not well known. These provide very relevant uncertainties that need to be studied in more detail to provide reliable conclusions regarding collision-based formation mechanisms of SMBH seeds.

\section{Summary and discussion}\label{Discussion}
In this work, we explored possible formation scenarios of SMSs in low-metallicity NSCs where mass loss from winds should be minimized. We explored how the interaction of collisions and accretion affects the formation of a central massive object for different physically motivated accretion scenarios: considering accretion rates that are constant in space and time, the Eddington accretion rate and the Bondi accretion rate. 

Our fiducial model consists of a cluster with $N=5000$, $M_{\mathrm{cl}}=M_{\mathrm{g}}=1.12\times 10^5 \MSun$, $R_{\mathrm{cl}}=R_{\mathrm{g}}=1$ pc, assuming a stellar mass distribution with Salpeter IMF with mass range $10-100 \MSun$. Then we studied different accretion scenarios and also the effect of varying different parameters e.g. the upper mass limit of the Salpeter IMF, the central densities of the cluster and the inclusion of mass loss due to collisions on the final mass of the MMO in the cluster.

In the case of a constant accretion rate, the number of collisions depends significantly on the magnitude of the accretion rate. Our simulations show that central massive objects of $\sim 10^{3-5} \MSun$ may form for accretion rates of $10^{-6}-10^{-4} \MSun$~yr$^{-1}$. If the SMSs formed in this way can directly collapse into seed black holes with similar mass, they can grow into the billion solar-mass black holes observed at $z\gtrsim 6$. In general, we find that the problem is quite sensitive to the initial conditions and the assumed recipe for the accretion, due to the highly chaotic and non-linear nature of the problem. The dominant factors that determine the mass growth are the mass of the MMO and the gas reservoir. Eddington accretion produces a central massive object of a few times $10^3 \MSun$ in our reference scenario, while Bondi accretion robustly produces final masses $\sim10^5 \MSun$. It is however important to note that the Bondi scenario assumes $\dot{M}_{\mathrm{BH}}\propto M_\ast^2$ and, as a result, the accretion rate remains low until a critical accretor mass is achieved leading to runaway growth. Only after a sufficiently long time of a few million years, the mass of the central object grows very rapidly. It is thus important to stress that such a scenario could only work if the gas is not previously expelled due to feedback, and if stellar evolution does not restrict the evolution to shorter timescales. In addition, Bondi accretion in general may be an extreme assumption, and we caution that some of the assumptions in our model will break down during the rapid phase of Bondi accretion due to the very rapid gas depletion.

We compared the collision and accretion timescales computed both analytically and from our simulations. Our calculations suggest that the collision timescale is much greater when calculated for the MMO. However, we emphasize that the collision rates we employed for this estimate did not account for the potentially enhanced collision probability in the presence of accretion, and our simulations show that the number of collisions increases for larger accretion rates. It further must be noted that when evaluating the timescale to have collisions between any two objects within the cluster, the collision timescale can become comparable to or shorter than accretion timescale. When considering the effects of an IMF,  the ``Spitzer Instability'' \citep{Spi69}  will lead to a shorter relaxation time in the core and an increased collisional cross section, both of which will increase the collision rate. We therefore expect collisions to play an important relevant role for the growth of the most massive object, consistent with what we find in the simulations as well. As the gas is depleted, collisions start to take over as the dominant mass growth mechanism for the MMO.  This is only accelerated by the effects of cluster contraction induced by the presence of the gas reservoir and the dissipational effects it introduces to the distribution of relative particle velocities and masses \citep{leigh14}.

We explored the dependence of our results on the details of the IMF, considering Salpeter-type IMFs with different upper-mass limits. For the parameter space we explored, when considering constant accretion rates, the results were insensitive to such a change, while in the case of Eddington accretion, a larger upper-mass limit favours larger final masses. In the case of Bondi-accretion, the time of occurrence of the first collision is more important than an initially larger mass, particularly when the mass change due to a collision is larger than the mass change due to a variation of the upper mass cutoff of the IMF.

We further explored the dependence on the central density of the cluster by varying its radius and keeping the total cluster mass constant. For models with constant accretion rates, a more compact cluster significantly favors the formation of a supermassive star, with the final mass varying by more than an order of magnitude, for variations of cluster radii from $0.3$ pc to 5 pc, as found by \citet{giersz15} for clusters without gas.  A very similar dependence is found for Eddington accretion. It is interesting to note that a central massive object of mass $\sim 10^4 \MSun$ could be formed for the case of Eddington accretion if the cluster has central radius $\sim 0.3 pc$ or less. In the Bondi case, the final mass is independent of the size of the cluster, but the required timescale to form the central massive object can vary substantially depending on when the first collision occurs.

A very relevant uncertainty in these models is the mass loss during collisions. We compared simulations employing different mass loss recipes, considering the prescription by \citet{Kat15}, where the effect of the mass loss is effectively negligible, as it becomes very small when the mass ratio between the collision partners is high. On the other hand, we noticed that a fixed mass loss of $3\%$ or $5\%$ during collisions has a more pronounced effect and can change the final mass by around an order of magnitude, as the mass loss may then become comparable to the mass gained via the collisions. It will thus be important to understand the expected mass loss from stellar models, as noted in the case of completely primordial clusters by \citet{Ali20}.

As mentioned already in the introduction, the future will provide relevant opportunities for hopefully observing and probing the formation of very massive objects in the early Universe. From the work pursued here (and considering the results of other studies), we expect that SMSs formed via moderate accretion rates of less than $10^{-3} \MSun$~yr$^{-1}$, the SMSs will appear with blue colours and high luminosities, while for larger accretion rates or in the regime when collisions occur extremely frequently, the SMSs may be bloated up and preferentially show red colours. It needs to be determined via future observations if such a bimodality actually occurs, or if there is a predominant way in which these types of objects may actually be found.

Understanding the evolution of very massive metal poor stars in general will be important for further progress on the topic, as the lifetime of these stars can be a relevant limiting factor for the models. For a 20 $\MSun$ single star it is roughly 5.6 Myr. Above $\sim 40\MSun$, the lifetime is almost constant ($\sim 3.5-5$ Myr) \citep{Hur00}, while SMSs evolve over a timescale (given by eq.~\ref{lifetimeSMS}) ranging from $6$ to $0.06$~Myrs for SMS masses of $10^{3-5}\MSun$. These results may significantly change in the presence of rotation \citep[e.g.][]{Leigh16}, which can act as a stabilizing factor and an accelerator for the collision rate by reducing the relative velocities between objects due to more correlated motions in rotating cluster environments.  It may also affect  the accretion rate of gas onto stars, when the first collisions actually occur and even the underlying stellar evolution of the collision products. We conclude that it will be important to investigate the stellar evolution of such objects in dynamical environments, considering accretion and/or collisions.

\section*{Acknowledgements}
We thank the anonymous referee for constructive comments on the manuscript. This work received funding from the Mitacs Globalink Research Award, Western University Science International Engagement Fund (SIEF), from ANID PIA ACT172033, the  Millenium Nucleus NCN19$\_$058 (TITANs) and the BASAL Centro de Excelencia en Astrofisica y Tecnologias Afines (CATA) grant PFB-06/2007. This research was made possible by the computational facilities provided by the Shared Hierarchical Academic Research Computing Network (SHARCNET: www.sharcnet.ca) and Compute Canada (www.computecanada.ca). We thank Tyrone Woods for valuable discussion.  N.W.C.L. gratefully acknowledges support from the Chilean government via Fondecyt Iniciac\'ion Grant \#11180005. This project was also supported by funds from the European Research Council (ERC) under the European Union’s Horizon 2020 research and innovation program under grant agreement No 638435 (GalNUC).

\section*{Data availability}
The data underlying this article will be shared on reasonable request to the corresponding author.

\bibliographystyle{mnras}
\bibliography{nuclearcluster} 

\begin{thebibliography}{}
\makeatletter
\relax
\def\mn@urlcharsother{\let\do\@makeother \do\$\do\&\do\#\do\^\do\_\do\%\do\~}
\def\mn@doi{\begingroup\mn@urlcharsother \@ifnextchar [ {\mn@doi@}
  {\mn@doi@[]}}
\def\mn@doi@[#1]#2{\def\@tempa{#1}\ifx\@tempa\@empty \href
  {http://dx.doi.org/#2} {doi:#2}\else \href {http://dx.doi.org/#2} {#1}\fi
  \endgroup}
\def\mn@eprint#1#2{\mn@eprint@#1:#2::\@nil}
\def\mn@eprint@arXiv#1{\href {http://arxiv.org/abs/#1} {{\tt arXiv:#1}}}
\def\mn@eprint@dblp#1{\href {http://dblp.uni-trier.de/rec/bibtex/#1.xml}
  {dblp:#1}}
\def\mn@eprint@#1:#2:#3:#4\@nil{\def\@tempa {#1}\def\@tempb {#2}\def\@tempc
  {#3}\ifx \@tempc \@empty \let \@tempc \@tempb \let \@tempb \@tempa \fi \ifx
  \@tempb \@empty \def\@tempb {arXiv}\fi \@ifundefined
  {mn@eprint@\@tempb}{\@tempb:\@tempc}{\expandafter \expandafter \csname
  mn@eprint@\@tempb\endcsname \expandafter{\@tempc}}}

\bibitem[\protect\citeauthoryear{{Abel}, {Bryan}  \& {Norman}}{{Abel}
  et~al.}{2002}]{Abe02}
{Abel} T.,  {Bryan} G.~L.,   {Norman} M.~L.,  2002, \mn@doi [Science]
  {10.1126/science.295.5552.93}, \href
  {https://ui.adsabs.harvard.edu/abs/2002Sci...295...93A} {295, 93}

\bibitem[\protect\citeauthoryear{{Agarwal}, {Khochfar}, {Johnson}, {Neistein},
  {Dalla Vecchia}  \& {Livio}}{{Agarwal} et~al.}{2012}]{Aga12}
{Agarwal} B.,  {Khochfar} S.,  {Johnson} J.~L.,  {Neistein} E.,  {Dalla
  Vecchia} C.,   {Livio} M.,  2012, \mn@doi [\mnras]
  {10.1111/j.1365-2966.2012.21651.x}, \href
  {https://ui.adsabs.harvard.edu/abs/2012MNRAS.425.2854A} {425, 2854}

\bibitem[\protect\citeauthoryear{{Agarwal}, {Cullen}, {Khochfar}, {Ceverino}
  \& {Klessen}}{{Agarwal} et~al.}{2019}]{Aga19}
{Agarwal} B.,  {Cullen} F.,  {Khochfar} S.,  {Ceverino} D.,   {Klessen} R.~S.,
  2019, \mn@doi [\mnras] {10.1093/mnras/stz1347}, \href
  {https://ui.adsabs.harvard.edu/abs/2019MNRAS.488.3268A} {488, 3268}

\bibitem[\protect\citeauthoryear{{Alexander} \& {Bar-Or}}{{Alexander} \&
  {Bar-Or}}{2017}]{Ale17}
{Alexander} T.,  {Bar-Or} B.,  2017, \mn@doi [Nature Astronomy]
  {10.1038/s41550-017-0147}, \href
  {https://ui.adsabs.harvard.edu/abs/2017NatAs...1E.147A} {1, 0147}

\bibitem[\protect\citeauthoryear{{Alexander} \& {Natarajan}}{{Alexander} \&
  {Natarajan}}{2014}]{Ale14}
{Alexander} T.,  {Natarajan} P.,  2014, \mn@doi [Science]
  {10.1126/science.1251053}, \href
  {https://ui.adsabs.harvard.edu/abs/2014Sci...345.1330A} {345, 1330}

\bibitem[\protect\citeauthoryear{{Alister Seguel}, {Schleicher}, {Boekholt},
  {Fellhauer}  \& {Klessen}}{{Alister Seguel} et~al.}{2020}]{Ali20}
{Alister Seguel} P.~J.,  {Schleicher} D.~R.~G.,  {Boekholt} T.~C.~N.,
  {Fellhauer} M.,   {Klessen} R.~S.,  2020, \mn@doi [\mnras]
  {10.1093/mnras/staa456}, \href
  {https://ui.adsabs.harvard.edu/abs/2020MNRAS.493.2352A} {493, 2352}

\bibitem[\protect\citeauthoryear{{Aykutalp}, {Wise}, {Spaans}  \&
  {Meijerink}}{{Aykutalp} et~al.}{2014}]{Ayk14}
{Aykutalp} A.,  {Wise} J.~H.,  {Spaans} M.,   {Meijerink} R.,  2014, \mn@doi
  [\apj] {10.1088/0004-637X/797/2/139}, \href
  {https://ui.adsabs.harvard.edu/abs/2014ApJ...797..139A} {797, 139}

\bibitem[\protect\citeauthoryear{{Ba{\~n}ados} et~al.,}{{Ba{\~n}ados}
  et~al.}{2018}]{Ban18}
{Ba{\~n}ados} E.,  et~al., 2018, \mn@doi [\nat] {10.1038/nature25180}, \href
  {https://ui.adsabs.harvard.edu/abs/2018Natur.553..473B} {553, 473}

\bibitem[\protect\citeauthoryear{{Baraffe}, {Heger}  \& {Woosley}}{{Baraffe}
  et~al.}{2001}]{Bar01}
{Baraffe} I.,  {Heger} A.,   {Woosley} S.~E.,  2001, \mn@doi [\apj]
  {10.1086/319808}, \href
  {https://ui.adsabs.harvard.edu/abs/2001ApJ...550..890B} {550, 890}

\bibitem[\protect\citeauthoryear{{Barkana} \& {Loeb}}{{Barkana} \&
  {Loeb}}{2001}]{Ren01}
{Barkana} R.,  {Loeb} A.,  2001, \mn@doi [\physrep]
  {10.1016/S0370-1573(01)00019-9}, \href
  {https://ui.adsabs.harvard.edu/abs/2001PhR...349..125B} {349, 125}

\bibitem[\protect\citeauthoryear{{Barrera}, {Leigh}, {Reinoso}, {Stutz}  \&
  {Schleicher}}{{Barrera} et~al.}{2020}]{Barrera2020}
{Barrera} C.,  {Leigh} N. W.~C.,  {Reinoso} B.,  {Stutz} A.~M.,   {Schleicher}
  D.,  2020, arXiv e-prints, \href
  {https://ui.adsabs.harvard.edu/abs/2020arXiv201100021B} {p. arXiv:2011.00021}

\bibitem[\protect\citeauthoryear{{Basu} \& {Das}}{{Basu} \&
  {Das}}{2019}]{Bas19}
{Basu} S.,  {Das} A.,  2019, \mn@doi [\apjl] {10.3847/2041-8213/ab2646}, \href
  {https://ui.adsabs.harvard.edu/abs/2019ApJ...879L...3B} {879, L3}

\bibitem[\protect\citeauthoryear{{Becerra}, {Marinacci}, {Bromm}  \&
  {Hernquist}}{{Becerra} et~al.}{2018}]{Bec18}
{Becerra} F.,  {Marinacci} F.,  {Bromm} V.,   {Hernquist} L.~E.,  2018, \mn@doi
  [\mnras] {10.1093/mnras/sty2210}, \href
  {https://ui.adsabs.harvard.edu/abs/2018MNRAS.480.5029B} {480, 5029}

\bibitem[\protect\citeauthoryear{{Begelman} \& {Volonteri}}{{Begelman} \&
  {Volonteri}}{2017}]{Beg17}
{Begelman} M.~C.,  {Volonteri} M.,  2017, \mn@doi [\mnras]
  {10.1093/mnras/stw2446}, \href
  {https://ui.adsabs.harvard.edu/abs/2017MNRAS.464.1102B} {464, 1102}

\bibitem[\protect\citeauthoryear{{Begelman}, {Volonteri}  \& {Rees}}{{Begelman}
  et~al.}{2006}]{Beg06}
{Begelman} M.~C.,  {Volonteri} M.,   {Rees} M.~J.,  2006, \mn@doi [\mnras]
  {10.1111/j.1365-2966.2006.10467.x}, \href
  {https://ui.adsabs.harvard.edu/abs/2006MNRAS.370..289B} {370, 289}

\bibitem[\protect\citeauthoryear{{Bender} et~al.,}{{Bender}
  et~al.}{2005}]{Ben05}
{Bender} R.,  et~al., 2005, \mn@doi [\apj] {10.1086/432434}, \href
  {https://ui.adsabs.harvard.edu/abs/2005ApJ...631..280B} {631, 280}

\bibitem[\protect\citeauthoryear{{Boekholt}, {Schleicher}, {Fellhauer},
  {Klessen}, {Reinoso}, {Stutz}  \& {Haemmerl{\'e}}}{{Boekholt}
  et~al.}{2018}]{Boe18}
{Boekholt} T.~C.~N.,  {Schleicher} D.~R.~G.,  {Fellhauer} M.,  {Klessen} R.~S.,
   {Reinoso} B.,  {Stutz} A.~M.,   {Haemmerl{\'e}} L.,  2018, \mn@doi [\mnras]
  {10.1093/mnras/sty208}, \href
  {https://ui.adsabs.harvard.edu/abs/2018MNRAS.476..366B} {476, 366}

\bibitem[\protect\citeauthoryear{{B{\"o}ker}, {Laine}, {van der Marel},
  {Sarzi}, {Rix}, {Ho}  \& {Shields}}{{B{\"o}ker} et~al.}{2002}]{Bok02}
{B{\"o}ker} T.,  {Laine} S.,  {van der Marel} R.~P.,  {Sarzi} M.,  {Rix} H.-W.,
   {Ho} L.~C.,   {Shields} J.~C.,  2002, \mn@doi [\aj] {10.1086/339025}, \href
  {https://ui.adsabs.harvard.edu/abs/2002AJ....123.1389B} {123, 1389}

\bibitem[\protect\citeauthoryear{{Bond}, {Arnett}  \& {Carr}}{{Bond}
  et~al.}{1984}]{Bon84}
{Bond} J.~R.,  {Arnett} W.~D.,   {Carr} B.~J.,  1984, \mn@doi [\apj]
  {10.1086/162057}, \href
  {https://ui.adsabs.harvard.edu/abs/1984ApJ...280..825B} {280, 825}

\bibitem[\protect\citeauthoryear{{Bondi}}{{Bondi}}{1952}]{Bon52}
{Bondi} H.,  1952, \mn@doi [\mnras] {10.1093/mnras/112.2.195}, \href
  {https://ui.adsabs.harvard.edu/abs/1952MNRAS.112..195B} {112, 195}

\bibitem[\protect\citeauthoryear{{Bonnell}, {Bate}, {Clarke}  \&
  {Pringle}}{{Bonnell} et~al.}{2001}]{Bon01}
{Bonnell} I.~A.,  {Bate} M.~R.,  {Clarke} C.~J.,   {Pringle} J.~E.,  2001,
  \mn@doi [\mnras] {10.1046/j.1365-8711.2001.04270.x}, \href
  {https://ui.adsabs.harvard.edu/abs/2001MNRAS.323..785B} {323, 785}

\bibitem[\protect\citeauthoryear{{Bromm} \& {Loeb}}{{Bromm} \&
  {Loeb}}{2003}]{Bro03}
{Bromm} V.,  {Loeb} A.,  2003, \mn@doi [\apj] {10.1086/377529}, \href
  {https://ui.adsabs.harvard.edu/abs/2003ApJ...596...34B} {596, 34}

\bibitem[\protect\citeauthoryear{Capelo, Volonteri, Dotti, Bellovary, Mayer  \&
  Governato}{Capelo et~al.}{2015}]{Cap15}
Capelo P.~R.,  Volonteri M.,  Dotti M.,  Bellovary J.~M.,  Mayer L.,
  Governato F.,  2015, \mn@doi [Monthly Notices of the Royal Astronomical
  Society] {10.1093/mnras/stu2500}, 447, 2123

\bibitem[\protect\citeauthoryear{{Carollo}, {Stiavelli}, {de Zeeuw}  \&
  {Mack}}{{Carollo} et~al.}{1997}]{Car97}
{Carollo} C.~M.,  {Stiavelli} M.,  {de Zeeuw} P.~T.,   {Mack} J.,  1997,
  \mn@doi [\aj] {10.1086/118654}, \href
  {https://ui.adsabs.harvard.edu/abs/1997AJ....114.2366C} {114, 2366}

\bibitem[\protect\citeauthoryear{{Chon} \& {Omukai}}{{Chon} \&
  {Omukai}}{2020}]{Cho20}
{Chon} S.,  {Omukai} K.,  2020, \mn@doi [\mnras] {10.1093/mnras/staa863}, \href
  {https://ui.adsabs.harvard.edu/abs/2020MNRAS.494.2851C} {494, 2851}

\bibitem[\protect\citeauthoryear{{Chon}, {Hosokawa}  \& {Yoshida}}{{Chon}
  et~al.}{2018}]{Cho18}
{Chon} S.,  {Hosokawa} T.,   {Yoshida} N.,  2018, \mn@doi [\mnras]
  {10.1093/mnras/sty086}, \href
  {https://ui.adsabs.harvard.edu/abs/2018MNRAS.475.4104C} {475, 4104}

\bibitem[\protect\citeauthoryear{{Corbett Moran}, {Grudi{\'c}}  \&
  {Hopkins}}{{Corbett Moran} et~al.}{2018}]{Mor18}
{Corbett Moran} C.,  {Grudi{\'c}} M.~Y.,   {Hopkins} P.~F.,  2018, arXiv
  e-prints, \href {https://ui.adsabs.harvard.edu/abs/2018arXiv180306430C} {p.
  arXiv:1803.06430}

\bibitem[\protect\citeauthoryear{{Dale} \& {Davies}}{{Dale} \&
  {Davies}}{2006}]{Dal06}
{Dale} J.~E.,  {Davies} M.~B.,  2006, \mn@doi [\mnras]
  {10.1111/j.1365-2966.2005.09937.x}, \href
  {https://ui.adsabs.harvard.edu/abs/2006MNRAS.366.1424D} {366, 1424}

\bibitem[\protect\citeauthoryear{{Davies}, {Miller}  \& {Bellovary}}{{Davies}
  et~al.}{2011}]{Dav11}
{Davies} M.~B.,  {Miller} M.~C.,   {Bellovary} J.~M.,  2011, \mn@doi [\apjl]
  {10.1088/2041-8205/740/2/L42}, \href
  {https://ui.adsabs.harvard.edu/abs/2011ApJ...740L..42D} {740, L42}

\bibitem[\protect\citeauthoryear{{Davis}, {Clarke}  \& {Freitag}}{{Davis}
  et~al.}{2010}]{Dav10}
{Davis} O.,  {Clarke} C.~J.,   {Freitag} M.,  2010, \mn@doi [\mnras]
  {10.1111/j.1365-2966.2010.16902.x}, \href
  {https://ui.adsabs.harvard.edu/abs/2010MNRAS.407..381D} {407, 381}

\bibitem[\protect\citeauthoryear{{Dayal}, {Rossi}, {Shiralilou}, {Piana},
  {Choudhury}  \& {Volonteri}}{{Dayal} et~al.}{2019}]{Day19}
{Dayal} P.,  {Rossi} E.~M.,  {Shiralilou} B.,  {Piana} O.,  {Choudhury} T.~R.,
   {Volonteri} M.,  2019, \mn@doi [\mnras] {10.1093/mnras/stz897}, \href
  {https://ui.adsabs.harvard.edu/abs/2019MNRAS.486.2336D} {486, 2336}

\bibitem[\protect\citeauthoryear{{Demircan} \& {Kahraman}}{{Demircan} \&
  {Kahraman}}{1991}]{Dem91}
{Demircan} O.,  {Kahraman} G.,  1991, \mn@doi [\apss] {10.1007/BF00639097},
  \href {https://ui.adsabs.harvard.edu/abs/1991Ap&SS.181..313D} {181, 313}

\bibitem[\protect\citeauthoryear{{Devecchi} \& {Volonteri}}{{Devecchi} \&
  {Volonteri}}{2009}]{Dev09}
{Devecchi} B.,  {Volonteri} M.,  2009, \mn@doi [\apj]
  {10.1088/0004-637X/694/1/302}, \href
  {https://ui.adsabs.harvard.edu/abs/2009ApJ...694..302D} {694, 302}

\bibitem[\protect\citeauthoryear{{Devecchi}, {Volonteri}, {Colpi}  \&
  {Haardt}}{{Devecchi} et~al.}{2010}]{Dev10}
{Devecchi} B.,  {Volonteri} M.,  {Colpi} M.,   {Haardt} F.,  2010, \mn@doi
  [\mnras] {10.1111/j.1365-2966.2010.17363.x}, \href
  {https://ui.adsabs.harvard.edu/abs/2010MNRAS.409.1057D} {409, 1057}

\bibitem[\protect\citeauthoryear{{Dijkstra}, {Ferrara}  \&
  {Mesinger}}{{Dijkstra} et~al.}{2014}]{Dji14}
{Dijkstra} M.,  {Ferrara} A.,   {Mesinger} A.,  2014, \mn@doi [\mnras]
  {10.1093/mnras/stu1007}, \href
  {https://ui.adsabs.harvard.edu/abs/2014MNRAS.442.2036D} {442, 2036}

\bibitem[\protect\citeauthoryear{{Dopcke}, {Glover}, {Clark}  \&
  {Klessen}}{{Dopcke} et~al.}{2011}]{Dop11}
{Dopcke} G.,  {Glover} S. C.~O.,  {Clark} P.~C.,   {Klessen} R.~S.,  2011,
  \mn@doi [\apjl] {10.1088/2041-8205/729/1/L3}, \href
  {https://ui.adsabs.harvard.edu/abs/2011ApJ...729L...3D} {729, L3}

\bibitem[\protect\citeauthoryear{{Fan} et~al.,}{{Fan} et~al.}{2001}]{Fan01}
{Fan} X.,  et~al., 2001, \mn@doi [\aj] {10.1086/324111}, \href
  {https://ui.adsabs.harvard.edu/abs/2001AJ....122.2833F} {122, 2833}

\bibitem[\protect\citeauthoryear{{Ferrara}, {Salvadori}, {Yue}  \&
  {Schleicher}}{{Ferrara} et~al.}{2014}]{Fer14}
{Ferrara} A.,  {Salvadori} S.,  {Yue} B.,   {Schleicher} D.,  2014, \mn@doi
  [\mnras] {10.1093/mnras/stu1280}, \href
  {https://ui.adsabs.harvard.edu/abs/2014MNRAS.443.2410F} {443, 2410}

\bibitem[\protect\citeauthoryear{{Ferrarese} et~al.,}{{Ferrarese}
  et~al.}{2006}]{Fer06}
{Ferrarese} L.,  et~al., 2006, \mn@doi [\apjl] {10.1086/505388}, \href
  {https://ui.adsabs.harvard.edu/abs/2006ApJ...644L..21F} {644, L21}

\bibitem[\protect\citeauthoryear{{Freese}, {Ilie}, {Spolyar}, {Valluri}  \&
  {Bodenheimer}}{{Freese} et~al.}{2010}]{Free10}
{Freese} K.,  {Ilie} C.,  {Spolyar} D.,  {Valluri} M.,   {Bodenheimer} P.,
  2010, \mn@doi [\apj] {10.1088/0004-637X/716/2/1397}, \href
  {https://ui.adsabs.harvard.edu/abs/2010ApJ...716.1397F} {716, 1397}

\bibitem[\protect\citeauthoryear{{Freese}, {Rindler-Daller}, {Spolyar}  \&
  {Valluri}}{{Freese} et~al.}{2016}]{Free16}
{Freese} K.,  {Rindler-Daller} T.,  {Spolyar} D.,   {Valluri} M.,  2016,
  \mn@doi [Reports on Progress in Physics] {10.1088/0034-4885/79/6/066902},
  \href {https://ui.adsabs.harvard.edu/abs/2016RPPh...79f6902F} {79, 066902}

\bibitem[\protect\citeauthoryear{{Freitag}}{{Freitag}}{2008}]{Fre08}
{Freitag} M.,  2008, in {Beuther} H.,  {Linz} H.,   {Henning} T.,  eds,
  Astronomical Society of the Pacific Conference Series Vol. 387, Massive Star
  Formation: Observations Confront Theory. p.~247 (\mn@eprint {arXiv}
  {0711.4057})

\bibitem[\protect\citeauthoryear{{Freitag}, {Guerkan}  \& {Rasio}}{{Freitag}
  et~al.}{2007}]{Fre07}
{Freitag} M.,  {Guerkan} M.~A.,   {Rasio} F.~A.,  2007, in {St. -Louis} N.,
  {Moffat} A. F.~J.,  eds,  Astronomical Society of the Pacific Conference
  Series Vol. 367, Massive Stars in Interactive Binaries. p.~707 (\mn@eprint
  {arXiv} {astro-ph/0410327})

\bibitem[\protect\citeauthoryear{{Fujii}, {Iwasawa}, {Funato}  \&
  {Makino}}{{Fujii} et~al.}{2007}]{Fuj07}
{Fujii} M.,  {Iwasawa} M.,  {Funato} Y.,   {Makino} J.,  2007, \mn@doi [\pasj]
  {10.1093/pasj/59.6.1095}, \href
  {https://ui.adsabs.harvard.edu/abs/2007PASJ...59.1095F} {59, 1095}

\bibitem[\protect\citeauthoryear{{Gallerani}, {Fan}, {Maiolino}  \&
  {Pacucci}}{{Gallerani} et~al.}{2017}]{Gal17}
{Gallerani} S.,  {Fan} X.,  {Maiolino} R.,   {Pacucci} F.,  2017, \mn@doi
  [\pasa] {10.1017/pasa.2017.14}, \href
  {https://ui.adsabs.harvard.edu/abs/2017PASA...34...22G} {34, e022}

\bibitem[\protect\citeauthoryear{{Georgiev}, {B{\"o}ker}, {Leigh},
  {L{\"u}tzgendorf}  \& {Neumayer}}{{Georgiev} et~al.}{2016}]{Geo16}
{Georgiev} I.~Y.,  {B{\"o}ker} T.,  {Leigh} N.,  {L{\"u}tzgendorf} N.,
  {Neumayer} N.,  2016, \mn@doi [\mnras] {10.1093/mnras/stw093}, \href
  {https://ui.adsabs.harvard.edu/abs/2016MNRAS.457.2122G} {457, 2122}

\bibitem[\protect\citeauthoryear{{Giersz}, {Leigh}, {Hypki}, {L{\"u}tzgendorf}
  \& {Askar}}{{Giersz} et~al.}{2015}]{giersz15}
{Giersz} M.,  {Leigh} N.,  {Hypki} A.,  {L{\"u}tzgendorf} N.,   {Askar} A.,
  2015, \mn@doi [\mnras] {10.1093/mnras/stv2162}, \href
  {https://ui.adsabs.harvard.edu/abs/2015MNRAS.454.3150G} {454, 3150}

\bibitem[\protect\citeauthoryear{{Glebbeek}, {Gaburov}, {de Mink}, {Pols}  \&
  {Portegies Zwart}}{{Glebbeek} et~al.}{2009}]{Gle09}
{Glebbeek} E.,  {Gaburov} E.,  {de Mink} S.~E.,  {Pols} O.~R.,   {Portegies
  Zwart} S.~F.,  2009, \mn@doi [\aap] {10.1051/0004-6361/200810425}, \href
  {https://ui.adsabs.harvard.edu/abs/2009A&A...497..255G} {497, 255}

\bibitem[\protect\citeauthoryear{{Glebbeek}, {Gaburov}, {Portegies Zwart}  \&
  {Pols}}{{Glebbeek} et~al.}{2013}]{Gle13}
{Glebbeek} E.,  {Gaburov} E.,  {Portegies Zwart} S.,   {Pols} O.~R.,  2013,
  \mn@doi [\mnras] {10.1093/mnras/stt1268}, \href
  {https://ui.adsabs.harvard.edu/abs/2013MNRAS.434.3497G} {434, 3497}

\bibitem[\protect\citeauthoryear{{Graham} \& {Spitler}}{{Graham} \&
  {Spitler}}{2009}]{Gra09}
{Graham} A.~W.,  {Spitler} L.~R.,  2009, \mn@doi [\mnras]
  {10.1111/j.1365-2966.2009.15118.x}, \href
  {https://ui.adsabs.harvard.edu/abs/2009MNRAS.397.2148G} {397, 2148}

\bibitem[\protect\citeauthoryear{{Greene}, {Strader}  \& {Ho}}{{Greene}
  et~al.}{2020}]{Gre20}
{Greene} J.~E.,  {Strader} J.,   {Ho} L.~C.,  2020, \mn@doi [\araa]
  {10.1146/annurev-astro-032620-021835}, \href
  {https://ui.adsabs.harvard.edu/abs/2020ARA&A..58..257G} {58, 257}

\bibitem[\protect\citeauthoryear{{Haemmerl{\'e}}, {Woods}, {Klessen}, {Heger}
  \& {Whalen}}{{Haemmerl{\'e}} et~al.}{2018}]{Hae18}
{Haemmerl{\'e}} L.,  {Woods} T.~E.,  {Klessen} R.~S.,  {Heger} A.,   {Whalen}
  D.~J.,  2018, \mn@doi [\mnras] {10.1093/mnras/stx2919}, \href
  {https://ui.adsabs.harvard.edu/abs/2018MNRAS.474.2757H} {474, 2757}

\bibitem[\protect\citeauthoryear{{Haemmerl{\'e}}, {Mayer}, {Klessen},
  {Hosokawa}, {Madau}  \& {Bromm}}{{Haemmerl{\'e}} et~al.}{2020}]{Hae20}
{Haemmerl{\'e}} L.,  {Mayer} L.,  {Klessen} R.~S.,  {Hosokawa} T.,  {Madau} P.,
    {Bromm} V.,  2020, \mn@doi [\ssr] {10.1007/s11214-020-00673-y}, \href
  {https://ui.adsabs.harvard.edu/abs/2020SSRv..216...48H} {216, 48}

\bibitem[\protect\citeauthoryear{{Haiman}}{{Haiman}}{2004}]{Hai04}
{Haiman} Z.,  2004, \mn@doi [\apj] {10.1086/422910}, \href
  {https://ui.adsabs.harvard.edu/abs/2004ApJ...613...36H} {613, 36}

\bibitem[\protect\citeauthoryear{{Harwit}}{{Harwit}}{1988}]{Har88}
{Harwit} M.,  1988, {Astrophysical Concepts}

\bibitem[\protect\citeauthoryear{{Haster}, {Antonini}, {Kalogera}  \&
  {Mandel}}{{Haster} et~al.}{2016}]{Has16}
{Haster} C.-J.,  {Antonini} F.,  {Kalogera} V.,   {Mandel} I.,  2016, \mn@doi
  [\apj] {10.3847/0004-637X/832/2/192}, \href
  {https://ui.adsabs.harvard.edu/abs/2016ApJ...832..192H} {832, 192}

\bibitem[\protect\citeauthoryear{{Hirano}, {Hosokawa}, {Yoshida}, {Umeda},
  {Omukai}, {Chiaki}  \& {Yorke}}{{Hirano} et~al.}{2014}]{Hir14}
{Hirano} S.,  {Hosokawa} T.,  {Yoshida} N.,  {Umeda} H.,  {Omukai} K.,
  {Chiaki} G.,   {Yorke} H.~W.,  2014, \mn@doi [\apj]
  {10.1088/0004-637X/781/2/60}, \href
  {https://ui.adsabs.harvard.edu/abs/2014ApJ...781...60H} {781, 60}

\bibitem[\protect\citeauthoryear{{Hirano}, {Hosokawa}, {Yoshida}  \&
  {Kuiper}}{{Hirano} et~al.}{2017}]{Hir17}
{Hirano} S.,  {Hosokawa} T.,  {Yoshida} N.,   {Kuiper} R.,  2017, \mn@doi
  [Science] {10.1126/science.aai9119}, \href
  {https://ui.adsabs.harvard.edu/abs/2017Sci...357.1375H} {357, 1375}

\bibitem[\protect\citeauthoryear{{Hosokawa}, {Omukai}  \& {Yorke}}{{Hosokawa}
  et~al.}{2012}]{Hos12}
{Hosokawa} T.,  {Omukai} K.,   {Yorke} H.~W.,  2012, \mn@doi [\apj]
  {10.1088/0004-637X/756/1/93}, \href
  {https://ui.adsabs.harvard.edu/abs/2012ApJ...756...93H} {756, 93}

\bibitem[\protect\citeauthoryear{{Hosokawa}, {Yorke}, {Inayoshi}, {Omukai}  \&
  {Yoshida}}{{Hosokawa} et~al.}{2013}]{Hos13}
{Hosokawa} T.,  {Yorke} H.~W.,  {Inayoshi} K.,  {Omukai} K.,   {Yoshida} N.,
  2013, \mn@doi [\apj] {10.1088/0004-637X/778/2/178}, \href
  {https://ui.adsabs.harvard.edu/abs/2013ApJ...778..178H} {778, 178}

\bibitem[\protect\citeauthoryear{{Hoyle} \& {Lyttleton}}{{Hoyle} \&
  {Lyttleton}}{1939}]{Hoy39}
{Hoyle} F.,  {Lyttleton} R.~A.,  1939, \mn@doi [Proceedings of the Cambridge
  Philosophical Society] {10.1017/S0305004100021368}, \href
  {https://ui.adsabs.harvard.edu/abs/1939PCPS...35..592H} {35, 592}

\bibitem[\protect\citeauthoryear{{Hoyle} \& {Lyttleton}}{{Hoyle} \&
  {Lyttleton}}{1940a}]{Hoy40}
{Hoyle} F.,  {Lyttleton} R.~A.,  1940a, \mn@doi [Proceedings of the Cambridge
  Philosophical Society] {10.1017/S0305004100017369}, \href
  {https://ui.adsabs.harvard.edu/abs/1940PCPS...36..325H} {36, 325}

\bibitem[\protect\citeauthoryear{{Hoyle} \& {Lyttleton}}{{Hoyle} \&
  {Lyttleton}}{1940b}]{Hoy240}
{Hoyle} F.,  {Lyttleton} R.~A.,  1940b, \mn@doi [Proceedings of the Cambridge
  Philosophical Society] {10.1017/S0305004100017461}, \href
  {https://ui.adsabs.harvard.edu/abs/1940PCPS...36..424H} {36, 424}

\bibitem[\protect\citeauthoryear{{Hurley}, {Pols}  \& {Tout}}{{Hurley}
  et~al.}{2000}]{Hur00}
{Hurley} J.~R.,  {Pols} O.~R.,   {Tout} C.~A.,  2000, \mn@doi [\mnras]
  {10.1046/j.1365-8711.2000.03426.x}, \href
  {https://ui.adsabs.harvard.edu/abs/2000MNRAS.315..543H} {315, 543}

\bibitem[\protect\citeauthoryear{{Inayoshi}, {Ostriker}, {Haiman}  \&
  {Kuiper}}{{Inayoshi} et~al.}{2018a}]{Ina18}
{Inayoshi} K.,  {Ostriker} J.~P.,  {Haiman} Z.,   {Kuiper} R.,  2018a, \mn@doi
  [\mnras] {10.1093/mnras/sty276}, \href
  {https://ui.adsabs.harvard.edu/abs/2018MNRAS.476.1412I} {476, 1412}

\bibitem[\protect\citeauthoryear{{Inayoshi}, {Li}  \& {Haiman}}{{Inayoshi}
  et~al.}{2018b}]{Ina218}
{Inayoshi} K.,  {Li} M.,   {Haiman} Z.,  2018b, \mn@doi [\mnras]
  {10.1093/mnras/sty1720}, \href
  {https://ui.adsabs.harvard.edu/abs/2018MNRAS.479.4017I} {479, 4017}

\bibitem[\protect\citeauthoryear{{Inayoshi}, {Visbal}  \& {Haiman}}{{Inayoshi}
  et~al.}{2019}]{Ina19}
{Inayoshi} K.,  {Visbal} E.,   {Haiman} Z.,  2019, arXiv e-prints, \href
  {https://ui.adsabs.harvard.edu/abs/2019arXiv191105791I} {p. arXiv:1911.05791}

\bibitem[\protect\citeauthoryear{{Ishii}, {Ueno}  \& {Kato}}{{Ishii}
  et~al.}{1999}]{Ish99}
{Ishii} M.,  {Ueno} M.,   {Kato} M.,  1999, \mn@doi [\pasj]
  {10.1093/pasj/51.4.417}, \href
  {https://ui.adsabs.harvard.edu/abs/1999PASJ...51..417I} {51, 417}

\bibitem[\protect\citeauthoryear{{Janka}}{{Janka}}{2002}]{Jan02}
{Janka} H.-T.,  2002, in {Gilfanov} M.,  {Sunyeav} R.,   {Churazov} E.,  eds,
  Lighthouses of the Universe: The Most Luminous Celestial Objects and Their
  Use for Cosmology. p.~357 (\mn@eprint {arXiv} {astro-ph/0202028}),
  \mn@doi{10.1007/10856495_56}

\bibitem[\protect\citeauthoryear{{Kaaz}, {Antoni}  \& {Ramirez-Ruiz}}{{Kaaz}
  et~al.}{2019}]{Kaa19}
{Kaaz} N.,  {Antoni} A.,   {Ramirez-Ruiz} E.,  2019, \mn@doi [\apj]
  {10.3847/1538-4357/ab158b}, \href
  {https://ui.adsabs.harvard.edu/abs/2019ApJ...876..142K} {876, 142}

\bibitem[\protect\citeauthoryear{{Katz}}{{Katz}}{2019}]{Kat19}
{Katz} H.,  2019, {Black hole formation in the first stellar clusters}.
pp 125--143, \mn@doi{10.1142/9789813227958_0007}

\bibitem[\protect\citeauthoryear{{Katz}, {Sijacki}  \& {Haehnelt}}{{Katz}
  et~al.}{2015}]{Kat15}
{Katz} H.,  {Sijacki} D.,   {Haehnelt} M.~G.,  2015, \mn@doi [\mnras]
  {10.1093/mnras/stv1048}, \href
  {https://ui.adsabs.harvard.edu/abs/2015MNRAS.451.2352K} {451, 2352}

\bibitem[\protect\citeauthoryear{{Kormendy} \& {Ho}}{{Kormendy} \&
  {Ho}}{2013}]{Cor13}
{Kormendy} J.,  {Ho} L.~C.,  2013, \mn@doi [\araa]
  {10.1146/annurev-astro-082708-101811}, \href
  {https://ui.adsabs.harvard.edu/abs/2013ARA&A..51..511K} {51, 511}

\bibitem[\protect\citeauthoryear{{Kroupa}, {Subr}, {Jerabkova}  \&
  {Wang}}{{Kroupa} et~al.}{2020}]{Kro20}
{Kroupa} P.,  {Subr} L.,  {Jerabkova} T.,   {Wang} L.,  2020, \mn@doi [\mnras]
  {10.1093/mnras/staa2276}, \href
  {https://ui.adsabs.harvard.edu/abs/2020MNRAS.498.5652K} {498, 5652}

\bibitem[\protect\citeauthoryear{{Latif} \& {Ferrara}}{{Latif} \&
  {Ferrara}}{2016}]{Lat17}
{Latif} M.~A.,  {Ferrara} A.,  2016, \mn@doi [\pasa] {10.1017/pasa.2016.41},
  \href {https://ui.adsabs.harvard.edu/abs/2016PASA...33...51L} {33, e051}

\bibitem[\protect\citeauthoryear{{Latif} \& {Schleicher}}{{Latif} \&
  {Schleicher}}{2019}]{Lat19}
{Latif} M.,  {Schleicher} D.,  2019, {Formation of the First Black Holes},
  \mn@doi{10.1142/10652.
}

\bibitem[\protect\citeauthoryear{{Latif}, {Schleicher}, {Schmidt}  \&
  {Niemeyer}}{{Latif} et~al.}{2013}]{Lat13}
{Latif} M.~A.,  {Schleicher} D.~R.~G.,  {Schmidt} W.,   {Niemeyer} J.,  2013,
  \mn@doi [\mnras] {10.1093/mnras/stt834}, \href
  {https://ui.adsabs.harvard.edu/abs/2013MNRAS.433.1607L} {433, 1607}

\bibitem[\protect\citeauthoryear{{Latif}, {Bovino}, {Grassi}, {Schleicher}  \&
  {Spaans}}{{Latif} et~al.}{2015}]{Lat15}
{Latif} M.~A.,  {Bovino} S.,  {Grassi} T.,  {Schleicher} D.~R.~G.,   {Spaans}
  M.,  2015, \mn@doi [\mnras] {10.1093/mnras/stu2244}, \href
  {https://ui.adsabs.harvard.edu/abs/2015MNRAS.446.3163L} {446, 3163}

\bibitem[\protect\citeauthoryear{{Latif}, {Omukai}, {Habouzit}, {Schleicher}
  \& {Volonteri}}{{Latif} et~al.}{2016}]{Lat16}
{Latif} M.~A.,  {Omukai} K.,  {Habouzit} M.,  {Schleicher} D.~R.~G.,
  {Volonteri} M.,  2016, \mn@doi [\apj] {10.3847/0004-637X/823/1/40}, \href
  {https://ui.adsabs.harvard.edu/abs/2016ApJ...823...40L} {823, 40}

\bibitem[\protect\citeauthoryear{{Latif}, {Khochfar}  \& {Whalen}}{{Latif}
  et~al.}{2020}]{Lat20}
{Latif} M.~A.,  {Khochfar} S.,   {Whalen} D.,  2020, \mn@doi [\apjl]
  {10.3847/2041-8213/ab7c61}, \href
  {https://ui.adsabs.harvard.edu/abs/2020ApJ...892L...4L} {892, L4}

\bibitem[\protect\citeauthoryear{{Leigh} \& {Geller}}{{Leigh} \&
  {Geller}}{2012}]{leigh12b}
{Leigh} N.,  {Geller} A.~M.,  2012, \mn@doi [\mnras]
  {10.1111/j.1365-2966.2012.21689.x}, \href
  {https://ui.adsabs.harvard.edu/abs/2012MNRAS.425.2369L} {425, 2369}

\bibitem[\protect\citeauthoryear{{Leigh}, {B{\"o}ker}  \& {Knigge}}{{Leigh}
  et~al.}{2012}]{leigh12}
{Leigh} N.,  {B{\"o}ker} T.,   {Knigge} C.,  2012, \mn@doi [\mnras]
  {10.1111/j.1365-2966.2012.21365.x}, \href
  {https://ui.adsabs.harvard.edu/abs/2012MNRAS.424.2130L} {424, 2130}

\bibitem[\protect\citeauthoryear{{Leigh}, {B{\"o}ker}, {Maccarone}  \&
  {Perets}}{{Leigh} et~al.}{2013a}]{leigh13}
{Leigh} N. W.~C.,  {B{\"o}ker} T.,  {Maccarone} T.~J.,   {Perets} H.~B.,
  2013a, \mn@doi [\mnras] {10.1093/mnras/sts554}, \href
  {https://ui.adsabs.harvard.edu/abs/2013MNRAS.429.2997L} {429, 2997}

\bibitem[\protect\citeauthoryear{{Leigh}, {Sills}  \& {B{\"o}ker}}{{Leigh}
  et~al.}{2013b}]{leigh13b}
{Leigh} N.,  {Sills} A.,   {B{\"o}ker} T.,  2013b, \mn@doi [\mnras]
  {10.1093/mnras/stt862}, \href
  {https://ui.adsabs.harvard.edu/abs/2013MNRAS.433.1958L} {433, 1958}

\bibitem[\protect\citeauthoryear{{Leigh}, {Mastrobuono-Battisti}, {Perets}  \&
  {B{\"o}ker}}{{Leigh} et~al.}{2014}]{leigh14}
{Leigh} N. W.~C.,  {Mastrobuono-Battisti} A.,  {Perets} H.~B.,   {B{\"o}ker}
  T.,  2014, \mn@doi [\mnras] {10.1093/mnras/stu622}, \href
  {https://ui.adsabs.harvard.edu/abs/2014MNRAS.441..919L} {441, 919}

\bibitem[\protect\citeauthoryear{{Leigh}, {Georgiev}, {B{\"o}ker}, {Knigge}  \&
  {den Brok}}{{Leigh} et~al.}{2015}]{leigh15}
{Leigh} N. W.~C.,  {Georgiev} I.~Y.,  {B{\"o}ker} T.,  {Knigge} C.,   {den
  Brok} M.,  2015, \mn@doi [\mnras] {10.1093/mnras/stv1012}, \href
  {https://ui.adsabs.harvard.edu/abs/2015MNRAS.451..859L} {451, 859}

\bibitem[\protect\citeauthoryear{{Leigh}, {Antonini}, {Stone}, {Shara}  \&
  {Merritt}}{{Leigh} et~al.}{2016}]{Leigh16}
{Leigh} N. W.~C.,  {Antonini} F.,  {Stone} N.~C.,  {Shara} M.~M.,   {Merritt}
  D.,  2016, \mn@doi [\mnras] {10.1093/mnras/stw2018}, \href
  {https://ui.adsabs.harvard.edu/abs/2016MNRAS.463.1605L} {463, 1605}

\bibitem[\protect\citeauthoryear{{Leigh}, {Geller}, {Shara}, {Garland},
  {Clees-Baron}  \& {Ahmed}}{{Leigh} et~al.}{2017}]{Lei17}
{Leigh} N. W.~C.,  {Geller} A.~M.,  {Shara} M.~M.,  {Garland} J.,
  {Clees-Baron} H.,   {Ahmed} A.,  2017, \mn@doi [\mnras]
  {10.1093/mnras/stx1704}, \href
  {https://ui.adsabs.harvard.edu/abs/2017MNRAS.471.1830L} {471, 1830}

\bibitem[\protect\citeauthoryear{{Luo}, {Shlosman}, {Nagamine}  \&
  {Fang}}{{Luo} et~al.}{2020}]{Luo20}
{Luo} Y.,  {Shlosman} I.,  {Nagamine} K.,   {Fang} T.,  2020, \mn@doi [\mnras]
  {10.1093/mnras/staa153}, \href
  {https://ui.adsabs.harvard.edu/abs/2020MNRAS.492.4917L} {492, 4917}

\bibitem[\protect\citeauthoryear{{Lupi}, {Colpi}, {Devecchi}, {Galanti}  \&
  {Volonteri}}{{Lupi} et~al.}{2014}]{Lup14}
{Lupi} A.,  {Colpi} M.,  {Devecchi} B.,  {Galanti} G.,   {Volonteri} M.,  2014,
  \mn@doi [\mnras] {10.1093/mnras/stu1120}, \href
  {https://ui.adsabs.harvard.edu/abs/2014MNRAS.442.3616L} {442, 3616}

\bibitem[\protect\citeauthoryear{{Maccarone} \& {Zurek}}{{Maccarone} \&
  {Zurek}}{2012}]{Macc12}
{Maccarone} T.~J.,  {Zurek} D.~R.,  2012, \mn@doi [\mnras]
  {10.1111/j.1365-2966.2011.20328.x}, \href
  {https://ui.adsabs.harvard.edu/abs/2012MNRAS.423....2M} {423, 2}

\bibitem[\protect\citeauthoryear{{Madau} \& {Rees}}{{Madau} \&
  {Rees}}{2001}]{Mad01}
{Madau} P.,  {Rees} M.~J.,  2001, \mn@doi [\apjl] {10.1086/319848}, \href
  {https://ui.adsabs.harvard.edu/abs/2001ApJ...551L..27M} {551, L27}

\bibitem[\protect\citeauthoryear{{Madau}, {Haardt}  \& {Dotti}}{{Madau}
  et~al.}{2014}]{Mad14}
{Madau} P.,  {Haardt} F.,   {Dotti} M.,  2014, \mn@doi [\apjl]
  {10.1088/2041-8205/784/2/L38}, \href
  {https://ui.adsabs.harvard.edu/abs/2014ApJ...784L..38M} {784, L38}

\bibitem[\protect\citeauthoryear{{Maeder} \& {Meynet}}{{Maeder} \&
  {Meynet}}{2000}]{Mae00}
{Maeder} A.,  {Meynet} G.,  2000, \mn@doi [\araa]
  {10.1146/annurev.astro.38.1.143}, \href
  {https://ui.adsabs.harvard.edu/abs/2000ARA&A..38..143M} {38, 143}

\bibitem[\protect\citeauthoryear{{Martins}, {Schaerer}, {Haemmerl{\'e}}  \&
  {Charbonnel}}{{Martins} et~al.}{2020}]{Mar20}
{Martins} F.,  {Schaerer} D.,  {Haemmerl{\'e}} L.,   {Charbonnel} C.,  2020,
  \mn@doi [\aap] {10.1051/0004-6361/201936963}, \href
  {https://ui.adsabs.harvard.edu/abs/2020A&A...633A...9M} {633, A9}

\bibitem[\protect\citeauthoryear{{Matsuoka} et~al.,}{{Matsuoka}
  et~al.}{2018}]{Mat18}
{Matsuoka} Y.,  et~al., 2018, \mn@doi [\pasj] {10.1093/pasj/psx046}, \href
  {https://ui.adsabs.harvard.edu/abs/2018PASJ...70S..35M} {70, S35}

\bibitem[\protect\citeauthoryear{{Matsuoka} et~al.,}{{Matsuoka}
  et~al.}{2019}]{Mat19}
{Matsuoka} Y.,  et~al., 2019, \mn@doi [\apj] {10.3847/1538-4357/ab3c60}, \href
  {https://ui.adsabs.harvard.edu/abs/2019ApJ...883..183M} {883, 183}

\bibitem[\protect\citeauthoryear{{Mayer}, {Fiacconi}, {Bonoli}, {Quinn},
  {Ro{\v{s}}kar}, {Shen}  \& {Wadsley}}{{Mayer} et~al.}{2015}]{May15}
{Mayer} L.,  {Fiacconi} D.,  {Bonoli} S.,  {Quinn} T.,  {Ro{\v{s}}kar} R.,
  {Shen} S.,   {Wadsley} J.,  2015, \mn@doi [\apj]
  {10.1088/0004-637X/810/1/51}, \href
  {https://ui.adsabs.harvard.edu/abs/2015ApJ...810...51M} {810, 51}

\bibitem[\protect\citeauthoryear{{Milosavljevi{\'c}}, {Couch}  \&
  {Bromm}}{{Milosavljevi{\'c}} et~al.}{2009}]{Mil09}
{Milosavljevi{\'c}} M.,  {Couch} S.~M.,   {Bromm} V.,  2009, \mn@doi [\apjl]
  {10.1088/0004-637X/696/2/L146}, \href
  {https://ui.adsabs.harvard.edu/abs/2009ApJ...696L.146M} {696, L146}

\bibitem[\protect\citeauthoryear{{Moeckel} \& {Clarke}}{{Moeckel} \&
  {Clarke}}{2011}]{Moe11}
{Moeckel} N.,  {Clarke} C.~J.,  2011, \mn@doi [\mnras]
  {10.1111/j.1365-2966.2010.17659.x}, \href
  {https://ui.adsabs.harvard.edu/abs/2011MNRAS.410.2799M} {410, 2799}

\bibitem[\protect\citeauthoryear{{Mortlock} et~al.,}{{Mortlock}
  et~al.}{2011}]{Mor11}
{Mortlock} D.~J.,  et~al., 2011, \mn@doi [\nat] {10.1038/nature10159}, \href
  {https://ui.adsabs.harvard.edu/abs/2011Natur.474..616M} {474, 616}

\bibitem[\protect\citeauthoryear{{Naoz}, {Yoshida}  \& {Gnedin}}{{Naoz}
  et~al.}{2013}]{Nao13}
{Naoz} S.,  {Yoshida} N.,   {Gnedin} N.~Y.,  2013, \mn@doi [\apj]
  {10.1088/0004-637X/763/1/27}, \href
  {https://ui.adsabs.harvard.edu/abs/2013ApJ...763...27N} {763, 27}

\bibitem[\protect\citeauthoryear{{Natarajan}}{{Natarajan}}{2021}]{Nat21}
{Natarajan} P.,  2021, \mn@doi [\mnras] {10.1093/mnras/staa3724}, \href
  {https://ui.adsabs.harvard.edu/abs/2021MNRAS.501.1413N} {501, 1413}

\bibitem[\protect\citeauthoryear{{Neumayer}, {Seth}  \& {B{\"o}ker}}{{Neumayer}
  et~al.}{2020}]{Neu20}
{Neumayer} N.,  {Seth} A.,   {B{\"o}ker} T.,  2020, \mn@doi [\aapr]
  {10.1007/s00159-020-00125-0}, \href
  {https://ui.adsabs.harvard.edu/abs/2020A&ARv..28....4N} {28, 4}

\bibitem[\protect\citeauthoryear{{Nguyen} et~al.,}{{Nguyen}
  et~al.}{2018}]{Ngu18}
{Nguyen} D.~D.,  et~al., 2018, \mn@doi [\apj] {10.3847/1538-4357/aabe28}, \href
  {https://ui.adsabs.harvard.edu/abs/2018ApJ...858..118N} {858, 118}

\bibitem[\protect\citeauthoryear{{Nguyen} et~al.,}{{Nguyen}
  et~al.}{2019}]{Ngu19}
{Nguyen} D.~D.,  et~al., 2019, \mn@doi [\apj] {10.3847/1538-4357/aafe7a}, \href
  {https://ui.adsabs.harvard.edu/abs/2019ApJ...872..104N} {872, 104}

\bibitem[\protect\citeauthoryear{{Oh} \& {Haiman}}{{Oh} \&
  {Haiman}}{2002}]{Oh02}
{Oh} S.~P.,  {Haiman} Z.,  2002, \mn@doi [\apj] {10.1086/339393}, \href
  {https://ui.adsabs.harvard.edu/abs/2002ApJ...569..558O} {569, 558}

\bibitem[\protect\citeauthoryear{{Omukai} \& {Palla}}{{Omukai} \&
  {Palla}}{2003}]{Omu03}
{Omukai} K.,  {Palla} F.,  2003, \mn@doi [\apj] {10.1086/374810}, \href
  {https://ui.adsabs.harvard.edu/abs/2003ApJ...589..677O} {589, 677}

\bibitem[\protect\citeauthoryear{{Omukai}, {Schneider}  \& {Haiman}}{{Omukai}
  et~al.}{2008}]{Omu08}
{Omukai} K.,  {Schneider} R.,   {Haiman} Z.,  2008, \mn@doi [\apj]
  {10.1086/591636}, \href
  {https://ui.adsabs.harvard.edu/abs/2008ApJ...686..801O} {686, 801}

\bibitem[\protect\citeauthoryear{{Onoue} et~al.,}{{Onoue} et~al.}{2019}]{ono19}
{Onoue} M.,  et~al., 2019, \mn@doi [\apj] {10.3847/1538-4357/ab29e9}, \href
  {https://ui.adsabs.harvard.edu/abs/2019ApJ...880...77O} {880, 77}

\bibitem[\protect\citeauthoryear{{Pacucci} \& {Loeb}}{{Pacucci} \&
  {Loeb}}{2020}]{Pac20}
{Pacucci} F.,  {Loeb} A.,  2020, \mn@doi [\apj] {10.3847/1538-4357/ab886e},
  \href {https://ui.adsabs.harvard.edu/abs/2020ApJ...895...95P} {895, 95}

\bibitem[\protect\citeauthoryear{{Pacucci}, {Volonteri}  \&
  {Ferrara}}{{Pacucci} et~al.}{2015}]{Pac15}
{Pacucci} F.,  {Volonteri} M.,   {Ferrara} A.,  2015, \mn@doi [\mnras]
  {10.1093/mnras/stv1465}, \href
  {https://ui.adsabs.harvard.edu/abs/2015MNRAS.452.1922P} {452, 1922}

\bibitem[\protect\citeauthoryear{{Pacucci}, {Natarajan}, {Volonteri},
  {Cappelluti}  \& {Urry}}{{Pacucci} et~al.}{2017}]{Pac17}
{Pacucci} F.,  {Natarajan} P.,  {Volonteri} M.,  {Cappelluti} N.,   {Urry}
  C.~M.,  2017, \mn@doi [\apjl] {10.3847/2041-8213/aa9aea}, \href
  {https://ui.adsabs.harvard.edu/abs/2017ApJ...850L..42P} {850, L42}

\bibitem[\protect\citeauthoryear{{Pelupessy}, {van Elteren}, {de Vries},
  {McMillan}, {Drost}  \& {Portegies Zwart}}{{Pelupessy} et~al.}{2013}]{Pel13}
{Pelupessy} F.~I.,  {van Elteren} A.,  {de Vries} N.,  {McMillan} S.~L.~W.,
  {Drost} N.,   {Portegies Zwart} S.~F.,  2013, \mn@doi [\aap]
  {10.1051/0004-6361/201321252}, \href
  {https://ui.adsabs.harvard.edu/abs/2013A&A...557A..84P} {557, A84}

\bibitem[\protect\citeauthoryear{{Piana}, {Dayal}, {Volonteri}  \&
  {Choudhury}}{{Piana} et~al.}{2021}]{Day21}
{Piana} O.,  {Dayal} P.,  {Volonteri} M.,   {Choudhury} T.~R.,  2021, \mn@doi
  [\mnras] {10.1093/mnras/staa3363}, \href
  {https://ui.adsabs.harvard.edu/abs/2021MNRAS.500.2146P} {500, 2146}

\bibitem[\protect\citeauthoryear{{Plummer}}{{Plummer}}{1911}]{Plu11}
{Plummer} H.~C.,  1911, \mn@doi [\mnras] {10.1093/mnras/71.5.460}, \href
  {https://ui.adsabs.harvard.edu/abs/1911MNRAS..71..460P} {71, 460}

\bibitem[\protect\citeauthoryear{{Portegies Zwart} \& {McMillan}}{{Portegies
  Zwart} \& {McMillan}}{2002}]{Zwa02}
{Portegies Zwart} S.~F.,  {McMillan} S. L.~W.,  2002, \mn@doi [\apj]
  {10.1086/341798}, \href
  {https://ui.adsabs.harvard.edu/abs/2002ApJ...576..899P} {576, 899}

\bibitem[\protect\citeauthoryear{{Portegies Zwart} \& {McMillan}}{{Portegies
  Zwart} \& {McMillan}}{2018}]{Por18}
{Portegies Zwart} S.,  {McMillan} S.,  2018, {Astrophysical Recipes; The art of
  AMUSE}, \mn@doi{10.1088/978-0-7503-1320-9.
}

\bibitem[\protect\citeauthoryear{{Portegies Zwart}, {Baumgardt}, {Hut},
  {Makino}  \& {McMillan}}{{Portegies Zwart} et~al.}{2004}]{Zwa04}
{Portegies Zwart} S.~F.,  {Baumgardt} H.,  {Hut} P.,  {Makino} J.,   {McMillan}
  S. L.~W.,  2004, \mn@doi [\nat] {10.1038/nature02448}, \href
  {https://ui.adsabs.harvard.edu/abs/2004Natur.428..724P} {428, 724}

\bibitem[\protect\citeauthoryear{{Portegies Zwart} et~al.,}{{Portegies Zwart}
  et~al.}{2009}]{Por09}
{Portegies Zwart} S.,  et~al., 2009, \mn@doi [\na]
  {10.1016/j.newast.2008.10.006}, \href
  {https://ui.adsabs.harvard.edu/abs/2009NewA...14..369P} {14, 369}

\bibitem[\protect\citeauthoryear{{Portegies Zwart}, {McMillan}, {van Elteren},
  {Pelupessy}  \& {de Vries}}{{Portegies Zwart} et~al.}{2013}]{Por13}
{Portegies Zwart} S.,  {McMillan} S.~L.~W.,  {van Elteren} E.,  {Pelupessy} I.,
    {de Vries} N.,  2013, \mn@doi [Computer Physics Communications]
  {10.1016/j.cpc.2012.09.024}, \href
  {https://ui.adsabs.harvard.edu/abs/2013CoPhC.184..456P} {184, 456}

\bibitem[\protect\citeauthoryear{{Regan} \& {Downes}}{{Regan} \&
  {Downes}}{2018}]{Reg18}
{Regan} J.~A.,  {Downes} T.~P.,  2018, \mn@doi [\mnras]
  {10.1093/mnras/sty1289}, \href
  {https://ui.adsabs.harvard.edu/abs/2018MNRAS.478.5037R} {478, 5037}

\bibitem[\protect\citeauthoryear{{Regan}, {Visbal}, {Wise}, {Haiman},
  {Johansson}  \& {Bryan}}{{Regan} et~al.}{2017}]{Reg17}
{Regan} J.~A.,  {Visbal} E.,  {Wise} J.~H.,  {Haiman} Z.,  {Johansson} P.~H.,
  {Bryan} G.~L.,  2017, \mn@doi [Nature Astronomy] {10.1038/s41550-017-0075},
  \href {https://ui.adsabs.harvard.edu/abs/2017NatAs...1E..75R} {1, 0075}

\bibitem[\protect\citeauthoryear{{Regan}, {Downes}, {Volonteri}, {Beckmann},
  {Lupi}, {Trebitsch}  \& {Dubois}}{{Regan} et~al.}{2019}]{Reg19}
{Regan} J.~A.,  {Downes} T.~P.,  {Volonteri} M.,  {Beckmann} R.,  {Lupi} A.,
  {Trebitsch} M.,   {Dubois} Y.,  2019, \mn@doi [\mnras]
  {10.1093/mnras/stz1045}, \href
  {https://ui.adsabs.harvard.edu/abs/2019MNRAS.486.3892R} {486, 3892}

\bibitem[\protect\citeauthoryear{{Regan}, {Wise}, {Woods}, {Downes}, {O'Shea}
  \& {Norman}}{{Regan} et~al.}{2020a}]{Reg20a}
{Regan} J.~A.,  {Wise} J.~H.,  {Woods} T.~E.,  {Downes} T.~P.,  {O'Shea} B.~W.,
    {Norman} M.~L.,  2020a, arXiv e-prints, \href
  {https://ui.adsabs.harvard.edu/abs/2020arXiv200808090R} {p. arXiv:2008.08090}

\bibitem[\protect\citeauthoryear{{Regan}, {Haiman}, {Wise}, {O'Shea}  \&
  {Norman}}{{Regan} et~al.}{2020b}]{Reg20b}
{Regan} J.~A.,  {Haiman} Z.,  {Wise} J.~H.,  {O'Shea} B.~W.,   {Norman} M.~L.,
  2020b, \mn@doi [The Open Journal of Astrophysics]
  {10.21105/astro.2006.14625}, \href
  {https://ui.adsabs.harvard.edu/abs/2020OJAp....3E...9R} {3, E9}

\bibitem[\protect\citeauthoryear{{Regan}, {Wise}, {O'Shea}  \&
  {Norman}}{{Regan} et~al.}{2020c}]{Reg20}
{Regan} J.~A.,  {Wise} J.~H.,  {O'Shea} B.~W.,   {Norman} M.~L.,  2020c,
  \mn@doi [\mnras] {10.1093/mnras/staa035}, \href
  {https://ui.adsabs.harvard.edu/abs/2020MNRAS.492.3021R} {492, 3021}

\bibitem[\protect\citeauthoryear{{Reinoso}, {Schleicher}, {Fellhauer},
  {Klessen}  \& {Boekholt}}{{Reinoso} et~al.}{2018}]{Rei18}
{Reinoso} B.,  {Schleicher} D.~R.~G.,  {Fellhauer} M.,  {Klessen} R.~S.,
  {Boekholt} T.~C.~N.,  2018, \mn@doi [\aap] {10.1051/0004-6361/201732224},
  \href {https://ui.adsabs.harvard.edu/abs/2018A&A...614A..14R} {614, A14}

\bibitem[\protect\citeauthoryear{{Rizzuto} et~al.,}{{Rizzuto}
  et~al.}{2021}]{Riz21}
{Rizzuto} F.~P.,  et~al., 2021, \mn@doi [\mnras] {10.1093/mnras/staa3634},
  \href {https://ui.adsabs.harvard.edu/abs/2021MNRAS.501.5257R} {501, 5257}

\bibitem[\protect\citeauthoryear{{Rossa}, {van der Marel}, {B{\"o}ker},
  {Gerssen}, {Ho}, {Rix}, {Shields}  \& {Walcher}}{{Rossa}
  et~al.}{2006}]{Ros06}
{Rossa} J.,  {van der Marel} R.~P.,  {B{\"o}ker} T.,  {Gerssen} J.,  {Ho}
  L.~C.,  {Rix} H.-W.,  {Shields} J.~C.,   {Walcher} C.-J.,  2006, \mn@doi
  [\aj] {10.1086/505968}, \href
  {https://ui.adsabs.harvard.edu/abs/2006AJ....132.1074R} {132, 1074}

\bibitem[\protect\citeauthoryear{{Sakurai}, {Hosokawa}, {Yoshida}  \&
  {Yorke}}{{Sakurai} et~al.}{2015}]{Saku15}
{Sakurai} Y.,  {Hosokawa} T.,  {Yoshida} N.,   {Yorke} H.~W.,  2015, \mn@doi
  [\mnras] {10.1093/mnras/stv1346}, \href
  {https://ui.adsabs.harvard.edu/abs/2015MNRAS.452..755S} {452, 755}

\bibitem[\protect\citeauthoryear{{Sakurai}, {Yoshida}, {Fujii}  \&
  {Hirano}}{{Sakurai} et~al.}{2017}]{Sak17}
{Sakurai} Y.,  {Yoshida} N.,  {Fujii} M.~S.,   {Hirano} S.,  2017, \mn@doi
  [\mnras] {10.1093/mnras/stx2044}, \href
  {https://ui.adsabs.harvard.edu/abs/2017MNRAS.472.1677S} {472, 1677}

\bibitem[\protect\citeauthoryear{{Sakurai}, {Yoshida}  \& {Fujii}}{{Sakurai}
  et~al.}{2019}]{Sak19}
{Sakurai} Y.,  {Yoshida} N.,   {Fujii} M.~S.,  2019, \mn@doi [\mnras]
  {10.1093/mnras/stz315}, \href
  {https://ui.adsabs.harvard.edu/abs/2019MNRAS.484.4665S} {484, 4665}

\bibitem[\protect\citeauthoryear{{Salpeter}}{{Salpeter}}{1955}]{Sal55}
{Salpeter} E.~E.,  1955, \mn@doi [\apj] {10.1086/145971}, \href
  {https://ui.adsabs.harvard.edu/abs/1955ApJ...121..161S} {121, 161}

\bibitem[\protect\citeauthoryear{{Schaller}, {Schaerer}, {Meynet}  \&
  {Maeder}}{{Schaller} et~al.}{1992}]{Sch92}
{Schaller} G.,  {Schaerer} D.,  {Meynet} G.,   {Maeder} A.,  1992, \aaps, \href
  {https://ui.adsabs.harvard.edu/abs/1992A&AS...96..269S} {96, 269}

\bibitem[\protect\citeauthoryear{{Schleicher}, {Palla}, {Ferrara}, {Galli}  \&
  {Latif}}{{Schleicher} et~al.}{2013}]{Sch13}
{Schleicher} D. R.~G.,  {Palla} F.,  {Ferrara} A.,  {Galli} D.,   {Latif} M.,
  2013, \mn@doi [\aap] {10.1051/0004-6361/201321949}, \href
  {https://ui.adsabs.harvard.edu/abs/2013A&A...558A..59S} {558, A59}

\bibitem[\protect\citeauthoryear{{Schleicher} et~al.,}{{Schleicher}
  et~al.}{2019}]{Sch19}
{Schleicher} D.~R.~G.,  et~al., 2019, Boletin de la Asociacion Argentina de
  Astronomia La Plata Argentina, \href
  {https://ui.adsabs.harvard.edu/abs/2019BAAA...61..234S} {61, 234}

\bibitem[\protect\citeauthoryear{{Sch{\"o}del}, {Feldmeier}, {Neumayer},
  {Meyer}  \& {Yelda}}{{Sch{\"o}del} et~al.}{2014}]{Sco14}
{Sch{\"o}del} R.,  {Feldmeier} A.,  {Neumayer} N.,  {Meyer} L.,   {Yelda} S.,
  2014, \mn@doi [Classical and Quantum Gravity]
  {10.1088/0264-9381/31/24/244007}, \href
  {https://ui.adsabs.harvard.edu/abs/2014CQGra..31x4007S} {31, 244007}

\bibitem[\protect\citeauthoryear{{Scott} \& {Graham}}{{Scott} \&
  {Graham}}{2013}]{sco13}
{Scott} N.,  {Graham} A.~W.,  2013, \mn@doi [\apj]
  {10.1088/0004-637X/763/2/76}, \href
  {https://ui.adsabs.harvard.edu/abs/2013ApJ...763...76S} {763, 76}

\bibitem[\protect\citeauthoryear{{Seth}, {Ag{\"u}eros}, {Lee}  \&
  {Basu-Zych}}{{Seth} et~al.}{2008}]{Set08}
{Seth} A.,  {Ag{\"u}eros} M.,  {Lee} D.,   {Basu-Zych} A.,  2008, \mn@doi
  [\apj] {10.1086/528955}, \href
  {https://ui.adsabs.harvard.edu/abs/2008ApJ...678..116S} {678, 116}

\bibitem[\protect\citeauthoryear{{Seth}, {Neumayer}  \& {B{\"o}ker}}{{Seth}
  et~al.}{2020}]{Set20}
{Seth} A.~C.,  {Neumayer} N.,   {B{\"o}ker} T.,  2020, in {Bragaglia} A.,
  {Davies} M.,  {Sills} A.,   {Vesperini} E.,  eds,  IAU Symposium Vol. 351,
  IAU Symposium. pp 13--18 (\mn@eprint {arXiv} {1908.00022}),
  \mn@doi{10.1017/S1743921319007117}

\bibitem[\protect\citeauthoryear{{Shapiro}}{{Shapiro}}{2005}]{Shapiro05}
{Shapiro} S.~L.,  2005, \mn@doi [\apj] {10.1086/427065}, \href
  {https://ui.adsabs.harvard.edu/abs/2005ApJ...620...59S} {620, 59}

\bibitem[\protect\citeauthoryear{{Shen} et~al.,}{{Shen} et~al.}{2019}]{She19}
{Shen} Y.,  et~al., 2019, \mn@doi [\apj] {10.3847/1538-4357/ab03d9}, \href
  {https://ui.adsabs.harvard.edu/abs/2019ApJ...873...35S} {873, 35}

\bibitem[\protect\citeauthoryear{{Shlosman}, {Choi}, {Begelman}  \&
  {Nagamine}}{{Shlosman} et~al.}{2016}]{Shl16}
{Shlosman} I.,  {Choi} J.-H.,  {Begelman} M.~C.,   {Nagamine} K.,  2016,
  \mn@doi [\mnras] {10.1093/mnras/stv2700}, \href
  {https://ui.adsabs.harvard.edu/abs/2016MNRAS.456..500S} {456, 500}

\bibitem[\protect\citeauthoryear{{Sills}, {Adams}, {Davies}  \& {Bate}}{{Sills}
  et~al.}{2002}]{Sil02}
{Sills} A.,  {Adams} T.,  {Davies} M.~B.,   {Bate} M.~R.,  2002, \mn@doi
  [\mnras] {10.1046/j.1365-8711.2002.05266.x}, \href
  {https://ui.adsabs.harvard.edu/abs/2002MNRAS.332...49S} {332, 49}

\bibitem[\protect\citeauthoryear{{Smith} \& {Bromm}}{{Smith} \&
  {Bromm}}{2019}]{Smi19}
{Smith} A.,  {Bromm} V.,  2019, \mn@doi [Contemporary Physics]
  {10.1080/00107514.2019.1615715}, \href
  {https://ui.adsabs.harvard.edu/abs/2019ConPh..60..111S} {60, 111}

\bibitem[\protect\citeauthoryear{{Spitzer}}{{Spitzer}}{1969}]{Spi69}
{Spitzer} Lyman J.,  1969, \mn@doi [\apjl] {10.1086/180451}, \href
  {https://ui.adsabs.harvard.edu/abs/1969ApJ...158L.139S} {158, L139}

\bibitem[\protect\citeauthoryear{{Stacy}, {Bromm}  \& {Lee}}{{Stacy}
  et~al.}{2016}]{Sta16}
{Stacy} A.,  {Bromm} V.,   {Lee} A.~T.,  2016, \mn@doi [\mnras]
  {10.1093/mnras/stw1728}, \href
  {https://ui.adsabs.harvard.edu/abs/2016MNRAS.462.1307S} {462, 1307}

\bibitem[\protect\citeauthoryear{{Stone}, {K{\"u}pper}  \& {Ostriker}}{{Stone}
  et~al.}{2017}]{Sto17}
{Stone} N.~C.,  {K{\"u}pper} A. H.~W.,   {Ostriker} J.~P.,  2017, \mn@doi
  [\mnras] {10.1093/mnras/stx097}, \href
  {https://ui.adsabs.harvard.edu/abs/2017MNRAS.467.4180S} {467, 4180}

\bibitem[\protect\citeauthoryear{{Sugimura}, {Hosokawa}, {Yajima}, {Inayoshi}
  \& {Omukai}}{{Sugimura} et~al.}{2018}]{Sug18}
{Sugimura} K.,  {Hosokawa} T.,  {Yajima} H.,  {Inayoshi} K.,   {Omukai} K.,
  2018, \mn@doi [\mnras] {10.1093/mnras/sty1298}, \href
  {https://ui.adsabs.harvard.edu/abs/2018MNRAS.478.3961S} {478, 3961}

\bibitem[\protect\citeauthoryear{{Surace} et~al.,}{{Surace}
  et~al.}{2018}]{Sur18}
{Surace} M.,  et~al., 2018, \mn@doi [\apjl] {10.3847/2041-8213/aaf80d}, \href
  {https://ui.adsabs.harvard.edu/abs/2018ApJ...869L..39S} {869, L39}

\bibitem[\protect\citeauthoryear{{Surace}, {Zackrisson}, {Whalen}, {Hartwig},
  {Glover}, {Woods}, {Heger}  \& {Glover}}{{Surace} et~al.}{2019}]{Sur19}
{Surace} M.,  {Zackrisson} E.,  {Whalen} D.~J.,  {Hartwig} T.,  {Glover}
  S.~C.~O.,  {Woods} T.~E.,  {Heger} A.,   {Glover} S.~C.~O.,  2019, \mn@doi
  [\mnras] {10.1093/mnras/stz1956}, \href
  {https://ui.adsabs.harvard.edu/abs/2019MNRAS.488.3995S} {488, 3995}

\bibitem[\protect\citeauthoryear{{Tagawa}, {Haiman}  \& {Kocsis}}{{Tagawa}
  et~al.}{2020}]{Tag20}
{Tagawa} H.,  {Haiman} Z.,   {Kocsis} B.,  2020, \mn@doi [\apj]
  {10.3847/1538-4357/ab7922}, \href
  {https://ui.adsabs.harvard.edu/abs/2020ApJ...892...36T} {892, 36}

\bibitem[\protect\citeauthoryear{{Takeo}, {Inayoshi}  \& {Mineshige}}{{Takeo}
  et~al.}{2020}]{Tak20}
{Takeo} E.,  {Inayoshi} K.,   {Mineshige} S.,  2020, arXiv e-prints, \href
  {https://ui.adsabs.harvard.edu/abs/2020arXiv200207187T} {p. arXiv:2002.07187}

\bibitem[\protect\citeauthoryear{{Tanaka} \& {Haiman}}{{Tanaka} \&
  {Haiman}}{2009}]{Tan09}
{Tanaka} T.,  {Haiman} Z.,  2009, \mn@doi [\apj]
  {10.1088/0004-637X/696/2/1798}, \href
  {https://ui.adsabs.harvard.edu/abs/2009ApJ...696.1798T} {696, 1798}

\bibitem[\protect\citeauthoryear{{Toyouchi}, {Hosokawa}, {Sugimura}, {Nakatani}
   \& {Kuiper}}{{Toyouchi} et~al.}{2019}]{Toy19}
{Toyouchi} D.,  {Hosokawa} T.,  {Sugimura} K.,  {Nakatani} R.,   {Kuiper} R.,
  2019, \mn@doi [\mnras] {10.1093/mnras/sty3012}, \href
  {https://ui.adsabs.harvard.edu/abs/2019MNRAS.483.2031T} {483, 2031}

\bibitem[\protect\citeauthoryear{{Trac}, {Sills}  \& {Pen}}{{Trac}
  et~al.}{2007}]{Tra07}
{Trac} H.,  {Sills} A.,   {Pen} U.-L.,  2007, \mn@doi [\mnras]
  {10.1111/j.1365-2966.2007.11709.x}, \href
  {https://ui.adsabs.harvard.edu/abs/2007MNRAS.377..997T} {377, 997}

\bibitem[\protect\citeauthoryear{{Umeda}, {Hosokawa}, {Omukai}  \&
  {Yoshida}}{{Umeda} et~al.}{2016}]{Ume16}
{Umeda} H.,  {Hosokawa} T.,  {Omukai} K.,   {Yoshida} N.,  2016, \mn@doi
  [\apjl] {10.3847/2041-8205/830/2/L34}, \href
  {https://ui.adsabs.harvard.edu/abs/2016ApJ...830L..34U} {830, L34}

\bibitem[\protect\citeauthoryear{{Vesperini}, {McMillan}, {D'Ercole}  \&
  {D'Antona}}{{Vesperini} et~al.}{2010}]{Ves10}
{Vesperini} E.,  {McMillan} S. L.~W.,  {D'Ercole} A.,   {D'Antona} F.,  2010,
  \mn@doi [\apjl] {10.1088/2041-8205/713/1/L41}, \href
  {https://ui.adsabs.harvard.edu/abs/2010ApJ...713L..41V} {713, L41}

\bibitem[\protect\citeauthoryear{{Vink}, {de Koter}  \& {Lamers}}{{Vink}
  et~al.}{2001}]{Vin01}
{Vink} J.~S.,  {de Koter} A.,   {Lamers} H.~J.~G.~L.~M.,  2001, \mn@doi [\aap]
  {10.1051/0004-6361:20010127}, \href
  {https://ui.adsabs.harvard.edu/abs/2001A&A...369..574V} {369, 574}

\bibitem[\protect\citeauthoryear{{Visbal}, {Haiman}  \& {Bryan}}{{Visbal}
  et~al.}{2014}]{Vis14}
{Visbal} E.,  {Haiman} Z.,   {Bryan} G.~L.,  2014, \mn@doi [\mnras]
  {10.1093/mnras/stu1794}, \href
  {https://ui.adsabs.harvard.edu/abs/2014MNRAS.445.1056V} {445, 1056}

\bibitem[\protect\citeauthoryear{{Vito} et~al.,}{{Vito} et~al.}{2019}]{Vit19}
{Vito} F.,  et~al., 2019, \mn@doi [\aap] {10.1051/0004-6361/201936217}, \href
  {https://ui.adsabs.harvard.edu/abs/2019A&A...630A.118V} {630, A118}

\bibitem[\protect\citeauthoryear{{Volonteri}}{{Volonteri}}{2010}]{Vol10}
{Volonteri} M.,  2010, \mn@doi [\aapr] {10.1007/s00159-010-0029-x}, \href
  {https://ui.adsabs.harvard.edu/abs/2010A&ARv..18..279V} {18, 279}

\bibitem[\protect\citeauthoryear{{Volonteri}}{{Volonteri}}{2012}]{Vol12}
{Volonteri} M.,  2012, \mn@doi [Science] {10.1126/science.1220843}, \href
  {https://ui.adsabs.harvard.edu/abs/2012Sci...337..544V} {337, 544}

\bibitem[\protect\citeauthoryear{{Volonteri} \& {Rees}}{{Volonteri} \&
  {Rees}}{2005}]{Vol05}
{Volonteri} M.,  {Rees} M.~J.,  2005, \mn@doi [\apj] {10.1086/466521}, \href
  {https://ui.adsabs.harvard.edu/abs/2005ApJ...633..624V} {633, 624}

\bibitem[\protect\citeauthoryear{{Volonteri} \& {Rees}}{{Volonteri} \&
  {Rees}}{2006}]{Vol06}
{Volonteri} M.,  {Rees} M.~J.,  2006, \mn@doi [\apj] {10.1086/507444}, \href
  {https://ui.adsabs.harvard.edu/abs/2006ApJ...650..669V} {650, 669}

\bibitem[\protect\citeauthoryear{{Volonteri}, {Silk}  \& {Dubus}}{{Volonteri}
  et~al.}{2015}]{Vol15}
{Volonteri} M.,  {Silk} J.,   {Dubus} G.,  2015, \mn@doi [\apj]
  {10.1088/0004-637X/804/2/148}, \href
  {https://ui.adsabs.harvard.edu/abs/2015ApJ...804..148V} {804, 148}

\bibitem[\protect\citeauthoryear{{Volonteri}, {Bogdanovi{\'c}}, {Dotti}  \&
  {Colpi}}{{Volonteri} et~al.}{2016}]{Vol16}
{Volonteri} M.,  {Bogdanovi{\'c}} T.,  {Dotti} M.,   {Colpi} M.,  2016, \mn@doi
  [IAU Focus Meeting] {10.1017/S1743921316005366}, \href
  {https://ui.adsabs.harvard.edu/abs/2016IAUFM..29B.285V} {29B, 285}

\bibitem[\protect\citeauthoryear{{Vorobyov} \& {Basu}}{{Vorobyov} \&
  {Basu}}{2006}]{Vor06}
{Vorobyov} E.~I.,  {Basu} S.,  2006, \mn@doi [\apj] {10.1086/507320}, \href
  {https://ui.adsabs.harvard.edu/abs/2006ApJ...650..956V} {650, 956}

\bibitem[\protect\citeauthoryear{{Vorobyov} \& {Basu}}{{Vorobyov} \&
  {Basu}}{2015}]{Vor15}
{Vorobyov} E.~I.,  {Basu} S.,  2015, \mn@doi [\apj]
  {10.1088/0004-637X/805/2/115}, \href
  {https://ui.adsabs.harvard.edu/abs/2015ApJ...805..115V} {805, 115}

\bibitem[\protect\citeauthoryear{{Wang} et~al.,}{{Wang} et~al.}{2019}]{Wan19}
{Wang} R.,  et~al., 2019, \mn@doi [\apj] {10.3847/1538-4357/ab4d4b}, \href
  {https://ui.adsabs.harvard.edu/abs/2019ApJ...887...40W} {887, 40}

\bibitem[\protect\citeauthoryear{{Willott} et~al.,}{{Willott}
  et~al.}{2010}]{Will10}
{Willott} C.~J.,  et~al., 2010, \mn@doi [\aj] {10.1088/0004-6256/139/3/906},
  \href {https://ui.adsabs.harvard.edu/abs/2010AJ....139..906W} {139, 906}

\bibitem[\protect\citeauthoryear{{Wise}, {Regan}, {O'Shea}, {Norman}, {Downes}
  \& {Xu}}{{Wise} et~al.}{2019}]{Wis19}
{Wise} J.~H.,  {Regan} J.~A.,  {O'Shea} B.~W.,  {Norman} M.~L.,  {Downes}
  T.~P.,   {Xu} H.,  2019, \mn@doi [\nat] {10.1038/s41586-019-0873-4}, \href
  {https://ui.adsabs.harvard.edu/abs/2019Natur.566...85W} {566, 85}

\bibitem[\protect\citeauthoryear{{Wolcott-Green}, {Haiman}  \&
  {Bryan}}{{Wolcott-Green} et~al.}{2017}]{Wol17}
{Wolcott-Green} J.,  {Haiman} Z.,   {Bryan} G.~L.,  2017, \mn@doi [\mnras]
  {10.1093/mnras/stx167}, \href
  {https://ui.adsabs.harvard.edu/abs/2017MNRAS.469.3329W} {469, 3329}

\bibitem[\protect\citeauthoryear{{Woods}, {Heger}, {Whalen}, {Haemmerl{\'e}}
  \& {Klessen}}{{Woods} et~al.}{2017}]{Woo17}
{Woods} T.~E.,  {Heger} A.,  {Whalen} D.~J.,  {Haemmerl{\'e}} L.,   {Klessen}
  R.~S.,  2017, \mn@doi [\apjl] {10.3847/2041-8213/aa7412}, \href
  {https://ui.adsabs.harvard.edu/abs/2017ApJ...842L...6W} {842, L6}

\bibitem[\protect\citeauthoryear{{Woods}, {Heger}  \& {Haemmerl{\'e}}}{{Woods}
  et~al.}{2020}]{Woo20}
{Woods} T.~E.,  {Heger} A.,   {Haemmerl{\'e}} L.,  2020, \mn@doi [\mnras]
  {10.1093/mnras/staa763}, \href
  {https://ui.adsabs.harvard.edu/abs/2020MNRAS.494.2236W} {494, 2236}

\bibitem[\protect\citeauthoryear{{Wu} et~al.,}{{Wu} et~al.}{2015}]{Wu15}
{Wu} X.-B.,  et~al., 2015, \mn@doi [\nat] {10.1038/nature14241}, \href
  {https://ui.adsabs.harvard.edu/abs/2015Natur.518..512W} {518, 512}

\bibitem[\protect\citeauthoryear{{Yoshida}, {Omukai}  \& {Hernquist}}{{Yoshida}
  et~al.}{2008}]{Yos08}
{Yoshida} N.,  {Omukai} K.,   {Hernquist} L.,  2008, \mn@doi [Science]
  {10.1126/science.1160259}, \href
  {https://ui.adsabs.harvard.edu/abs/2008Sci...321..669Y} {321, 669}

\makeatother
\end{thebibliography}





\bsp	
\label{lastpage}
\end{document}